\documentclass[12pt,a4paper,aps,preprint,superscriptaddress,nofootinbib,longbibliography]{revtex4-1} %
\pdfoutput=1
\usepackage{graphicx}
\usepackage{cancel}
\usepackage{amssymb}
\usepackage{textcomp}
\usepackage{amsmath}
\usepackage{bm}
\usepackage{times}
\usepackage{epsfig}
\usepackage{epstopdf}
\usepackage{colortbl}
\usepackage[dvipsnames]{xcolor} 
\usepackage{multirow}
\usepackage{hyperref}
\usepackage[utf8]{inputenc}
\usepackage{slashed}
\allowdisplaybreaks[4]
 \usepackage[titletoc]{appendix}

\usepackage[compact]{titlesec}
\titlespacing{\section}{0pt}{2ex}{1ex}
\titlespacing{\subsection}{0pt}{1ex}{0ex}
\titlespacing{\subsubsection}{0pt}{0.5ex}{0ex}

\def\lsim{\mathrel{\raise.3ex\hbox{$<$\kern-.75em\lower1ex\hbox{$\sim$}}}}
\def\gsim{\mathrel{\raise.3ex\hbox{$>$\kern-.75em\lower1ex\hbox{$\sim$}}}}

\newcommand{\T}{{\rm T}}

\newcommand{\calO}{{\cal O}}
\newcommand{\calL}{{\cal L}}

\begin{document}
\hspace*{\fill}

\title{An EFT toolbox for baryon and lepton number violating dinucleon to dilepton decays}

\author{Xiao-Gang He}\email{hexg@phys.ntu.edu.tw}
\affiliation{Tsung-Dao Lee Institute, and School of Physics and Astronomy, Shanghai Jiao Tong University, 200240, China}
\affiliation{Department of Physics, National Taiwan University, Taipei 106, Taiwan}
\affiliation{Physics Division, National Center for Theoretical Sciences, Hsinchu 300, Taiwan}

\author{Xiao-Dong Ma}\email{maxid@sjtu.edu.cn}
\affiliation{Tsung-Dao Lee Institute, and School of Physics and Astronomy, Shanghai Jiao Tong University, 200240, China}
\affiliation{Department of Physics, National Taiwan University, Taipei 106, Taiwan}
\affiliation{School of Nuclear Science and Technology, Lanzhou University, Lanzhou 730000, China}

\begin{abstract}

In this paper we systematically consider the baryon ($B$) and lepton ($L$) number violating dinucleon to dilepton decays ($pp\to \ell^+\ell^{\prime+}, pn\to \ell^+\bar\nu^\prime, nn\to \bar\nu\bar\nu^\prime$) with $\Delta B=\Delta L=-2$ in the framework of effective field theory. We start by constructing a basis of dimension-12 (dim-12) operators mediating such processes in the low energy effective field theory (LEFT) below the electroweak scale. Then we consider their standard model effective field theory (SMEFT) completions upwards and their chiral realizations in baryon chiral perturbation theory (B$\chi$PT) downwards. We work to the first nontrivial orders in each effective field theory, collect along the way the matching conditions, and express the decay rates in terms of the Wilson coefficients associated with the dim-12 operators in the SMEFT and the low energy constants pertinent to B$\chi$PT. We find the current experimental limits push the associated new physics scale larger than $1-3$ TeV, which is still accessible to the future collider searches. Through weak isospin symmetry, we find the current experimental limits on the partial lifetime of transitions $pp\to \ell^+\ell^{\prime+}, pn\to \ell^+\bar\nu^\prime$ imply stronger limits on $nn\to\bar\nu\bar\nu^\prime$ than their existing lower bounds, which are improved by $2-3$ orders of magnitude. Furthermore, assuming charged mode transitions are also dominantly generated by the similar dim-12 SMEFT interactions, the experimental limits on $pp\to e^+e^+,e^+\mu^+,\mu^+\mu^+$ lead to stronger limits on $pn\to \ell^+_\alpha\bar\nu_\beta$ with $\alpha,\beta=e,\mu$ than their existing bounds. Conversely, the same assumptions help us to set a lower bound on the lifetime of the experimentally unsearched mode $pp\to e^+\tau^+$ from that of $pn\to e^+\bar\nu_\tau$, i.e., $\Gamma^{-1}_{pp\to e^+\tau^+}\gtrsim 2\times 10^{34}~\rm yr$.

\end{abstract}

\maketitle
\thispagestyle{empty}

{\small\tableofcontents}

\setcounter{page}{1}

\section{Introduction}
\label{sec1}

The observed matter-antimatter asymmetry of the universe requires the violation of baryon number, which as one of the three Sakharov conditions for a successful baryogenesis mechanism~\cite{Sakharov:1967dj}; at the same time, the extremely possible Majorana nature of neutrinos breaks the lepton number. Both facts lead to the existence of a class of new physics scenarios beyond the minimal standard model (SM) in which the baryon and/or lepton numbers are violated explicitly or spontaneously to a certain degree, this is because the baryon ($B$) and lepton ($L$) numbers are accidental global symmetries in the SM but are violated through the quantum anomaly at an unobservable level~\cite{tHooft:1976rip}.\footnote{The difference $B-L$ is still an exact symmetry survived from the anomaly cancellation in the SM.} On the other hand, the violation of baryon and/or lepton numbers would be expected in the grand unified theories (GUT)~\cite{Georgi:1974sy,Babu:1993we}. Thus, the search of rare baryon and/or lepton number violation signals becomes more and more important than ever before for the pursuit of new physics (NP). 

Usually, the experimentally accessible baryon number violation signatures can be categorized into two classes in terms of the net baryon number being changed by one unit ($\Delta B=1$) or two units ($\Delta B=2$).
For the $\Delta B=1$ case, the relevant processes are the single free or bound nucleon decays, which could change the lepton numbers by $\Delta L=\pm 1$ or $\pm 3$ units, like the most well-known proton decay mode $p\to e^+\pi^0$, etc, see Ref.~\cite{Heeck:2019kgr} and references therein for a thorough discussion on these $\Delta B=1$ processes. Such processes have been searched for experimentally for a long time but with null results~\cite{Zyla:2020zbs}, which, however, tightly constrain the NP scenarios and push the NP scale around the GUT scale unaccessible directly for the current and future collider experiments.  
 
On the other hand, for the $\Delta B=2$ case, the net lepton numbers can be changed by either $\Delta L =0$ or $\Delta L =\pm2$ units.\footnote{For the processes with other larger even units of lepton numbers ($\Delta L =\pm4, \pm 6, \cdots$), their effect would be expected to be severely suppressed due to higher dimensionality of the relevant interactions~\cite{Kobach:2016ami} and thus omitted here.} They are interesting because they can be searched for experimentally with clean signatures and high precision, as well as because there exist NP scenarios in which their contributions are dominant but that of $\Delta B=1$ processes like proton decay are suppressed~\cite{Nussinov:2001rb,Arnold:2012sd,Dev:2015uca, Gardner:2018azu,Girmohanta:2019fsx}. The neutron-antineutron oscillation ($n{\rm-}\bar n$) is a representative example for the $\Delta L =0$ case, which has attracted a lot of attention in recent years both theoretically and experimentally, see the review~\cite{Phillips:2014fgb} and references therein for a summary on the state of the art of this process. For the $\Delta L =\pm2$ case, the interesting processes are the dinucleon to dilepton decays in nuclei, including $pp\to \ell^+\ell^{\prime+}$, $pn\to \ell^+\bar\nu^\prime$ and $nn\to \bar\nu\bar\nu^\prime$ with $\ell(\ell')=e,\mu,\tau$ and $\bar\nu(\bar\nu^\prime)=\bar\nu_e,\bar\nu_\mu, \bar\nu_\tau$, respectively.\footnote{There are also $\Delta L =0$ dinucleon decays with the final states being a pair of leptons conserving lepton flavor and number or two mesons, like $pn\to \ell^+\nu$, $nn\to \nu\bar\nu$ and  $pp\to \pi^+\pi^+, K^+K^+$~\cite{Girmohanta:2019xya,Girmohanta:2019cjm}.} Such processes get less attention than the above mentioned $\Delta B=1$ and $n{\rm-}\bar n$ oscillation processes~\cite{Arnold:2012sd,Bramante:2014uda, Girmohanta:2019fsx,Girmohanta:2020eav}, but they may open new avenues for the baryon number violation signals due to their distinct theoretical origin and clear experimental signatures.\footnote{A timely brief status report of the $\Delta B=2$ physics can be found in~\cite{Babu:2020nnh}.} The most stringent lower limits on the partial lifetime of those dinucleon to dilepton decays in oxygen $^{16}$O and carbon $^{12}$C nuclei are reported by the earlier Frejus and KamLAND experiments~\cite{Berger:1991fa,Araki:2005jt} and the recent Super-K results~~\cite{Sussman:2018ylo,Takhistov:2015fao}, and are collected in Tab.~\ref{tab:exp} for our latter use. 
\begin{table}
\centering
\resizebox{\linewidth}{!}{
\begin{tabular}{|l |c| l |c| l |c|}
\hline
Decay mode & Lifetime limit & Decay mode &Lifetime limit   & Decay mode & Lifetime limit
\\\hline
$pp\to e^+e^+$ & $4.2\times 10^{33}$~{\rm yr}~\cite{Sussman:2018ylo}
& $pn\to e^+\bar\nu^\prime$ & $2.6\times 10^{32}$~{\rm yr}~\cite{Takhistov:2015fao}
& $nn\to \bar\nu\bar\nu^\prime$ & $1.4\times 10^{30}$~{\rm yr}~\cite{Araki:2005jt}
\\
$pp\to e^+\mu^+$ & $4.4\times 10^{33}$~{\rm yr}~\cite{Sussman:2018ylo}
& $pn\to \mu^+\bar\nu^\prime$ & $2.2\times 10^{32}$~{\rm yr}~\cite{Takhistov:2015fao}
& &
\\
$pp\to \mu^+\mu^+$ & $4.4\times 10^{33}$~{\rm yr}~\cite{Sussman:2018ylo}
&  $pn\to \tau^+\bar \nu^\prime$ & $2.9\times 10^{31}$~{\rm yr}~\cite{Takhistov:2015fao}
&  & 
\\
$pp\to e^+\tau^+$ & -
& & & & 
\\\hline
\end{tabular}
}
\caption{ The lower bound on the partial lifetime of the dinucleon to dilepton transitions. Where the limits for the charged modes $(pp)$ and $(pn)$ were set per oxygen nucleus $^{16}$O in water Cherenkov Super-K experiment at $90\%$ confidence level~\cite{Takhistov:2015fao,Sussman:2018ylo} while the limit on the neutral mode $(nn)$ is obtained from the carbon nucleus $^{12}$C in KamLAND experiment~\cite{Araki:2005jt}. The bounds on modes involving anti-neutrinos are also valid for the cases with neutrinos since they are treated as invisible states in the experimental search.  }
\label{tab:exp}
\end{table}

\begin{figure}
\centering
\includegraphics[width=17cm]{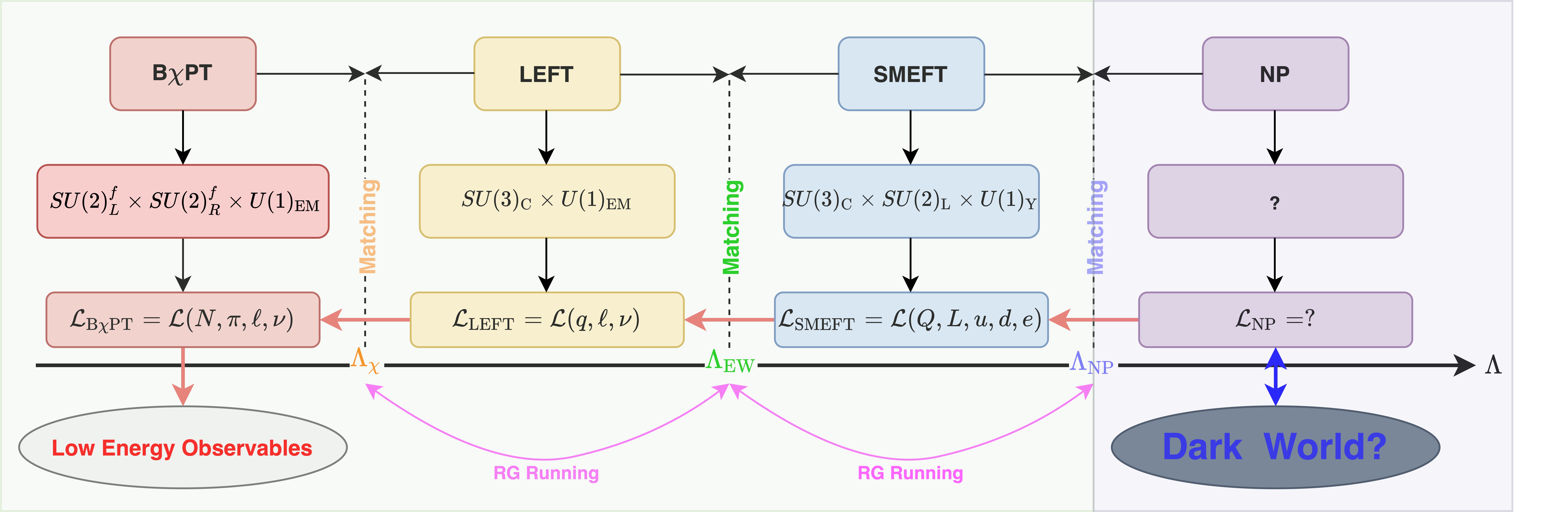}
\caption{The flow chart for an EFT calculation of the low energy observables.}
\label{Fig1}
\end{figure}

Confronted with the relatively less studies on those baryon and lepton number violating $\Delta B=\Delta L=-2$ decays $pp\to \ell^+\ell^{\prime+}$, $pn\to \ell^+\bar\nu^\prime$ and $nn\to \bar\nu\bar\nu^\prime$, it is our goal in this paper to make a comprehensive analysis via the model-independent framework of effective field theory (EFT). Our approach is pictorially explained in Fig.~\ref{Fig1} which shows the series of EFTs relevant to the decays, including the relevant degrees of freedom and symmetries in each EFT, and the matching and renormalization group running procedures among different EFTs.  We start by constructing a basis mediating such processes using the first two flavors of quarks $(u,d)$ and the charged and neutral leptons $(\ell,\nu)$  that enjoys the QCD and QED gauge symmetries $SU(3)_{\rm C}\times U(1)_{\rm EM}$ in the low energy effective field theory (LEFT) below the electroweak scale $\Lambda_{\rm EW}$. At leading order in the LEFT, the contribution arises from effective interactions of dimension-12 (dim-12) $\Delta B=\Delta L=-2$ operators that involve six quark fields and two lepton fields $(qqqqqqll)$, where $q$ denotes $u,d$ while $l$ stands for $\ell, \nu$. By crossing symmetry, those operators also parametrize the general interactions  contributing to the baryon and lepton number violating conversion processes $e^-p\to e^+\bar p\;,\;\bar\nu \bar n$ and  $e^- n\to \bar \nu\bar p$ in the electron-deuteron ($e$-d) scattering~\cite{Gardner:2018azu}, the neutron exotic decay $n\to \bar pe^+\bar\nu$~\cite{He:2021sbl} and the hydrogen-antihydrogen (${\rm H} {\rm -}\bar {\rm H}$) oscillation~\cite{Feinberg:1978sd}. 

To translate the experimental constraints at low energy to those on NP at a high scale $\Lambda_{\rm NP}$, we have to climb up the ladder of energy scales in Fig.~\ref{Fig1}. If there are no new particles with a mass at or below $\Lambda_{\rm EW}$ , the standard model effective field theory (SMEFT) defined between some NP scale $\Lambda_{\rm NP}$ and the electroweak scale $\Lambda_{\rm EW}$ is a good starting point to parametrize the UV NP in a model-independent way. Based on this logic, we then consider the leading order SMEFT completions of the LEFT interactions in question and it happens that the relevant operators also first appear at dim 12, and without counting lepton flavors we obtain 29 independent operators contributing to the decays. Along the way we perform a tree-level matching between the SMEFT and LEFT interactions at $\Lambda_{\rm EW}$, and it turns out that the interaction structures simplify significantly due to the constraints of the SM gauge symmetry $SU(3)_{\rm C}\times SU(2)_{\rm L}\times U(1)_{\rm Y}$  and many LEFT Wilson coefficients vanish at this leading order. 

To calculate the transition rate, at low energy which we take to be the chiral symmetry breaking scale $\Lambda_{\chi}$, we perform the non-perturbative matching of the LEFT interactions of quarks to those of nucleons and mesons using the two-flavor baryon chiral perturbation theory (B$\chi$PT) formalism~\cite{Gasser:1987rb,Jenkins:1990jv,Scherer:2002tk}. In order to realize this, we first organize the LEFT operators into irreducible representations under the chiral group $SU(2)_{\rm L}^f\times SU(2)_{\rm R}^f$ of chiral $u, d$ quarks and then construct the corresponding hadronic operators using the spurion techniques with the non-perturbative QCD effect being encoded in the so-called low energy constants (LECs). These LECs may be extracted by chiral symmetry from other measured processes, or computed in lattice theory (LQCD), or estimated based on the naive dimensional analysis~\cite{Manohar:1983md, Gavela:2016bzc, Weinberg:1989dx}. 

Last we collect all pieces together to express the transition rate as the function of the Wilson coefficients of effective interactions in the SMEFT and the LECs. The merit in such an approach is that the uncertainties incurred in the result may be estimated systematically. Taking the current experimental limits into consideration, we find the NP scale is pushed around 1-3 TeV, which is accessible to the future collider searches to  complementarily study the relevant NP. If we further assume these transitions are dominantly generated by the similar dim-12 SMEFT operators, then the three different types of transitions are correlated with each other via the weak isospin symmetry, and the experimental limits on the partial lifetime of one type of transitions may be translated into stronger limits on that of another type than their existing experimental lower bounds. In addition, this correlation helps us to set a lower bound on the lifetime of the experimentally unsearched decay mode $pp\to e^+\tau^+$ from that of $pn\to e^+\bar\nu_\tau$, and we find the partial lifetime is constrained to be $\Gamma^{-1}_{pp\to e^+\tau^+}\gtrsim2\times 10^{34}~\rm yr$. 

This paper is organized as follows.  In section~\ref{quarklevel:ope}, we will first list at quark level the operators for dinucleon and dilepton transitions. We then establish the basis of dim-12 operators contributing to the dinucleon to dilepton transitions with $\Delta B=\Delta L=-2$ in the LEFT in subsection ~\ref{LEFT:ope}, and we will do a similar job to establish the leading order dim-12 operators contributing to the same transitions in the SMEFT in subsection~\ref{SMEFT:ope}. In subsection~\ref{SMEFT:matching}, we give tree-level matching relations between LEFT and SMEFT operators. Section~\ref{sec:chiral} is devoted to the chiral realization of the six-part of these dim-12 operators at $\Lambda_{\chi}$. After some brief explanation how to perform matches to chiral theory from quark level operators in subsection~\ref{chiral:basics}, in subsections~\ref{chiral:irrep} and~\ref{chiral:matching}, we discuss the chiral irreducible representation decompositions and matching operators. In section~\ref{sec:cal}, we calculate the decay rate from the series of EFTs and make predictions for the sensitivity of such processes for future experimental searches. We summarize our main results in section~\ref{sec:con}. In Appendices~\ref{app:colortensor} to~\ref{app:commatching} we list the operator basis in each EFT obtained and chiral decompositions and matching leading to the main results described in the text.

\section{Operator basis for $pp\to \ell^+\ell^{\prime+},pn\to \ell^+\bar\nu^\prime, nn\to \bar\nu\bar\nu^\prime$ transitions}
\label{quarklevel:ope}
Since all operators we are interested in violate baryon and lepton numbers and are thus non-Hermitian, we only list one half of them relevant to this work, and the other half is easily obtained by Hermitian conjugation. To have dinucleon to dilepton transitions of $\Delta B=\Delta L=-2$, the quark level operators first appear at dim 12 and contain six quark fields made out of the up and down quarks $(u,d)$ and two leptons ($\ell,\nu$). By the Fierz transformation, these operators can be factorized as the pure quark sectors multiplied  by the proper lepton currents. At the hadron level, according to the initial state nucleons or the final state leptons, by the electric charge conservation, we have three different types of transitions: $pp\to \ell^+\ell^{\prime+}$, $pn\to \ell^+\bar\nu^\prime$ and $nn\to \bar\nu\bar\nu^\prime$, respectively. The corresponding lepton currents for each type of transitions are defined as follows
\begin{align}
pp\to \ell\ell^\prime:~~
&j_{S,\pm}^{\ell\ell^\prime}=(\ell^\T CP_\pm \ell^\prime)\;, 
&&j_{V}^{\ell\ell^\prime,\mu}=(\ell_R^\T C\gamma^\mu \ell_L^\prime)\;,
&&j_{T,\pm}^{\ell\ell^\prime,\mu\nu}=(\ell^\T C\sigma^{\mu\nu}P_{\pm}\ell^\prime)\;,
\label{lep:curpp} 
\\
pn\to \ell\bar\nu^\prime:~~
&j_{S}^{\ell\nu^\prime}=(\ell^\T_L C \nu_L^\prime)\;, 
&&j_{V}^{\ell\nu^\prime,\mu}=(\ell^\T_R C\gamma^\mu \nu_L^\prime)\;,
&&j_{T}^{\ell\nu^\prime,\mu\nu}=(\ell^\T_L C\sigma^{\mu\nu}\nu_L^\prime)\;,
\label{lep:curpn} 
\\
nn\to \bar\nu\bar\nu^\prime:~~
&j_{S}^{\nu\nu^\prime}=(\nu^\T_L C\nu^\prime_L)\;, 
&&j_{T}^{\nu\nu^\prime,\mu\nu}=(\nu^\T_L C\sigma^{\mu\nu}\nu^\prime_L)\;,
\label{lep:curnn} 
\end{align}
where $C$ is the charge conjugation matrix satisfying $C^{\rm T}=C^\dagger=-C$ and $C^2=-1$, the charged lepton field is denoted by $\ell,\ell^\prime\in\{e,\mu,\tau\}$ and the SM left-handed neutrinos by $\nu_L,\nu^\prime_L\in\{\nu_e,\nu_\mu,\nu_\tau\}$.\footnote{We do not include the right-handed neutrinos since they are absent in the SM framework.} The chiral projection operators are abbreviated as $P_{\pm}\equiv P_{R,L}=(1\pm\gamma_5)/2$ and therefore $\ell_{R,L}=P_{R,L}\ell=P_\pm\ell$. Here we see that for the diproton and dineutron transitions $pp\to\ell^+\ell^{\prime+}$, $nn\to\bar\nu\bar\nu^{\prime}$, the scalar lepton current is symmetric under the exchange of the two leptons while the tensor lepton current is anti-symmetric.  Especially, for the transitions with identical charged leptons $pp\to e^+e^+, \mu^+\mu^+$ the tensor lepton current vanishes, and the non-vanishing scalar and vector currents can be equivalently parametrized by the four-component Dirac fields as 
\begin{align}
j_{S}&=(\ell^\T C\ell)\;, 
&j_{S,5}&=(\ell^\T C\gamma_5\ell)\;, 
&j_{V,5}^\mu&=(\ell^\T C\gamma^\mu\gamma_5 \ell)\;,
\end{align}
and the vector current $(\ell^\T C\gamma^\mu\ell)=0$ from the same reason as the tensor current. 

For the dinucleon to dilepton transitions, from the above classification, we may first write down a minimal hadron level effective Lagrangian consisting of nucleon currents $N^\T\Gamma N^\prime$ ($N,N^\prime\in\{p,n\}$) as well as the lepton currents in Eqs.~(\ref{lep:curpp}-\ref{lep:curnn}),
\begin{align}
{\cal L}_{NN^\prime}=\sum_{a} C^{(pp)}_a\calO^{(pp)}_a +\sum_{b}C^{(pn)}_b\calO^{(pn)}_b+\sum_{c} C^{(nn)}_c\calO^{(nn)}_c \;,
\label{chpt:LO}
\end{align}
where $a,b,c$ are just labels to distinguish different operators, and $C^{(NN^\prime)}_{a,b,c}$ are the Wilson coefficients associated with the relevant hadronic operators $\calO^{(NN^\prime)}_{a,b,c}$. Those operators $\calO^{(NN^\prime)}_i$ are dim-6 and composed of dinucleon and the above lepton currents, and can be written as
\begin{align}\nonumber
pp\to\ell^+\ell^{\prime+}:~
&{\calO}^{(pp)S}_{L}=(p^\T Cp)(\ell^\T_L C\ell^\prime_L)\;,
&&{\calO}^{(pp)S}_{5L}=(p^\T C\gamma_5p)(\ell^\T_L C\ell^\prime_L)\;,
\\\nonumber
&{\calO}^{(pp)S}_{R}=(p^\T Cp)(\ell^\T_R C\ell^\prime_R)\;,
&&{\calO}^{(pp)S}_{5R}=(p^\T C\gamma_5p)(\ell^\T_R C\ell^\prime_R)\;,
\\
&{\calO}^{(pp)V}=(p^\T C\gamma_\mu\gamma_5 p)(\ell^\T_R C\gamma^\mu\ell^\prime_L)\;,
\label{ppll}
\\\nonumber
{pn\to\ell^+\bar\nu^\prime}:~
&{\calO}^{(pn)S}_{L}=(p^\T Cn)(\ell^\T_L C\nu^\prime_L)\;,
&&{\calO}^{(pn)S}_{5L}=(p^\T C\gamma_5n)(\ell^\T_L C\nu_L^\prime)\;,
\\\nonumber
&{\calO}^{(pn)V}_L=(p^\T C\gamma_\mu n)(\ell^\T_R C\gamma^\mu\nu_L^\prime)\;,
&&{\calO}^{(pn)V}_{5L}=(p^\T C\gamma_\mu\gamma_5 n)(\ell^\T_R C\gamma^\mu\nu_L^\prime)\;,
\\
&{\calO}^{(pn)T}=(p^\T C\sigma_{\mu\nu}n)(\ell^\T_L C\sigma_{\mu\nu}\nu_L^\prime)\;, \label{pnlnu}
\\
{nn\to\bar\nu\bar\nu^\prime}:~
&{\calO}^{(nn)S}_{L}=(n^\T Cn)(\nu^\T_L C\nu_L^\prime)\;,
&&{\calO}^{(nn)S}_{5L}=(n^\T C\gamma_5 n)(\nu^\T_L C\nu_L^\prime)\;,
\label{nnnunu}
\end{align}
where there are no operators with a vector or tensor diproton/dineutron current since such currents vanish, i.e., $(N^\T C\gamma_\mu N)=(N^\T C\sigma_{\mu\nu} N)=(N^\T C\sigma_{\mu\nu}\gamma_5 N)=0$ for $N\in\{ p,n\}$. 

Since the dinucleon fields in the above operators must originate from the $u, d$ quarks, it is thus necessary to start from the classification of relevant operators at the quark level. There are two approaches in constructing relevant quark level operators responsible for dinucleon to dilepton transitions. One is to use effective degrees of freedom below the electroweak scale to obtain all relevant operators respecting $SU(3)_{\rm C}\times U(1)_{\rm EM}$, this is the LEFT approach. And the other is to use effective degrees of freedom of the SM to obtain all relevant operators respecting the SM gauge symmetry $SU(3)_{\rm C}\times SU(2)_{\rm L}\times U(1)_{\rm Y}$, the SMEFT. We will obtain the operators in both approaches.

In both the LEFT and SMEFT approaches the color $SU(3)_{\rm C}$ symmetry must be respected, the six quarks in all operators obtained must form color singlets. We denote a general six-quark field configuration as 
\begin{align}
{\cal O}^{ijklmn}=q^i_Iq^j_Jq^k_Kq^l_Lq^m_Mq^n_N\;,
\end{align}
where the superscripts $\{i,j,k,l,m,n\}$ are the color indices in fundamental representation of $SU(3)_{\rm C}$ while the subscripts $\{I,J,K,L,M,N\}$ encode the flavor and chiral information for each quark field. To form a color invariant operator, the color indices must be contracted by a color tensor $T_{ijklmn}$ such that ${\cal O}^{ijklmn}T_{ijklmn}$ is invariant under $SU(3)_{\rm C}$.
This color symmetry can be achieved by contracting the color indices $\{i,j,k,l,m,n\}$ in the quark fields with the following five independent color tensors 
\begin{align}\nonumber
T^{SSS}_{\{ij\}\{kl\}\{mn\}}&=\epsilon_{ikm}\epsilon_{jln}+\epsilon_{ikn}\epsilon_{jlm}+\epsilon_{ilm}\epsilon_{jkn}+\epsilon_{iln}\epsilon_{jkm}\;,
\\\nonumber
T^{SAA}_{\{ij\}[kl][mn]}&=\epsilon_{imn}\epsilon_{jkl}+\epsilon_{ikl}\epsilon_{jmn}\;,\quad
T^{SAA}_{\{kl\}[mn][ij]}=\epsilon_{ijk}\epsilon_{mnl}+\epsilon_{ijl}\epsilon_{mnk}\;,
\\
T^{SAA}_{\{mn\}[ij][kl]}&=\epsilon_{ijm}\epsilon_{kln}+\epsilon_{ijn}\epsilon_{klm}
\;,\quad
T^{AAA}_{[ij][kl][mn]}=\epsilon_{ijm}\epsilon_{kln}-\epsilon_{ijn}\epsilon_{klm}\;.
\end{align}
We put the details for the color tensor construction and their subtleties in Appendix~\ref{app:colortensor}.

\subsection{LEFT operators}
\label{LEFT:ope}
In the LEFT framework, the effective degrees of freedom are the SM light quarks $(u,d,s,c,b)$, charged and neutral leptons $(e,\mu,\tau,\nu_e,\nu_\mu,\nu_\tau)$, and the effective interactions are governed by the higher dimensional local operators ${\cal Q}_i^{d}$ built out of those fields and satisfying the symmetry $SU(3)_{\rm C}\times U(1)_{\rm EM}$. The LEFT Lagrangian ${\cal L}_{\rm LEFT}$ is organized in terms of the canonical mass dimension of the local operators  
\begin{align}
{\cal L}_{\rm LEFT}={\cal L}_{{\rm dim}\leq4}+\sum_{{\dim 5},i}{\hat C_{5,i}\over \Lambda}{\cal Q}^5_i+\sum_{{\dim 6},i}{\hat C_{6,i}\over \Lambda^2}{\cal Q}^6_i+\cdots + \sum_{{\dim 12},i}{\hat C_{12,i}\over \Lambda^8}{\cal Q}^{12}_i + \cdots\;,
\end{align}
where ${\cal L}_{{\rm dim}\leq4}$ is the renormalizable terms, and the Wilson coefficients $\hat C_{d,i}$ together with the heavy scale $\Lambda$ encode informations about the presumed fundamental physics. The systematic enumeration of operator bases up to dim 9 have been figured out in~\cite{Jenkins:2017jig,Liao:2020zyx,Li:2020tsi,Murphy:2020cly}. In our case, the relevant operators first appear at dim 12 and can be parametrized as the product of six-quark sectors and a proper lepton current given in Eqs.~(\ref{lep:curpp}-\ref{lep:curnn}).

For the quark sectors, we can repeatedly apply the Fierz identities (FI) to reach as many quark scalar bilinear currents as possible. In this way, by the Lorentz symmetry, for the operators with a  lepton scalar current their quark sectors can be factorized as three quark scalar currents, and for the operators with a lepton vector (tensor) current their quark sectors can be factorized as two quark scalar currents along with a single quark vector (tensor) current. As a non-trivial example, at the end of Appendix~\ref{app:LEFTreduction},  we will employ the FIs to show how the operators with a lepton tensor current in quark scalar-vector-vector bilinear structure are shifted into those with a scalar-scalar-tensor structure outlined here.
Lastly, the color $SU(3)_{\rm C}$ invariance can be done by contracting the free quark color indices using the independent color tensors constructed earlier. In doing so, one should be careful with operators containing several identical quark fields since the color relations in Eq.~\eqref{relation:TSA1} and the FIs in Appendix~\ref{app:LEFTreduction} may further restrict their independency. Combining the above points and excluding potential redundant operators, and for $n_f$ ( $n_f=3$ in the real case) flavors  of lepton fields $(\ell, \nu)$, the final bases of the dim-12 operators mediating the transitions $pp\to \ell^+\ell^{\prime+},pn\to \ell^+\bar\nu^\prime, nn\to \bar\nu\bar\nu^\prime$ are summarized one by one as follows: 

\noindent
{\small$\scriptscriptstyle\blacksquare$ \small\bf\boldmath Dim-12 operators contributing to $pp\to\ell^+\ell^{\prime+}$}\\ 
One can attach different lepton currents to operators already formed by six quarks to form the operators. 
We will discuss for each type in the following. For a scalar current $j_{S,\pm}^{\ell\ell^\prime}$, one just attaches it to some color singlet and Lorentz scalar six quark operators. An example of this class of operators, the ${\cal Q}_{1LLL,a}^{(pp)S,\pm}$, is given in the following
\begin{align}\nonumber
{\cal Q}_{1LLL,a}^{(pp)S,\pm}&=(u_L^{i\T} Cu_L^j)(u_L^{k\T} Cd_L^l)(u_L^{m\T} Cd_L^n)j_{S,\pm}^{\ell\ell^\prime}T^{SSS}_{\{ij\}\{kl\}\{mn\}}\;.
\label{ope:scalarf}
\end{align}
We find that there are 28 independent operators without referring to lepton flavors which are listed in \eqref{ope:scalarf} in Appendix~\ref{app:LEFTope}. 

For the operators with a vector lepton current $j_{V}^{\ell\ell^\prime,\mu}$, there are 19 independent operators, and for the operators with a tensor lepton current $j_{T}^{\ell\ell^\prime,\mu\nu}$, we find there are 16 independent operators. Therefore there are total $63$ operators which we give them in Eqs.~(\ref{ope:scalarf}-\ref{pp:tensorf}) in Appendix~\ref{app:LEFTope}. For $n_f$ flavors of charged leptons, there are $41n_f^2+6n_f$ operators. 

As a cross-check, we also confirmed our above results (and following ones) by the Hilbert series method~\cite{Hanany:2010vu, Lehman:2015via, Henning:2015daa,Henning:2015alf}. In Appendix~\ref{app:LEFTope}, by a non-trivial example, we also show how to reduce the redundant operators using the Fierz and Schouten identities. 

When restricting to the same flavor leptons with $\ell^\prime=\ell$, we find there are $28+19=47$ independent operators from the scalar and vector lepton currents, because the tensor lepton current vanishes for identical fields. Besides the transitions $pp\to e^+e^+, \mu^+\mu^+$ , such operators also contribute to the ${\rm H}{\rm -}\bar{\rm H}$ oscillation process and have been enumerated long ago by Caswell, Milutinovic and Senjanovic in Ref.~\cite{Caswell:1982qs}. However, we find 13 out of total 60 operators in their counting are redundant and all of them belong to the class with the vector lepton current $j_{V}^{\ell\ell,\mu}$. In Appendix~\ref{app:LEFTreduction}, we show explicitly the redundancy of the basis in~\cite{Caswell:1982qs} and give the correspondence of their basis (after excluding the redundant ones) with the basis in Eqs.~(\ref{ope:scalarf},~\ref{ope:vectorf}).

\noindent
{\small$\scriptscriptstyle\blacksquare$ \small\bf\boldmath Dim-12 operators contributing to $pn\to\ell^+\nu^\prime$}\\
For the operators with a scalar current $j_{S}^{\ell\nu^\prime}$, we find there are 14 independent operators without counting lepton flavors. For the operators with a vector current $j_{V}^{\ell\nu^\prime,\mu}$, there are 24 independent operators. And for the operators with a tensor current $j_{T}^{\ell\nu^\prime,\mu\nu}$, there are 13 independent operators.

In total, there are 51 independent operators for the transition $pn\to \ell^+\bar\nu^\prime$ which are listed in Eqs.~(\ref{pn:scalarf}-\ref{pn:tensorf}) in Appendix~\ref{app:LEFTope}.
For $n_f$ flavors of charged leptons and neutrinos,  there are $51n_f^2$ operators. These operators are also responsible for the conversions $e^- p\to \bar \nu \bar n$ and $e^- n\to \bar\nu \bar p$ in the electron-deuteron scattering. In addition, we find they can contribute to the unique neutron decay mode with the baryon number being changed by two units $n\to \bar p e^+\bar \nu$~\cite{He:2021sbl}.

\noindent
{\small$\scriptscriptstyle\blacksquare$ \small\bf\boldmath Dim-12 operators contributing to $nn\to\bar\nu\bar\nu^\prime$}\\
For the operators with a scalar current $j_{S}^{\nu\nu^\prime}$, we find there are 14 independent operators. And for the operators with a tensor neutrino current $j_{T}^{\nu\nu^\prime,\mu\nu}$, we find there are only 8 independent operators. In total, there are 22 independent operators which are listed in Eqs.~(\ref{nn:scalarf},~\ref{nn:tensorf}) in Appendix~\ref{app:LEFTope}.  For $n_f$ flavors of neutrinos there are $11n_f^2+3n_f$ operators. Excluding the scalar neutrino currents, one can easily identity the remaining 7 quark operators in Eq.~(\ref{nn:scalarf}) plus their parity partners are just the 14 operators contributing to the neutron-antineutron oscillation~\cite{Caswell:1982qs}. 

\subsection{SMEFT completions}
\label{SMEFT:ope}

To connect with the NP scenarios at a higher scale than $\Lambda_{\rm EW}$, the SMEFT can serve as a suitable bridge between the LEFT interactions and the unknown NP as shown in Fig.~\ref{Fig1}. It parametrizes the high scale UV NP in a model-independent way and therefore is a good starting point for the systematic EFT analysis of low energy processes. In this section, we consider the leading order SMEFT completions for the $\Delta B=\Delta L=-2$ dinucleon to dilepton transition operators discussed above. It happens that the relevant SMEFT operators also first appear at dim 12 at leading order, and contain six quark fields and two lepton fields. By the similar logic as the construction of the LEFT operators, we first factorize the operators as the convolution of the six-quark part and the proper lepton bilinear current through the Fierz rearrangement. Furthermore, since the nucleons made out of the up and down quarks, we only focus on the first generation of quark fields but without restriction for the generation of the lepton fields. We denote the SM left-handed lepton and quark doublet fields as $L(1,2,-1/2)$, $Q(3,2,1/6)$ and right-handed up-type quark, down-type quark and charged lepton fields as $u_R(3,1,2/3)$, $d_R(3,1,-1/3)$, $e_R(1,1,-1)$. We employ the front Latin letters $(a,b,c,d,e,f)$ for the $SU(2)_{\rm L}$ indices and the middle ones $(i,j,k,l,m,n)$ for the color $SU(3)_{\rm C}$ indices in fundamental representations, respectively. Similar to the classification of operators in the LEFT, we classify the relevant dim-12 SMEFT operators in terms of the lepton currents, and the final results are summarized in Appendix~\ref{app:SMEFTope}.

Here we briefly comment the procedures to reach the independent operators in Appendix~\ref{app:SMEFTope}. 1). Based on the $U(1)_{\rm Y}$ invariance, one can easily identify the allowed field configurations with six quarks and two leptons. 2). For each field configuration, we first use the Fierz transformation to fix the Lorentz structure of the operator so that it takes the quark-lepton factorized form ${\cal O}_q\times j_L$ in which the lepton current $j_L$ can be either scalar, vector, or tensor type. For the scalar/vector/tensor lepton current, the Fierz transformation can be used further to organize the corresponding six-quark part to be scalar-scalar-scalar/scalar-scalar-vector/scalar-scalar-tensor bilinear structures as we did in section~\ref{LEFT:ope}. 3). Followed by step 2), we consider the electroweak $SU(2)_{\rm L}$ invariance which can be done by implementing the contractions using the rank-2 Levi-Civita tensor $\epsilon_{ab}$. In doing so, the SI identity $\epsilon_{ab}\epsilon_{cd}=\epsilon_{ac}\epsilon_{bd}+\epsilon_{ad}\epsilon_{cb}$ has to be considered carefully for the multiple $SU(2)_{\rm L}$ contractions so as to reduce the redundant operators. 4). Last, the color $SU(3)_{\rm C}$ invariance can be done by contracting the free color indices using the independent color tensors discussed in section~\ref{app:colortensor} and Appendix~\ref{app:colortensor}. If there are multiple identical quark fields, the color relations in Eq.~\eqref{relation:TSA1} and the FIs in Appendix~\ref{app:LEFTreduction} must be taken into account to reduce the operators into the minimal basis given in Appendix~\ref{app:SMEFTope}. 5). We also count the number of independent operators in each configuration using the Hilbert series method~\cite{Henning:2015alf} and confirmed our result. 

The following is an example of using SM building blocks to build the SMEFT dinucleon to dilepton operators which are different from those in the LEFT,
\begin{align}
{\cal O}_{Q^6L^2}^{S,(S)}&=(Q^{i\T}_a CQ^j_b)(Q^{k\T}_c CQ^l_d)(Q^{m\T}_e CQ^n_f)(L^\T_g CL_h^\prime)\epsilon_{ab}\epsilon_{cd}\epsilon_{eg}\epsilon_{fh}T^{SAA}_{\{mn\}[kl][ij]}\;.
\end{align}
This time, $u_L$ and $d_L$ must appear at the same time so that the SM gauge symmetries are respected. Expanding $Q$ into its $u_L$ and $d_L$ components, one obtains
\begin{align}\nonumber
{\cal O}_{Q^6L^2}^{S,(S)}
&=4(u^{i\T}_{L} Cu^j_{L})  (u^{k\T}_{L} Cd^l_{L})(u^{m\T}_{L} Cd^n_{L})T^{SAA}_{\{ij\}[kl][mn]} j_{S,-}^{\ell\ell^\prime}
\\\nonumber
&-8(u^{i\T}_{L} Cd^j_{L})  (u^{k\T}_{L} Cd^l_{L})(u^{m\T}_{L} Cd^n_{L})T^{SAA}_{\{ij\}[kl][mn]} j_S^{\ell\nu^\prime}
\\\nonumber
&+4(d^{i\T}_{L} Cd^j_{L})  (u^{k\T}_{L} Cd^l_{L})(u^{m\T}_{L} Cd^n_{L})T^{SAA}_{\{ij\}[kl][mn]} j_S^{\nu\nu^\prime}
\\
&=4{\cal O}_{1LLL,b}^{(pp)S,-}-8{\cal O}_{1LLL,b}^{(pn)S}+4{\cal O}_{1LLL,b}^{(nn)S}\;.
\label{ope:Oq6L2}
\end{align}
\indent Therefore it is expected that the SMEFT approach will have less independent operators than that can be constructed in the LEFT approach. Without counting the lepton flavors, for operators with a scalar lepton current, we find there are 12 independent operators.  For operators with a vector lepton current, we find there are 7 independent operators. And for operators with a tensor lepton current, we find there are 10  independent operators. In total, there are 29 operators which are listed in Appendix~\ref{app:SMEFTope}. For $n_f$ flavors of leptons there are $18n_f^2+n_f$ operators. Except the sub-GeV scale dinucleon to dilepton processes studied in this work, these SMEFT operators are crucial for the model-independent study of the $\Delta B=\Delta L=2$ signals on colliders, for example, the process $pp\to\ell^+\ell^{\prime+}+4~{\rm jets}$ at LHC~\cite{Bramante:2014uda} and $e^- p\to \ell^++5~{\rm jets}$ in the future electron-proton colliders like LHeC. 

In literature, Refs.~\cite{Girmohanta:2019fsx,Girmohanta:2020eav} also provide a bunch of dim-12 operators contributing to dinucleon to dilepton transitions in the SMEFT. We find the operators given in~\cite{Girmohanta:2019fsx} are neither complete nor independent as a basis. Specifically, the 28 operators\footnote{Ref.~\cite{Girmohanta:2019fsx} also considered operators with the SM singlet right-handed neutrinos, here we only focus on the SMEFT subsets.} listed in~\cite{Girmohanta:2019fsx} can be covered by 21 operators in our counting and 8 of them are redundant. In addition, the 9 operators ${\cal O}_{u^3d^3L^21,2}^{S,(A)}$, ${\cal O}_{u^2d^2Q^2L^22}^{S,(A)}$, ${\cal O}_{u^2d^2Q^2L^22}^{T,(A)}$, ${\cal O}_{udQ^4L^23}^{T,(A)}$, ${\cal O}_{Q^6L^2}^{T,(S)}$, ${\cal O}_{u^4d^2e^2}^{T,(A)}$, ${\cal O}_{u^3dQ^2e^21,2}^{T,(A)}$ in our basis are missed in their list. In Appendix~\ref{app:SMEFTreduction}, we translate their operators as linear combinations of our above operators so that one can easily recognize the redundancy and incompleteness of the operators in~\cite{Girmohanta:2019fsx}.

\subsection{Relations between SMEFT and LEFT operators}
\label{SMEFT:matching}

As already mentioned earlier that some of the SMEFT operators contain several LEFT operators, i.e. ${\cal O}_{Q^6L^2}^{S,(S)} = 4{\cal O}_{1LLL,b}^{(pp)S,-}-8{\cal O}_{1LLL,b}^{(pn)S}+4{\cal O}_{1LLL,b}^{(nn)S}$. The SMEFT approach will have less independent operators than that can be constructed in the LEFT approach at the same order. To have better idea on how these operators are related to each other, in Tab.~\ref{tab:matchingLEFT}, we perform a tree-level matching of the dim-12 SMEFT operators listed in Appendix~\ref{app:SMEFTope} to the dim-12 LEFT operators listed in Appendix~\ref{app:LEFTope} at the electroweak scale $\Lambda_{\rm EW}$. We have approximated the CKM factor $V_{ud}\approx1$ arising from the mismatch of the flavor and mass eigenstates of the left-handed down quark $d_L$. From Tab.~\ref{tab:matchingLEFT}, it is obvious that the operators with the singlet charged lepton scalar/tensor current $(e_R^\T C\Gamma e_R^\prime)$ can only exclusively contribute to the transition $pp\to\ell^+\ell^{\prime+}$, while the operators with an anti-symmetric scalar lepton current `S,(A)' and the operators with a symmetric tensor current `T,(S)' can only contribute to the transition $pn\to\ell^+\nu^\prime$. The remaining operators with a symmetric scalar lepton current `S,(S)' or with an anti-symmetric tensor current `T,(A)' could contribute to both the three different transition channels. Last but not least, one must be careful that the SMEFT Wilson coefficients with a superscript `(A)' vanish for identical lepton fields since the relevant operators are anti-symmetric for the two lepton fields.

From Appendix~\ref{app:SMEFTope} and Tab.~\ref{tab:matchingLEFT}, we see there are many operators which can be constructed in the LEFT, but not in the SMEFT at leading dim-12 order, for instance, operators ${\cal Q}_{4LLR}^{(pp)S,\pm}$, ${\cal Q}_{4LLR}^{(pn)S}$ and ${\cal Q}_{4LLR}^{(nn)S}$ for the three channels respectively. Those operators are not $SU(2)_{\rm L}\times U(1)_{\rm Y}$ invariant and can only be generated by the higher dim-14 and/or dim-16 SMEFT $SU(2)_{\rm L}\times U(1)_{\rm Y}$ invariant operators consisting of dim-12 fermion part $(qqqqqqll)$ together with additional Higgs doublets. The physical effects from such operators are suppressed with additional factors like $v^2/\Lambda_{\rm NP}^2$ and/or $v^4/\Lambda_{\rm NP}^4$ relative to the dim-12 SMEFT operators and will be neglected in our numerical analysis.

\begin{table}
\centering
\resizebox{\linewidth}{!}{
\begin{tabular}{|c| l | l | l |}
\hline
~SMEFT operators~
& \multicolumn{1}{|c |}{$pp\to \ell\ell^\prime$} & \multicolumn{1}{|c |}{$pn\to \ell\bar\nu^\prime$} & \multicolumn{1}{|c |}{$nn\to \bar\nu\bar\nu^\prime$}
\\\hline
$\calO_{u^3d^3L^21}^{S,(A)}$
& \multicolumn{1}{|c |}{-}
& $C_{1RRR,a}^{(pn)S}=-2C_{u^3d^3L^21}^{S,(A)}$ 
&  \multicolumn{1}{|c |}{-}
\\
$\calO_{u^3d^3L^22}^{S,(A)}$
&  \multicolumn{1}{|c |}{-}
& $C_{1RRR,b}^{(pn),}=-2C_{u^3d^3L^22}^{S,(A)}$
& \multicolumn{1}{|c |}{-}
\\
$\calO_{u^2d^2Q^2L^21}^{S,(S)}$
&$C_{3RRL,a}^{(pp)S,-}=C_{u^2d^2Q^2L^21}^{S,(S)}$~~~~~~~~~~~
& $C_{3RRL,a}^{(pn)S}=-2C_{u^2d^2Q^2L^21}^{S,(S)}$~~~~~~~~~~~
& $C_{3RRL,a}^{(nn)S}=C_{u^2d^2Q^2L^21}^{S,(S)}$~~~~~~~~~~~
\\
$\calO_{u^2d^2Q^2L^22}^{S,(A)}$
&  \multicolumn{1}{|c |}{-}
& $C_{3RRL,b}^{(pn)S}=-4C_{u^2d^2Q^2L^22}^{S,(A)}$
& \multicolumn{1}{|c |}{-}
\\
$\calO_{u^2d^2Q^2L^23}^{S,(S)}$
& $C_{3RRL,b}^{(pp)S,-}=C_{u^2d^2Q^2L^23}^{S,(S)}$
& $C_{3RRL,c}^{(pn)S}=-2C_{u^2d^2Q^2L^23}^{S,(S)}$
& $C_{3RRL,b}^{(nn)S}=C_{u^2d^2Q^2L^23}^{S,(S)}$
\\
$\calO_{udQ^4L^21}^{S,(A)}$
&  \multicolumn{1}{|c |}{-}
& $C_{3LLR,c}^{(pn)S}=-8C_{udQ^4L^21}^{S,(A)}$
&  \multicolumn{1}{|c |}{-}
\\
$\calO_{udQ^4L^22}^{S,(S)}$
& $C_{2LLR,b}^{(pp)S,-}=2C_{udQ^4L^22}^{S,(S)}$
& $C_{3LLR,b}^{(pn)S}=-4C_{udQ^4L^22}^{S,(S)}$
& $C_{2LLR,b}^{(nn)S}=2C_{udQ^4L^22}^{S,(S)}$
\\
$\calO_{Q^6L^2}^{S,(S)}$
& $C_{1LLL,b}^{(pp)S,-}=4C_{Q^6L^2}^{S,(S)}$
& $C_{1LLL,b}^{(pn)S}=-8C_{Q^6L^2}^{S,(S)}$
& $C_{1LLL,b}^{(nn)S}=4C_{Q^6L^2}^{S,(S)}$
\\
$\calO_{u^4d^2e^21}^{S,(S)}$
& $C_{1RRR,a}^{(pp)S,+}=C_{u^4d^2e^21}^{S,(S)}$
&  \multicolumn{1}{|c |}{-} &  \multicolumn{1}{|c |}{-}
\\
$\calO_{u^4d^2e^22}^{S,(S)}$
& $C_{1RRR,b}^{(pp)S,+}=C_{u^4d^2e^22}^{S,(S)}$
&  \multicolumn{1}{|c |}{-} &  \multicolumn{1}{|c |}{-}
\\
$\calO_{u^3dQ^2e^2}^{S,(S)}$
& $C_{2RRL,b}^{(pp)S,+}=2C_{u^3dQ^2e^2}^{S,(S)}$
&  \multicolumn{1}{|c |}{-} &  \multicolumn{1}{|c |}{-}
\\
$\calO_{u^2Q^4e^2}^{S,(S)}$ 
& $C_{3LLR,b}^{(pp)S,+}=4C_{u^2Q^4e^2}^{S,(S)}$
&  \multicolumn{1}{|c |}{-} &  \multicolumn{1}{|c |}{-}
\\\hline
$\calO_{u^3d^2QeL1}^{V}$
& $C_{1RR,a}^{(pp)V}=-C_{u^3d^2QeL1}^{V}$
&  $C_{1RR,a}^{(pn)V}=C_{u^3d^2QeL1}^{V}$
&  \multicolumn{1}{|c |}{-}
\\
$\calO_{u^3d^2QeL2}^{V}$
& $C_{1RR,b}^{(pp)V}=C_{u^3d^2QeL2}^{V}$
&  $C_{1RR,b}^{(pn)V}=-C_{u^3d^2QeL2}^{V}$
&  \multicolumn{1}{|c |}{-}
\\
$\calO_{u^3d^2QeL3}^{V}$
& $C_{1RR,c}^{(pp)V}=-C_{u^3d^2QeL3}^{V}$
& $C_{1RR,c}^{(pn)V}=C_{u^3d^2QeL3}^{V}$
&  \multicolumn{1}{|c |}{-}
\\
$\calO_{u^2dQ^3eL1}^{V}$
& $C_{4LR,c}^{(pp)V}=2C_{u^2dQ^3eL1}^{V}$
& $C_{4RL,b}^{(pn)V}=-2C_{u^2dQ^3eL1}^{V}$
&  \multicolumn{1}{|c |}{-}
\\
$\calO_{u^2dQ^3eL2}^{V}$
& $C_{4LR,d}^{(pp)V}=2C_{u^2dQ^3eL2}^{V}$
& $C_{4RL,d}^{(pn)V}=2C_{u^2dQ^3eL2}^{V}$
&  \multicolumn{1}{|c |}{-}
\\
$\calO_{u^2dQ^3eL3}^{V}$
& $C_{4LR,e}^{(pp)V}=-2C_{u^2dQ^3eL3}^{V}$
& $C_{4RL,e}^{(pn)V}=-2C_{u^2dQ^3eL3}^{V}$
& \multicolumn{1}{|c |}{-}
\\
$\calO_{uQ^5eL}^{V}$
& $C_{1LL,c}^{(pp)V}=4C_{uQ^5eL}^{V}$
& $C_{2LL,c}^{(pn)V}=-4C_{uQ^5eL}^{V}$
&  \multicolumn{1}{|c |}{-}
\\\hline
$\calO_{u^2d^2Q^2L^21}^{T,(S)}$
&  \multicolumn{1}{|c |}{-} 
& $C_{3RRL,a}^{(pn)T}=-4C_{u^2d^2Q^2L^21}^{T,(S)}$
& \multicolumn{1}{|c |}{-}
\\
$\calO_{u^2d^2Q^2L^22}^{T,(A)}$
& $C_{3RRL}^{(pp)T,-}=C_{u^2d^2Q^2L^22}^{T,(A)}$
& $C_{3RRL,b}^{(pn)T}=-2C_{u^2d^2Q^2L^22}^{T,(A)}$
& $C_{3RRL}^{(nn)T}=-C_{u^2d^2Q^2L^22}^{T,(A)}$
\\
$\calO_{u^2d^2Q^2L^23}^{T,(S)}$
&  \multicolumn{1}{|c |}{-}
& $C_{3RRL,c}^{(pn)T}=-4C_{u^2d^2Q^2L^23}^{T,(S)}$
&  \multicolumn{1}{|c |}{-}
\\
$\calO_{udQ^4L^21}^{T,(A)}$
& $C_{2LLR,b}^{(pp)T,-}=2C_{udQ^4L^21}^{T,(A)}$
& $C_{3LLR,c}^{(pn)T}=-4C_{udQ^4L^21}^{T,(A)}$
& $C_{2LLR,b}^{(nn)T}=-2C_{udQ^4L^21}^{T,(A)}$
\\
$\calO_{udQ^4L^22}^{T,(S)}$
&  \multicolumn{1}{|c |}{-}
& $C_{3LLR,b}^{(pn)T}=-8C_{udQ^4L^22}^{T,(S)}$
&  \multicolumn{1}{|c |}{-}
\\
$\calO_{udQ^4L^23}^{T,(A)}$
& $C_{2LLR,c}^{(pp)T,-}=-2C_{udQ^4L^23}^{T,(A)}$
& $C_{3LLR,d}^{(pn)T}=-4C_{udQ^4L^23}^{T,(A)}$
& $C_{2LLR,c}^{(nn)T}=-2C_{udQ^4L^23}^{T,(A)}$
\\
$\calO_{Q^6L^2}^{T,(S)}$
&  \multicolumn{1}{|c |}{-}
& $C_{1LLL,b}^{(pn)T}=-16C_{Q^6L^2}^{T,(S)}$
&  \multicolumn{1}{|c |}{-}
\\
$\calO_{u^4d^2e^2}^{T,(A)}$
& $C_{1RRR}^{(pp)T,+}=C_{u^4d^2e^2}^{T,(A)}$
&  \multicolumn{1}{|c |}{-} &  \multicolumn{1}{|c |}{-}
\\
$\calO_{u^3dQ^2e^21}^{T,(A)}$
& $C_{2RRL,a}^{(pp)T,+}=2C_{u^3dQ^2e^21}^{T,(A)}$
&  \multicolumn{1}{|c |}{-} & \multicolumn{1}{|c |}{-}
\\
$\calO_{u^3dQ^2e^22}^{T,(A)}$
& $C_{2RRL,c}^{(pp)T,+}=2C_{u^3dQ^2e^22}^{T,(A)}$
& \multicolumn{1}{|c |}{-}& \multicolumn{1}{|c |}{-}
\\\hline
\end{tabular}
}
\caption{The SMEFT dinucleon to dilepton operators and their matching onto the LEFT at the $\Lambda_{\rm EW}$. Where the notation for the Wilson coefficients is similar to the corresponding operators with the replacement of ${\cal O}$ by $C$, e.g., $C_{u^3dQ^2e^22}^{T,(A)}$ for $\calO_{u^3dQ^2e^22}^{T,(A)}$, We do not show the explicit flavors of leptons in the above matching but can be easily recognized through the corresponding operators.}
\label{tab:matchingLEFT}
\end{table}

\newpage
\section{Chiral realizations} 
\label{sec:chiral}

\subsection{Some basics of Chiral matching}
\label{chiral:basics}

After establishing the operator basis for dinucleon to dilepton transitions in the LEFT and SMEFT, the next step is to calculate the transition matrix elements and the decay rates. However, the hadronic matrix elements between the initial dinucleon state and the QCD vacuum are not a trivial task due to their non-perturbative QCD nature. In order to obtain those matrix elements with a controllable uncertainty, fortunately, one can employ the successful effective chiral perturbation theory of the low energy QCD and the spurion field techniques to shift the quark level interactions into those interactions among hadrons and leptons.     

In the QCD sector, the approximate chiral symmetry $G_\chi=SU(2)_{\rm L}^f\times SU(2)_R^f$ of the two-flavor QCD Lagrangian under the limit of massless up and down quarks is spontaneously broken into its isospin subgroup $SU(2)_{\rm V}$ by the quark condensation $\langle \bar qq\rangle$ at the scale $\Lambda_\chi$ .\footnote{Usually the strange quark $s$ can also be included in this framework to consider the larger group breaking pattern $SU(3)_L^f\times SU(3)_R^f\to SU(3)_V$~\cite{Gasser:1983yg, Gasser:1984gg}. For our purpose, it is enough to only focus on the two-flavor case in which the chiral symmetry breaking effect is relatively smaller.} The interaction of the resultant pseudo Nambu-Goldstone pion fields at low energy ($p<\Lambda_\chi$) is described by the $\chi$PT which inherits the QCD chiral symmetry~\cite{Gasser:1983yg, Gasser:1984gg}, and the baryon extended $\chi$PT termed as B$\chi$PT is our main focus in this section. The (B)$\chi$PT Lagrangian is organized in terms of the power of the soft momentum $p$ relative to $\Lambda_\chi$. Introducing the proper external sources transforming under the chiral group, the global chiral symmetry can be promoted to be a local one. Therefore, the (B)$\chi$PT Ward identities can be easily formulated and the interaction of hadrons with other light particles such as leptons and photon can be included. In the following we will use the two-flavor B$\chi$PT formalism and the spurion field techniques to construct an equivalent effective chiral Lagrangian for the dim-12 interactions in section~\ref{LEFT:ope}.

Before performing the detailed non-perturbative matching for the LEFT interactions, we can expect the six-quark part of those dim-12 operators will be transformed into proper nucleon current together with pions and derivatives. For our purpose of capturing the leading order contributions to the dinucleon to dilepton transitions, it is enough to consider the dim-6 terms composed of a nucleon bilinear current and a lepton bilinear current without any pions and derivatives in Eqs.~(\ref{ppll}-\ref{nnnunu}). The hadron level Wilson coefficients $C^{(NN^\prime)}_{a,b,c}$ will be determined below by the chiral matching to the quark level operators. Once these Wilson coefficients are obtained, it is straightforward to obtain the transition amplitudes and henceforth the decay rates, which will be postponed in section~\ref{sec:cal} for a detailed analysis. Now we turn to the chiral matching and find the relationship between the Wilson coefficients of the operators in Eqs.~(\ref{ppll}-\ref{nnnunu}) and those in the LEFT/SMEFT. 

We start from the basics of the $\chi$PT.  For the light two-flavor quarks $q=(u,~d)^\T$,  the QCD-like Lagrangian with extended external sources is parametrized as
\begin{align}
\mathcal{L}=\mathcal{L}_{\text{QCD}}^{M=0}+\overline{q_L}l_\mu \gamma^\mu q_L+\overline{q_R}r_\mu \gamma^\mu q_R-\left[\overline{q_R}(s-ip)q_L-\overline{q_R}(t_l^{\mu\nu}\sigma_{\mu\nu})q_L+\text{h.c.}\right]\;,
\label{qcdLag}
\end{align}
where the flavor space $2\times2$ matrices $\{l_\mu=l_\mu^\dagger,~r_\mu=r_\mu^\dagger,~s=s^\dagger,~p=p^\dagger,~t_r^{\mu\nu}=t_l^{\mu\nu\dagger}\}$ are the external sources related to the corresponding quark currents. Under the global chiral transformation $q_{L}\to \hat L q_{L}$ and $q_{R} \rightarrow \hat R q_{R}$ with $(\hat L,\hat R)\in G_\chi$, the pure QCD part $\mathcal{L}_{\text{QCD}}^{M=0}$ is invariant. The introduction of the external sources with proper transformation properties can promote the global chiral symmetry to be a local one. In this way, the whole Lagrangian $\mathcal{L}$ can be made invariant under the local chiral transformation $q_{L}\to L(x) q_{L}\equiv Lq_{L}$ and $q_{R} \rightarrow R(x) q_{R}\equiv Rq_{R}$ together with the following transformations of the external sources
\begin{align}
&\chi \to R\chi L^{\dagger}\;,
&&l_{\mu} \to L l_{\mu} L^{\dagger}+i L \partial_{\mu} L^{\dagger}\;,
&&r_{\mu}\to R r_{\mu} R^{\dagger}+i R \partial_{\mu} R^{\dagger}\;,
&&t_l^{\mu\nu} \to Rt_l^{\mu\nu} L^{\dagger}\;,
\end{align}
where $\chi\equiv 2B(s+ip)$ with $B=-\langle\bar q q\rangle/(2F_\pi^2)\approx 2.8~\rm GeV$. $F_\pi$ is the pion decay constant, and the quark condensate $\langle\bar q q\rangle$ can be treated as an order parameter to measure the strength of the spontaneous chiral symmetry breaking. 

For the Lagrangian in Eq.~\ref{qcdLag}, the equivalent chiral Lagrangian at low energy can be constructed by identifying the relevant degrees of freedom and to write down the most general chiral invariant Lagrangian ordered in terms of the number of soft momenta. The relevant degrees of freedom are just the light hadrons (pseudo-scalar pions and nucleons) and all possible non-QCD states (like leptons and photon) encoded in the external sources. Define the pseudo Nambu-Goldstone matrix $U$ as
\begin{align}
&U=u^2\;,
&&u={\rm exp}\left({i\Pi \over 2F_0}\right)\;,
&&\Pi=\pi^a\tau^a=
\begin{pmatrix}
\pi^0 & \sqrt{2}\pi^+\\
\sqrt{2}\pi^- & -\pi^0
\end{pmatrix}\;.
\end{align}
Then, under the chiral transformation $(L,R)\in G_\chi$, they transform as $U \to  R U L^\dagger$ and $u\to Ruh^\dagger=huL^\dagger$ with the compensator matrix $h\in SU(2)_{V}$ as a function of $U,L,R$. Furthermore, we define the chiral vielbein as
\begin{align}
u_\mu=i\left(u^\dagger(\partial_\mu-ir_\mu)u-u(\partial_\mu-il_\mu)u^\dagger\right)\;,
~~u_\mu^\dagger=u_\mu\;,
\end{align}
which transforms as $u_\mu \to h u_\mu h^\dagger$ under the chiral group. The power counting of these building blocks in terms of the soft momentum $p$ is 
\begin{align}
&u=\calO(p^0)\;,
&&u_\mu=\calO(p^1)\;,
&&\chi=\calO(p^2)\;.
\end{align} 
Then the leading order mesonic chiral Lagrangian is at $\calO(p^2)$ and takes the form
\begin{align}
&\mathcal{L}_2=\frac{F^2_0}{4}{\rm Tr}[u_\mu u^\mu+\chi_+]\;,
&&\chi_+=u^\dagger\chi u^\dagger+u\chi^\dagger u\;,
\label{eq:L2}
\end{align}
where $F_0$ is the pion decay constant in the chiral limit. Here we see the tensor external source does not enter into the leading chiral Lagrangian but rather first appears at $\calO(p^4)$ \cite{Cata:2007ns}.

Next, we include the nucleons in this framework. Denote the nucleon doublet as $\Psi=(p, n)^\T$ which transforms as $\Psi\to h\Psi$ under the chiral transformation. The covariant derivative of the nucleon doublet is
\begin{align}
&D_\mu \Psi=(\partial_\mu+\Gamma_\mu)\Psi\;,
&&\Gamma_\mu=\frac{1}{2}\left(u^\dagger(\partial_\mu-ir_\mu)u+u(\partial_\mu-il_\mu)u^\dagger\right)\;,
\end{align}
where $\Gamma_\mu$ is the chiral connection which helps $D_\mu \Psi$ to have the same transformation rule as $\Psi$. The power counting for the nucleon field is $\Psi=\calO(p^0)$, and also $D_\mu \Psi=\calO(p^0)$, this latter result is because the nucleon mass $m_N$ is comparable with the expansion scale $\Lambda_\chi$. However, $(i\slashed{D}-m_N) \Psi=\calO(p^1)$, then the leading order baryonic chiral Lagrangian takes
\begin{align}
\calL_{\pi N}^{(1)}=\bar{\Psi}\left(i\slashed{D}-m_N+{g_A\over 2} \gamma^\mu\gamma_5u_\mu\right)\Psi\;,
\end{align}
where $g_A$ is the axial-nucleon coupling constant. For the chiral matching of the dim-12 operators below, 
we will treat $D_\mu \Psi$ as a higher term than $\Psi$ through the naive dimensional analysis, and neglect their contribution to the leading order chiral realization of the relevant dim-12 operators. 
 A possible way out for saving the power counting rule of the nucleons is via the heavy baryon chiral perturbation formalism (HB$\chi$PT)~\cite{Jenkins:1990jv} but with the sacrifice of Lorentz covariance. The HB$\chi$PT formalism is beyond our current scope and one can check Ref.~\cite{Bijnens:2017xrz} for the treatment of neutron-antineutron oscillation. A brief comment concerning the relation between the Lorentz covariant operators and their HB$\chi$PT reduction is given at the end of this section.

\subsection{Decomposition of irreducible chiral symmetry}
\label{chiral:irrep}

In the matching onto B$\chi$PT for the effective operators in our case, the lepton current together with the associated Wilson coefficient of a dim-12  operator in the LEFT behaves as a fixed external source, thus we only have to cope with the six-quark sector of the operator. One of the key steps for the chiral matching is to identify irreducible chiral representations. We describe the procedures in the following. Suppose the quark sector has been decomposed into a sum of irreducible representations/tensors (irreps) of the chiral group
\begin{align}
P=\theta^{uvwxyz}(q^{i\T}_{\chi_1,u}C\Gamma_1 q^j_{\chi_2, v})(q^{k\T}_{\chi_3,w}C\Gamma_2 q^l_{\chi_4, x})(q^{m\T}_{\chi_5,y}C\Gamma_3 q^n_{\chi_6, z})T_{ijklmn}^{\rm color}\;,
\end{align}
where $q_{\chi_i}$ are the chiral quark doublets $(u, d)^\T_{\chi_i}$ with $\chi_i$ being the proper chiral projectors $P_\pm$, $\Gamma_i$ are the Dirac gamma matrices, $T_{ijklmn}^{\rm color}$ a general color tensor discussed in section~\ref{quarklevel:ope} and Appendix~\ref{app:colortensor}. The flavor indices $\{u,v,w,x,y,z\}$ as dummy indices are summed over and take 1 or 2 for the up quark $u$ or down quark $d$ respectively. The set of pure numbers $\theta^{uvwxyz}$ depends on the irrep under consideration. $\theta$ is promoted as a spurion field that transforms properly together with chiral transformations of the quarks under $G_\chi$,  so that $P$ looks like a chiral invariant.

The chiral counterparts of operator $P$ are constructed out of the spurion field $\theta$ plus the hadronic degrees of freedom $\{\Psi, D_\mu, u, u_\mu, \chi, \cdots \}$ and share the same symmetry transformation properties as that of $P$, which include the chiral symmetry, the Lorentz and the global baryon/lepton transformation properties. Since $P$ is chiral invariant and violates baryon number by two units, the matched operators must also be chiral invariant and contain exactly one spurion field $\theta$  and two nucleon fields $\Psi$s. Based on the chiral power counting property of the hadronic degrees of freedom, the obtained operators are ordered in terms of the number of soft momenta $p$ and the dominant terms are those with least power of $p$. Last, for each independent operator we associate it with an unknown LEC which accommodates the non-perturbative QCD dynamics. These LECs can be determined by fitting to the data, or calculated using the LQCD method, or estimated based on the naive dimensional analysis. In addition, for the LEFT operators belonging to the same chiral irrep, the chiral symmetry implies their chiral counterparts at a given chiral order share the same LEC. Here we remark that the above procedures have been used previously to the non-perturbative matching of the dim-9 operators mediating the nuclear and kaon neutrinoless double beta decay processes~\cite{Liao:2019gex,Graesser:2016bpz} as well as the neutron-antineutron oscillations~\cite{Bijnens:2017xrz}.

With the above procedures, we can now match the operator basis in the LEFT in section~\ref{LEFT:ope} onto B$\chi$PT at leading order of the chiral expansion, i.e., at $\calO(p^0)$.\footnote{The higher order terms can be constructed in the same style as was done in~\cite{Bijnens:2017xrz}.} We first transform the LEFT operator basis into a chiral basis in which each operator itself belongs to some irrep of the chiral group $G_\chi$. The chiral bases are shown in Tab.~\ref{tab:chiralirrep_pp}, Tab.~\ref{tab:chiralirrep_pn} and Tab.~\ref{tab:chiralirrep_nn} in Appendix~\ref{app:chiralope} for the operators contributing to the transition $pp\to\ell^+\ell^{\prime+}$, $pn\to\ell^+\bar\nu^\prime$ and $nn\to\bar\nu\bar\nu^\prime$, respectively. In the tables, we list their relations with the LEFT operators in the first and second column and show their chiral irreps in the third column (where the subscripts $(a,b,c)$ behind some irrep are used to distinguish different irreps with the same chiral type)\footnote{For the operators belonging to the same irrep, they must have the similar chiral, Lorentz and color structures so that they can be related to each other through the action of the chiral transformation.} and the corresponding chiral spurion fields in last column. Except the gray sectors, which already include the parity conjugates, all the rest ones have their parity conjugates with $L\leftrightarrow R$ (and an additional exchange of $+\leftrightarrow -$ for the tensor operators in Tab.~\ref{tab:chiralirrep_pp}). The parity conjugate of chiral operator $P_i$ is denoted by $\tilde P_i$ once it is needed. For those chiral operators expressed as a linear combination of two or more LEFT operators, their equivalent definitions are given in Appendix~\ref{app:chiralope} through fully symmetrizing all free quark flavors with the same chirality. The relations between the Wilson coefficients of the chiral basis as those of the LEFT operators can be determined easily. The general expression for the spurion fields takes the form 
\begin{align}
&\theta^{u_{1L}\cdots u_{nL}v_{1R}\cdots v_{mR}}_{(i_1\cdots i_n)(j_1\cdots j_m)}=\theta^{v_{1R}\cdots v_{mR}u_{1L}\cdots u_{nL}}_{(j_1\cdots j_m)(i_1\cdots i_n)}=[\delta_{(i_1}^{u_1}\delta_{i_2}^{u_2}\cdots\delta_{i_n)}^{u_n}][\delta_{(j_1}^{v_1}\delta_{j_2}^{v_2}\cdots\delta_{j_m)}^{v_m}]\;,
\end{align}
where we take the symmetrization notation with the round brackets $(\cdots)$. Symmetrization with respect to a group of indices is defined by placing these indices between round brackets $(\cdots)$, so we have 
\begin{align}
\delta_{(i_1}^{u_1}\delta_{i_2}^{u_2}\cdots\delta_{i_n)}^{u_n}= {1\over n!}\left[\delta_{i_1}^{u_1}\delta_{i_2}^{u_2}\cdots\delta_{i_n}^{u_n}+ (n!-1){\rm~permutations~of~}(i_1,\cdots, i_n)\right]\;, 
\end{align}
where we take the normalization as in~\cite{Rinaldi:2018osy}. For example, for the operators with a scalar lepton current belonging to the chiral $({\bf 3}_L,{\bf 1}_R)$ and $({\bf 5}_L,{\bf 3}_R)$ irreps in Tab.~\ref{tab:chiralirrep_pp}, we have
\begin{align} \nonumber
\theta^{u_Lv_L}_{(11)}&= \theta^{u_Lv_R}_{11}=\delta^u_1\delta^v_1\;,
\\ \nonumber
\theta^{u_Lv_Lw_Lx_Ly_Rz_R}_{(1112)(12)}&=
{1\over 8}[\delta^u_1\delta^v_1\delta^w_1\delta^x_2+\delta^u_1\delta^v_1\delta^w_2\delta^x_1+\delta^u_1\delta^v_2\delta^w_1\delta^x_1+\delta^u_2\delta^v_1\delta^w_1\delta^x_1][\delta^y_1\delta^z_2+\delta^y_2\delta^z_1]\;,
\\ \nonumber
\theta^{u_Lv_Lw_Lx_Ly_Rz_R}_{(1122)(11)}&=
{1\over 6}[\delta^u_1\delta^v_1\delta^w_2\delta^x_2+\delta^u_1\delta^v_2\delta^w_1\delta^x_2+\delta^u_1\delta^v_2\delta^w_2\delta^x_1+1\leftrightarrow 2]\delta^y_1\delta^z_1\;,
\\
\theta^{u_Lv_Lw_Lx_Ly_Rz_R}_{(1111)(22)}&=
\delta^u_1\delta^v_1\delta^w_1\delta^x_1\delta^x_2\delta^z_2\;.
\end{align}

From Tabs.~\ref{tab:chiralirrep_pp}-\ref{tab:chiralirrep_nn}, we see that for the operators with a scalar lepton current, there are six types of irreps under the chiral group: $({\bf 3}_L,{\bf 1}_R)$, $({\bf 5}_L,{\bf 3}_R)$, $({\bf 7}_L,{\bf 1}_R)$ plus their parity conjugates. The three operators $P_{1,a}^{(pp)S},P_{1,a}^{(pn)S},P_{1,a}^{(nn)S}$ belong to the same irrep $({\bf 7}_L,{\bf 1}_R)$ and relate to each other by the chiral symmetry, and similar situation appears in the irrep $({\bf 5}_L,{\bf 3}_R)$ in which the nine operators are related to each other. However, for the irrep type $({\bf 3}_L,{\bf 1}_R)$, there are three different irreps distinguished by the subscripts $a,b,c$ and each one contains three operators. For the operators with a vector lepton current, there are also six types of irreps: $({\bf 2}_L,{\bf 2}_R)$, $({\bf 4}_L,{\bf 2}_R)$, $({\bf 4}_L,{\bf 4}_R)$, $({\bf 6}_L,{\bf 2}_R)$, $({\bf 2}_L,{\bf 4}_R)$ and $({\bf 2}_L,{\bf 6}_R)$. One should be careful that there are four different irreps for the type $({\bf 2}_L,{\bf 2}_R)$ since the parity conjugate of the irrep $({\bf 2}_L,{\bf 2}_R)|_a$ is different from itself and will be denoted as $({\bf 2}_L,{\bf 2}_R)|_d$ . Last, for the operators with a tensor lepton current, there are still six types of irreps: $({\bf 1}_L,{\bf 1}_R)$, $({\bf 3}_L,{\bf 1}_R)$, $({\bf 3}_L,{\bf 3}_R)$, $({\bf 5}_L,{\bf 1}_R)$, $({\bf 1}_L,{\bf 3}_R)$ and $({\bf 1}_L,{\bf 5}_R)$. Where the type $({\bf 3}_L,{\bf 3}_R)$ contains four different irreps due to the parity conjugates of $({\bf 3}_L,{\bf 3}_R)|_{a,b}$ are distinct from themselves for the operators in Tab.~\ref{tab:chiralirrep_pp}. Another interesting fact is that the LEFT operators belonging to the same chiral irrep will not mix with each other under the QCD renormalization and have the same anomalous dimensions since the QCD preserves the chiral symmetry and quark flavors. In addition, the QCD renormalization for the operators related to each other by parity is also the same. For the operators with a scalar lepton current, the 1-loop QCD renormalization is identical to the dim-9 operators contributing to the $n{\rm -}\bar n$ oscillation and can be found in~\cite{Caswell:1982qs,Rinaldi:2019thf}. But there is no result for the operators with a vector or tensor lepton current yet, and we will neglect the QCD renormalization effect for the current work due to the involvement of considerable effort. However, from the 1-loop anomalous dimension matrix result for the dim-9 $n{\rm-}\bar n$ oscillation operators given in~\cite{Rinaldi:2019thf}, we can estimate the 1-loop QCD correction for those operators with a lepton scalar current. From the electroweak scale $\Lambda_{\rm EW}$ to the scale $\Lambda_{\chi}$, we find the running effect could have substantial impact on some scalar lepton current operators but the influence on the derived NP scale is at most ${\cal O}(30~\%)$ due to the high power dependence ($C_{i}\propto \Lambda^{-8}_{{\rm NP}}$). We will systematically explore in the future work their renormalization effect.

\subsection{Chiral matching for the operators}
\label{chiral:matching}

Based on the chiral irrep, we reorganize the chiral building blocks in terms of the power of soft momentum $p$ and the explicit chiral left or right doublet indices such that they have only one or two free indices, i.e., the building blocks are constructed to take the forms: $X_{u_{\chi_1}}$ and $X_{u_{\chi_1}v_{\chi_2}}$ with $\chi_{i}=L,R$.\footnote{The building blocks with three or more free indices are not independent and can be reduced into a product of $X_{u_{\chi_1}}$s and $X_{u_{\chi_1}v_{\chi_2}}$s.} They transform as $X_{u_{\chi_1}}\to (g_{\chi_1})_{u_{\chi_1} \hat u_{\chi_1}}X_{\hat u_{\chi_1}}$ and $X_{u_{\chi_1}v_{\chi_2}}\to (g_{\chi_1})_{u_{\chi_1} \hat u_{\chi_1}}(g_{\chi_2})_{u_{\chi_2} \hat u_{\chi_2}}X_{\hat u_{\chi_1}\hat v_{\chi_2}}$ under chiral transformation $g_{\chi_i}\in SU(2)_{\rm L,R}^f$. Therefore, the first few building blocks with lower chiral order are constructed as follows~\cite{Bijnens:2017xrz}: 
\begin{align}\nonumber
\calO(p^0):~
&(Ui\tau^2)_{x_Ry_L}\;, 
&&(u\Psi)_{x_R}\;,
&&(u^\dagger\Psi)_{x_L}\;,
\\\nonumber
\calO(p^1):~
&(uu_\mu ui\tau^2)_{x_Ry_L}\;,
&&(u u_\mu u^\dagger i\tau^2)_{x_Ry_R}\;,
&&(u^\dagger u_\mu u i\tau^2)_{x_Ly_L}\;,
\\
\calO(p^2):~
&(\chi i\tau^2)_{x_Ry_L}\;,
&&(\chi^\dagger i\tau^2)_{x_Ly_R}\;,
&&\cdots\;,
\end{align}
where for $\calO(p^2)$ we just show a few examples, and the full list should include terms with two $u_\mu$s, field strength tensors for the vector external sources, etc. Due to the fact that $x\tau^2=\tau^2x^*$ for $x\in SU(2)$, the other possible ${\cal O}(p^0)$ and ${\cal O}(p^1)$ building blocks are not independent: $(U^\dagger i\tau^2)_{x_Ly_R}=-(Ui\tau^2)_{y_Rx_L}$ and $(u^\dagger u_\mu u^\dagger i\tau^2)_{x_Ly_R}=-(uu_\mu ui\tau^2)_{y_Rx_L}$. Note that the two ${\cal O}(p^1)$ objects with the same chirality are anti-symmetric under the exchange of the two indices. In the above, we neglect those $\calO(p^0)$ terms with derivatives acting on the nucleon field like $(uD_\mu \Psi)_{x_R}$ since, at leading ${\cal O}(p^{0})$ order of the matched operators, they are actually redundant and can be transformed into those non-derivative operators plus higher order terms (${\cal O}(p^{\geq 1})$) via the equation of motion (EoM) of nucleon fields and integration by parts (IBP) relations. At the same time they can not yield the operators in Eq.~\eqref{chpt:LO} used for the analysis in this paper. For instance, we consider the lepton vector current operators, one possible leading ${\cal O}(p^{0})$ order operator with derivative acting on the nucleon field takes $(\Psi^{\T}_{a}iD^{\mu}\Psi_{b})P(u) j_{V,\mu}$ with $P(u)$ a polynomial of pion field $u$. It can be reduced as follows 
\begin{align*}
&2\Psi^{\T}_{a}CiD^{\mu}\Psi_{b}P(u) j_{V,\mu}
\\
&=\Psi^{\T}_{a}C(i\slashed{D}\gamma^{\mu}+\gamma^{\mu}i\slashed{D})\Psi_{b}P(u) j_{V,\mu}
\\
&\overset{\rm EoM}{=}\Psi^{\T}_{a}C(i\slashed{D}\gamma^{\mu}+m_{N}\gamma^{\mu})\Psi_{b}P(u) j_{V,\mu}+{\cal O}(p^{1})
\\
&\overset{\rm IBP}{=}-\Psi^{\T}_{a}C(\overleftarrow{\slashed{D}}\gamma^{\mu}-m_{N}\gamma^{\mu})\Psi_{b}P(u) j_{V,\mu}-\Psi^{\T}_{a}C\gamma^{\nu}\gamma^{\mu}\Psi_{b}\partial_{\nu}[P(u) j_{V,\mu}] +{\cal O}(p^{1})\;,
\end{align*}
where in the last line, the first term can be further reduced by the EoM to be a derivative-free operator, the second term is again ${\cal O}(p^{1})$ due to the derivative acting on the pion and external source. The other ${\cal O}(p^{0})$ terms with derivatives acting on the nucleon field can be reduced in a similar fashion. 

Without consideration of the operators involving covariant derivatives acting on the nucleon fields, the leading order matching results for all the relevant chiral irreps are shown in Tab.~\ref{tab:matchingchpt}. Where the spurion fields are easily identified from Tabs.~\ref{tab:chiralirrep_pp}-\ref{tab:chiralirrep_nn} for each specific operator, and $g_{i}$ are the unknown LECs parametrizing non-perturbative QCD effect. One should keep in mind that for each independent irrep there is a corresponding LEC. 
 
Here we again take the operator ${\cal O}_{Q^6L^2}^{S,(S)}$ as an example to show the relevant spurion fields and the chiral matching result. From Eq.~\eqref{ope:Oq6L2}, the six-quark part of the matched three LEFT operators ${\cal O}_{1LLL,b}^{(pp)S,-}, {\cal O}_{1LLL,b}^{(nn)S},{\cal O}_{1LLL,b}^{(pn)S}$ can be rewritten as
\begin{align}\nonumber
&(u^{i\T}_{L} Cu^j_{L})  (u^{k\T}_{L} Cd^l_{L})(u^{m\T}_{L} Cd^n_{L})T^{SAA}_{\{ij\}[kl][mn]} =(\delta^u_1\delta^v_1)T_{u_Lv_L}\equiv \theta^{u_Lv_L}_{(11)}T_{u_Lv_L}\;,
\\\nonumber
& (u^{i\T}_{L} Cd^j_{L})  (u^{k\T}_{L} Cd^l_{L})(u^{m\T}_{L} Cd^n_{L})T^{SAA}_{\{ij\}[kl][mn]} ={1\over 2}(\delta^u_1\delta^v_2+\delta^u_2\delta^v_1)T_{u_Lv_L}\equiv \theta^{u_Lv_L}_{(12)}T_{u_Lv_L}\;,
\\
&(d^{i\T}_{L} Cd^j_{L})  (u^{k\T}_{L} Cd^l_{L})(u^{m\T}_{L} Cd^n_{L})T^{SAA}_{\{ij\}[kl][mn]} =(\delta^u_2\delta^v_2)T_{u_Lv_L}\equiv \theta^{u_Lv_L}_{(11)}T_{u_Lv_L}\;,
\end{align}
where we have defined the spurion fields as 
\begin{align}
&\theta^{u_Lv_L}_{(11)}=\delta^u_1\delta^v_1\;,
&&\theta^{u_Lv_L}_{(12)}={1\over 2}(\delta^u_1\delta^v_2+\delta^u_2\delta^v_1)\;,
&&\theta^{u_Lv_L}_{(11)}= \delta^u_2\delta^v_2\;,
\end{align}
and $T_{u_Lv_L}=T_{v_Lu_L}$ is a three-dimensional irrep tensor under the group $SU(2)_{\rm L}^f$ but a singlet under $SU(2)_{\rm R}^f$, and takes the form
\begin{align}
T_{u_Lv_L}={1\over 4}\epsilon_{wx}\epsilon_{yz}(q^{i\T}_{Lu} Cq^j_{Lv})  (q^{k\T}_{Lw} Cq^l_{Lx})(q^{m\T}_{Ly} Cq^n_{Lz})T^{SAA}_{\{ij\}[kl][mn]} 
\overset{G_\chi}{\to} L_u^x L_v^{y}T_{xy}\in ({\bf 3}_L,{\bf 1}_R)\;.
\end{align}
According to the previous procedures, the leading order chiral realization of $T_{u_Lv_L}$ is at ${\cal O}(p^0)$ and formed by two $(u^\dagger \Psi)_{u_L}$s to have the same baryon number and chiral structure. In addition, the Lorentz covariance further restricts $T_{u_Lv_L}$ to be a scalar and have the general form $(u^\dagger)_{u_La}(u^\dagger)_{v_La} \Psi_{b}\left[ g_{3\times 1}+\hat g_{3\times 1} \gamma_5  \right]\Psi_b$ as shown in Tab.~\ref{tab:matchingchpt}. Then the complete matching for ${\cal O}_{Q^6L^2}^{S,(S)}$ together with its Wilson coefficient becomes 
{\footnotesize
\begin{align}
C_{Q^6L^2}^{S,(S)}{\cal O}_{Q^6L^2}^{S,(S)}\to 4C_{Q^6L^2}^{S,(S)}\left( \theta^{u_Lv_L}_{(11)}j_{S,-}^{\ell\ell^\prime}- 2\theta^{u_Lv_L}_{(12)}j_{S}^{\ell\nu^\prime}+\theta^{u_Lv_L}_{(22)}j_{S}^{\nu\nu^\prime}\right)(u^\dagger)_{u_La}(u^\dagger)_{v_Lb} \Psi_{a}\left[ g_{3\times 1}+\hat g_{3\times 1} \gamma_5  \right]\Psi_b\;.
\end{align}}
\indent Keeping the $\calO(p^0)$ terms in Tab.~\ref{tab:matchingchpt} and expanding them to zeroth order in the pion fields, we have
\begin{align}\nonumber
O^S_{3\times1,i}&\to \theta^{u_Lv_L}_{(\alpha\beta)}[\Psi^\T_{u_L}C(g_{3\times 1,i}+\hat g_{3\times 1,i}\gamma_5)\Psi_{v_L}]\;,
\\\nonumber
O^S_{5\times3}&\to\theta^{u_Lv_Lw_Lx_Ly_Rz_R}_{(\alpha\beta\gamma\rho)(\sigma\tau)}
\epsilon_{y_Rw_L}\epsilon_{z_Rx_L}[\Psi^\T_{u_L}C(g_{5\times 3}+\hat g_{5\times 3}\gamma_5)\Psi_{v_L}]\;,
\\\nonumber
O^{V,\mu}_{2\times2,i}&\to \theta^{u_Lv_R}_{\alpha\beta}[\Psi^\T_{u_L}C\gamma^\mu(g_{2\times 2,i}+\hat g_{2\times 2,i}\gamma_5)\Psi_{v_R}]\;,
\\\nonumber
O^{V,\mu}_{4\times2,i}&\to g_{4\times 2,i}\theta^{u_Lv_Lw_Lx_R}_{(\alpha\beta\gamma)\rho}\epsilon_{x_Rw_L}[\Psi^\T_{u_L}C\gamma^\mu\gamma_5 \Psi_{v_L}]\;,
 \\\nonumber
O^{V,\mu}_{4\times4}&\to \theta^{u_Lv_Lw_Lx_Ry_Rz_R}_{(\alpha\beta\gamma)(\rho\sigma\tau)}\epsilon_{y_Rv_L}\epsilon_{z_Rw_L} [\Psi^\T_{u_L}C\gamma^\mu(g_{4\times 4}+\hat g_{4\times 4}\gamma_5) \Psi_{x_R}]\;,
\\\nonumber
O^{T,\mu\nu}_{1\times1,i}&={1\over2}\epsilon^{ab}[\Psi^\T_{a}C\sigma^{\mu\nu}(g_{1\times 1,i}+\hat g_{1\times 1,i}\gamma_5)\Psi_b]\;,
 \\
O^{T,\mu\nu}_{3\times3,i}&\to \theta^{u_Lv_Lw_Rx_R}_{(\alpha\beta)(\gamma\rho)}\epsilon_{x_Rv_L}[\Psi^\T_{u_L}C\sigma^{\mu\nu}(g_{3\times 3,i}+\hat g_{3\times 3,i}\gamma_5)\Psi_{w_R}]\;,
\label{chiralresult}
\end{align}
and the similar expressions for the parity conjugates $\tilde O^S_{1\times3,i}$, $\tilde O^S_{3\times5}$ and $\tilde O^{V,\mu}_{4\times2,i}$. Taking the specific expressions of the spurion fields in~Tabs.~\ref{tab:chiralirrep_pp}-\ref{tab:chiralirrep_nn}  into consideration, we can obtain the matching results for the Wilson coefficients of the operators in Eqs.~(\ref{ppll}-\ref{nnnunu}) as the function of the LECs and the LEFT/SMEFT Wilson coefficients, the full matching results from the LEFT and SMEFT operators are listed in  Appendix~\ref{app:commatching}. 

In the following, we show as an example the matching results from the SMEFT operators $C_{Q^6L^2}^{S,(S)}$ and $C_{Q^6L^2}^{T,(S)}$. In terms of their LEFT counterparts ${\cal Q}_{1LLL}^{(pp,pn,nn)S}$ and ${\cal Q}_{1LLL}^{(pn)T}$ , we have 
\begin{align}\nonumber
&C_{R(L)}^{(pp)S}=g_{3\times1,a}\left({3\over5}C_{1LLL,a}^{(pp)S,\pm}+C_{1LLL,b}^{(pp)S,\pm} \right)+\cdots\;,
&&C_{5R(L)}^{(pp)S} =C_{R(L)}^{(pp)S}|_{g\to \hat g}\;, 
\\\nonumber
&C_L^{(pn)S}= g_{3\times1,a}\left({9\over5}C_{1LLL,a}^{(pn)S}+C_{1LLL,b}^{(pn)S}\right)+\cdots\;,
&&C_{5L}^{(pn)S} =C_L^{(pn)S}|_{g\to \hat g}\;,
\\\nonumber
&C^{(pn)T}=g_{1\times 1,a}^r\left( {1\over3}C_{1LLL,a}^{(pn)T}+C_{1LLL,b}^{(pn)T}\right)+\cdots\;,
\\
&C_L^{(nn)S}=g_{3\times1,a}\left({3\over5}C_{1LLL,a}^{(nn)S}+C_{1LLL,b}^{(nn)S}\right)+\cdots\;,
&&C_{5L}^{(nn)S} =C_L^{(nn)S}|_{g\to \hat g}\;,
\end{align}
where $\cdots$ stand for contributions from other operators. For the tensor case $C_L^{(pn),T}$, we have used the identity $\sigma^{\mu\nu}\gamma_5=i\epsilon^{\mu\nu\rho\sigma}\sigma_{\rho\sigma}/2$ to eliminate the operator $(p^\T C\sigma_{\mu\nu}\gamma_5n)(\ell_R^\T C\sigma^{\mu\nu}\nu_L)$ in favor of $(p^\T C\sigma_{\mu\nu}n)(\ell_R^\T C\sigma^{\mu\nu}\nu_L)$ with the shifted LECs $g_{i\times i,x}^r=g_{i\times i ,x}-\hat g_{i\times i,x}$. After neglecting the QCD running effect and replacing the LEFT Wilson coefficients by the SMEFT ones $C_{Q^6L^2}^{S,(S)}$ and $C_{Q^6L^2}^{T,(S)}$ as the way shown in Tab.~\ref{tab:matchingLEFT}, we find the above results are simplified to become
\begin{align}\nonumber
&C_L^{(pp)S}=4g_{3\times1,a}C_{Q^6L^2}^{S,(S)}+\cdots\;,
&&C_{5L}^{(pp)S}=C_{L}^{(pp)S}|_{g\to \hat g}\;,
\\\nonumber
&C_L^{(pn)S}=-8g_{3\times1,a}C_{Q^6L^2}^{S,(S)}+\cdots\;,
&&C_{5L}^{(pn)S} =C_L^{(pn)S}|_{g\to \hat g}\;, 
\\\nonumber
&C^{(pn)T}=-16g_{1\times 1,a}^rC_{Q^6L^2}^{T,(S)}+\cdots\;,
\\
&C_L^{(nn)S}=4g_{3\times1,a}C_{Q^6L^2}^{S,(S)}+\cdots\;,
&&C_{5L}^{(nn)S}=C_L^{(nn)S}|_{g\to \hat g}\;.
\end{align}
Once the hadronic LEC $g_{i}$ is known one can obtain the dinucleon and dilepton transitions. In the following section we will discuss how this can be done and obtain constraints on the LEFT and SMEFT operators. 

Before doing that, let us have some discussion about the LECs. By the parity invariance of QCD, we expect the LECs of an operator and its parity conjugate are the same up to a sign determined by the parity transformation property of the quark and the corresponding hadron level operators, i.e., $ g_{i\times j}=\pm g_{j\times i}$. Particularly, for the scalar current case, we have
\begin{align}
&g_{3\times 1,i}=-g_{1\times 3,i}\;,
&&{\hat g}_{3\times 1,i}=+\hat g_{1\times 3,i}\;,
&&g_{3\times 5}=-g_{5\times 3}\;,
&& {\hat g}_{3\times 5}=+\hat g_{5\times 3}\;.
\end{align}
The numerical value of $g_{3\times 1,i}$ and $g_{3\times 5}$ can be determined by the LQCD results for the $n{\rm-}\bar n$ oscillation matrix elements~\cite{Rinaldi:2019thf}. This is because the quark sectors of the 14 operators with a scalar lepton current contributing to $nn\to\bar\nu\bar\nu^\prime$ transitions are exactly the 14 dim-9 operators mediating $n{\rm-}\bar n$ oscillation.  Neglecting the lepton current, the scalar chiral operators in the irreps $({\bf 1}_L,{\bf 3}_R)$ and $({\bf 5}_L,{\bf 3}_R)$ in Tab.~\ref{tab:chiralirrep_nn} have the following correspondence with the dim-9 chiral operators for the $n{\rm-}\bar n$ oscillation~\cite{Rinaldi:2019thf}
\begin{align}\nonumber
&Q_1=-4\tilde P_{1,b}^{(nn)S}\;,
&&Q_2=-4\tilde P_{2,b}^{(nn)S}\;,
&&Q_3=-4 P_{3,b}^{(nn)S}\;,
\\
&Q_5=P_{4}^{(nn)S}\;,
&&Q_6=-4P_{2,a}^{(nn)S}\;,
&&Q_7=-4 P_{3,a}^{(nn)S}\;. 
\end{align}
After comparing the LQCD results on the $n{\rm-}\bar n$ matrix elements from $Q_i$ and our chiral matching results for $P_i^{(nn)S}$ in Eq.~\eqref{chiralresult}, we find 
\begin{align}
&g_{1\times3,a,c}\sim -6\times 10^{-6}~\rm GeV^6\;, 
&&g_{1\times3,b}\sim 1\times 10^{-5}~\rm GeV^6\;,  
&&g_{5\times3}\sim 5\times 10^{-6}~\rm GeV^6\;,
\label{coupling}
\end{align}
at the scale $\Lambda\sim 2$~GeV or so and the uncertainty in~\cite{Rinaldi:2019thf} is neglected. 

Except the above LECs, the rest of the hadronic couplings, such as the $\hat g_i$ and the ones related to the vector and tensor current operators, have not been determined. For these LECs, we will use dimensional analysis as a guide to illustration. Since the transition from quarks to hadrons is through the non-perturbative QCD dynamics, the only relevant scale $\Lambda_{\rm QCD} \sim 200$ MeV will come into play. To make the dimensionality correct, one can take as a rough estimate the couplings to be of order $\Lambda_{\rm QCD}^6\sim 6.4\times10^{-5}~\rm GeV^6$. This is larger than the numbers in~\eqref{coupling}. But considering the large uncertainties involved, we can take it as a guide for estimate.

Alternatively, these LECs can be estimated via the naive dimensional analysis by keeping track of $4\pi$ factors~\cite{Manohar:1983md, Gavela:2016bzc} and relating the hadronic matrix element to the chiral symmetry breaking scale $\Lambda_\chi = 1190$ MeV. One introduces ``reduced'' couplings for the hadron and quark level operators and to match them~\cite{Weinberg:1989dx}.  For a coupling constant $g$ appearing in an interaction of dimensionality $D$ in mass which containing $N$ field operators, the reduced coupling is $(4\pi)^{2-N} \Lambda_\chi^{D-4} g$. For our case, the hadronic operators involve two fields ($N=2$) with a coupling $g_N$ as given in Eqs.~(\ref{ppll}-\ref{nnnunu}). The quark operators involve six quarks ($N=6$) with a coupling $g_q$, therefore we would obtain  $g_N/\Lambda_\chi =  g_q (\Lambda_\chi)^5/(4\pi)^4$ by matching. Setting $g_q = 1$, one would have the hadronic coupling to be of order  $ (\Lambda_\chi)^6 /(4\pi)^4 \sim 11 \times10^{-5}$ which is about 2 times the above dimensional estimate. For our numerical estimate of the undetermined LECs in the next section, we assume their value to be the similar order as $\Lambda_{\rm QCD}^6$ .

Last, we comment the heavy baryon $\chi$PT formalism~\cite{Jenkins:1990jv}, which is a consistent framework for the power counting of nucleon fields. In this framework, for our case, the anti-nucleon mode is integrated out and the remaining heavy nucleon doublet is defined as ${\cal N}_v(x)=e^{+imv\cdot x}P_{v+}\Psi$ with $P_{v\pm}\equiv{1\over2}(1\pm \slashed{v})$, where $v$ is a reference velocity satisfying $v^2=1$ and usually taken as $v=(1,\bf 0)$. The chiral power counting for $D_\mu {\cal N}_v(x)$ is ${\cal O}(p^1)$ as promised in this formalism. To leading order of chiral matching, there should have no derivatives acting on the nucleon fields. Since
\begin{align}
&{\cal N}^\T_v C {\cal N}_v
={\cal N}^\T_v P_{v+}^\T C P_{v+}{\cal N}_v
={\cal N}^\T_v CP_{v-} P_{v+}{\cal N}_v
=0\;,
\end{align}
we find that the matched operators using heavy nucleon fields can be directly obtained from the matched operators using the relativistic nucleon fields in Tab.~\ref{tab:matchingchpt} by replacing the nucleon field $\Psi$ by ${\cal N}_v$ together with the omission of the operators with a scalar nucleon current. The use of relativistic formalism is its explicit Lorentz invariance and chiral symmetry, which are convenient for the loop calculations.

\section{Dinucleon and dilepton transition rate}
\label{sec:cal}

Combining the previous sections for the effective interactions from the SMEFT, to LEFT, then to B$\chi$PT, in this section we will collect all pieces together and calculate the dinucleon to dilepton decay rate.
Denote collectively $NN^\prime\in\{pp, pn, nn\}$ and $l_\alpha l_\beta\in \{\ell^+\ell^{\prime+},\ell^+\bar\nu^{\prime},\bar\nu\bar\nu^\prime\}$, then the decay rate for dinucleon $NN^\prime$ to dilepton $l_\alpha l_\beta$ transition in nucleus can be estimated in the following way~\cite{Goity:1994dq}
{\small
\begin{align}
\Gamma_{NN^\prime\to l_\alpha l_\beta}={1\over (2\pi)^3\sqrt{\rho_N \rho_{N^\prime} }}\int d^3k_1d^3k_2\rho_N(k_1)\rho_{N^\prime}(k_2)v_{\rm rel.}(1-\mathbf{v_1\cdot v_2})\sigma(NN^\prime\to l_\alpha l_\beta)\;,
\end{align}}
where $\rho_N(k)$ is the nucleon density distribution in momentum space and $\rho_N$ is the average nucleon density defined as $\rho_N=\int d^3k \rho_N(k)/(\sqrt{2\pi})^3$.  $\mathbf{v_1}(\mathbf{v_2})$ is the velocity of the nucleon $N(N^\prime)$. The total cross-section for the free nucleon scattering process $N(k_1)N^\prime(k_2)\to l_\alpha(p_1)l_\beta(p_2)$ is
\begin{align}
\sigma(NN^\prime\to l_\alpha l_\beta)={1\over S}{1\over 4E_1E_2v_{\rm rel.}}\int d\Pi_2\overline{\left|{\cal M}_{NN^\prime\to l_\alpha l_\beta}\right|}^{2}, \label{cross}
\end{align}
where $E_{1}(E_{2})$ is the energy of the initial state nucleon $N (N^\prime)$, and $S$ is a symmetry factor and equals 2 for identical final leptons $l_\alpha=l_\beta=\{e^+,\mu^+,\bar\nu_e,\bar\nu_\mu,\bar\nu_\tau\}$, otherwise $S=1$. $d\Pi_2$ is the relativistically invariant two-body phase space.  

The dinucleon collisions occur at low relative velocity $v$, they may be affected by some other SM interaction resulting in modification of the cross sections. For example for $pp \to \ell^+_\alpha\ell^+_\beta$, there is a repulsive force between the two protons due to electrodynamics which reduces the cross section. The effect of exchange photons between protons is best captured by the Sommerfeld effect~\cite{Hisano:2004ds,He:2009ra}. Because the repulsive nature of the electromagnetic force there is a reduction of the cross section, the original cross section $\sigma$ is modified to $\tilde \sigma = \sigma SF$ with $SF$ given by $(\alpha_{\rm em} \pi/v)/({\rm exp}[\alpha_{\rm em} \pi/v] -1)$. This reduction factor $SF$ could be very severe if $v$ is very small. For the case in question, the typical $v$ of about $0.1$ leads to $SF \approx 0.9$. Had $v$ be 0.01, $SF$ is further reduced to 0.26. Therefore the case we are considering the reduction is not severe. One expects such effects for $np$ and $nn$ cases will be smaller. We still use Eq.~\eqref{cross} as our order of magnitude estimate.

To a good approximation for the  oxygen nuclei $^{16}$O, we treat the nucleons to be quasi-free and neglect the small effects due to the nucleon Fermi motion and nuclear binding energy, as well as the above Sommerfeld suppression effect. The average nuclear matter density $\rho_N$ approximately equals $0.25~{\rm fm}^{-3}$ for either proton or neutron. Then the transition rate reduces into
\begin{align}
\Gamma_{NN^\prime\to l_\alpha l_\beta}={1\over S}{\rho_N\over 4 m_N^2}\overline{\left|{\cal M}_{NN^\prime\to l_\alpha l_\beta}\right|}^{2}\Pi_2\;,
\end{align}
where we have neglected the mass difference between proton and neutron and taken both to be $m_N=(m_p+m_n)/2$.  The two-body final state phase factor $\Pi_{2}$ takes
\begin{align}
\Pi_{2}={1\over 8\pi}[\lambda(1,\delta_\alpha,\delta_\beta)]^{1/2}\;,~
\delta_\alpha={m_\alpha^2\over 4m_N^2}\;,~
\lambda(x,y,z)=x^2+y^2+z^2-2(xy+yz+zx)\;. 
\end{align}
Working on the center of mass frame of the two-nucleon system and neglecting the nucleons' velocity, from the effective interaction in Eq.~\eqref{chpt:LO}, then the spin-averaged squared amplitudes are
\begin{align}\nonumber
\overline{\left|{\cal M}_{pp\to \ell^+_\alpha\ell^{+}_\beta}\right|}^{2}&=32m_N^4\left[ S^2\left(1-\delta_\alpha-\delta_\beta\right) \big(\big|C_{5L}^{(pp)S}\big|^2+\big|C_{5R}^{(pp)S}\big|^2\big)\right.
\\\nonumber
&\left.
+\left(\delta_\alpha+\delta_\beta -(\delta_\alpha-\delta_\beta)^2 \right)\big|C^{(pp)V}\big|^2
-4S^2\sqrt{\delta_\alpha\delta_\beta}{\rm Re}\big[C_{5L}^{(pp)S}C_{5R}^{(pp)S*}\big]
\right.
\\
&\left.
+2S\big(\left(1+\delta_\alpha-\delta_\beta\right)\sqrt{\delta_\beta}{\rm Re} \big[C^{(pp)V}C_{5R}^{(pp)S*}\big]
-(L, \delta_\alpha\leftrightarrow R, \delta_\beta)\big)\right]+{\cal O}(v^2)
\;,
\label{squaredpp}
\\\nonumber
\overline{\left|{\cal M}_{pn\to \ell^+_\alpha\bar\nu_\beta}\right|}^{2}&=
8m_N^4(1-\delta_\alpha)\left[\big|C_{5L}^{(pn)S}\big|^2+\delta_\alpha\big|C_{5L}^{(pn)V}\big|^2
+(2+\delta_\alpha)\big|C_{L}^{(pn)V}\big|^2+4(1+2\delta_\alpha)
\right.
\\
&\left.
\times\big|C^{(pn)T}\big|^2
-2\sqrt{\delta_\alpha}{\rm Re}\big[C_{5L}^{(pn)S} C_{5L}^{(pn)V*}-6C_{L}^{(pn)V}C^{(pn)T*}\big]
\right]+{\cal O}(v^2)\;,
\\
\overline{\left|{\cal M}_{nn\to \bar\nu_\alpha\bar\nu_\beta}\right|}^{2}&=32S^2m_N^4\big|C_{5L}^{(nn)S}\big|^2+{\cal O}(v^2)\;,
\label{squarednn}
\end{align}
where we see that the contribution to $pp\to\ell^+_\alpha\ell^+_\beta \ (nn\to\bar\nu\bar\nu^\prime)$ transition from the operators ${\cal O}^{(pp)S}_{L,R} \ ({\cal O}^{(nn)S}_{L})$ vanishes in that $pp \ (nn)$ annihilation through such operators is $p$-wave ($\propto v^2$), whereas the contribution from the vector operator ${\cal O}^{(pp)V}$ is helicity-suppressed ($\propto \delta_{\alpha,\beta}$). For $pn\to\bar\nu\bar\nu^\prime$, the vanishing from ${\cal O}^{(pn)S}_{L}$ has a similar reason.

The partial lifetime characterizing the matter instability is the inverse of the rate $(\tau/{\cal B}_i)_i\equiv\Gamma^{-1}_i$, where ${\cal B}_i$ is a branching ratio. Taking the experimental lower limits on the partial lifetime in Tab.~\ref{tab:exp} into consideration, and by the relation $\Gamma^{-1}_i=(\tau/{\cal B}_i)_i\geq \tau_{\rm exp}$, we can obtain the constraints on the coefficients in Eqs.~(\ref{squaredpp}-\ref{squarednn}). Assuming one term dominates at a time, then the result is shown in Tab.~\ref{tab:constraint}, where the upper limit on the Wilson coefficients is classified in terms of the final state leptons. We see the most stringent limit is for the operator ${\cal O}_{5L,R}^{(pp)S}|_{ee,e\mu,\mu\mu}$ in which $C_{5L,R}^{(pp)S}|_{ee,e\mu,\mu\mu}\leq 6.8,9.4, 6.6\times10^{-32}~\rm GeV^{-2}$, this is because the strong experimental limits on these decay modes.  
\begin{table}
\centering
\resizebox{\linewidth}{!}{
\begin{tabular}{|c|c|c|c|c|c|}
\hline
WC & [${10^{-30}\rm~GeV^{-2}}$] & WC & [${10^{-30}\rm ~GeV^{-2}}$] & WC &  [${10^{-30}\rm ~GeV^{-2}}$]
\\\hline
$C_{L,R}^{(pp)S}$   & -
&$C_{5L,R}^{(pp)S}|_{ee,e\mu,\mu\mu}$ & $0.068,~0.094,~0.067$
&$C^{(pp)V}|_{ee,e\mu,\mu\mu}$ & $350,~1.7,~1.7$
\\\hline
$C_{L}^{(pn)S}$   & -
&$C_{5L}^{(pn)S}|_{e\nu,\mu\nu,\tau\nu}$ & $0.77,~0.84,~22$
&$C_{L}^{(pn)V}|_{e\nu,\mu\nu,\tau\nu}$ & $0.55,~0.59,~13$
\\
-&-
&$C_{5L}^{(pn)V}|_{e\nu,\mu\nu,\tau\nu}$  & $2800,~15,~23$
&$C^{(pn)T}|_{e\nu,\mu\nu,\tau\nu}$ & $0.39,~0.42,~6.6$
\\\hline
$C_{L}^{(nn)S}$   & -
&$C_{5L}^{(nn)S}|_{\nu_\alpha\nu_\alpha,\nu_\alpha\nu_{\beta\neq\alpha} }$ & $3.7,~5.3$
 & &
\\\hline
\end{tabular}
}
\caption{The upper limit on the Wilson coefficients (WC) of the dim-6 hadronic operators in Eqs.~(\ref{ppll}-\ref{nnnunu}). Where we take the current experimental lower limit on the dinucleon to dilepton transitions in Tab.~\ref{tab:exp} to set the bound.  }
\label{tab:constraint}
\end{table}
\begin{table}[!t]
\centering
\resizebox{\linewidth}{!}{
\begin{tabular}{|c| c | c | c |}
\hline
\multirow{2}{*}{~~SMEFT WCs~~}
& \multicolumn{1}{|c |}{$pp\to e^+e^+,~e^+\mu^+,~\mu^+\mu^+$} & \multicolumn{1}{|c |}{$pn\to e^+\nu,~\mu^+\bar\nu,~\tau^+\bar\nu$} & \multicolumn{1}{|c |}{$nn\to \bar\nu_\alpha\bar\nu_\alpha,~\bar\nu_\alpha\bar\nu_{\beta\neq\alpha}$}
\\
& \multicolumn{1}{|c |}{$\Lambda_{\rm NP}\equiv |C_i|^{-{1\over 8}}~[\rm TeV]$} & \multicolumn{1}{|c |}{$\Lambda_{\rm NP}\equiv |C_i|^{-{1\over 8}}~[\rm TeV]$} & \multicolumn{1}{|c |}{$\Lambda_{\rm NP}\equiv |C_i|^{-{1\over 8}}~[\rm TeV]$}
\\\hline
$C_{u^3d^3L^21}^{S,(A)}$
& \multicolumn{1}{|c |}{-}
&$2.04,~2.02,~1.34\times \left[{\hat g_{1\times3,a}\over \Lambda_{\rm QCD}^6}\right]^{1\over 8}$
&  \multicolumn{1}{|c |}{-}
\\
$C_{u^3d^3L^22}^{S,(A)}$
&  \multicolumn{1}{|c |}{-}
&$1.89,~1.87,~1.25\times \left[{\hat g_{1\times3,a}\over \Lambda_{\rm QCD}^6}\right]^{1\over 8}$
& \multicolumn{1}{|c |}{-}
\\
$C_{u^2d^2Q^2L^21,3}^{S,(S)}$
&~~~~~~~$2.35,~2.26,~2.36\times \left[{\hat g_{3\times1,c}\over \Lambda_{\rm QCD}^6}\right]^{1\over 8}$~~~~~~~
&~~~~~~~$1.89,~1.87,~1.25\times \left[{\hat g_{3\times1,c}\over \Lambda_{\rm QCD}^6}\right]^{1\over 8}$~~~~~~~
&~~~~~~~ $1.43,~1.37\times \left[{\hat g_{3\times1,c}\over \Lambda_{\rm QCD}^6}\right]^{1\over 8}$~~~~~~~
\\
$C_{u^2d^2Q^2L^22}^{S,(A)}$
&  \multicolumn{1}{|c |}{-}
& $2.07,~2.04,~1.36\times \left[{\hat g_{1\times3,b}\over \Lambda_{\rm QCD}^6}\right]^{1\over 8}$
& \multicolumn{1}{|c |}{-}
\\
$C_{udQ^4L^21}^{S,(A)}$
&  \multicolumn{1}{|c |}{-}
& $2.25,~2.23,~1.48\times \left[{\hat g_{1\times3,c}\over \Lambda_{\rm QCD}^6}\right]^{1\over 8}$
&  \multicolumn{1}{|c |}{-}
\\
$C_{udQ^4L^22}^{S,(S)}$
& $2.57,~2.46,~2.57\times \left[{\hat g_{3\times1,b}\over \Lambda_{\rm QCD}^6}\right]^{1\over 8}$
& $2.07,~2.04,~1.36\times \left[{\hat g_{3\times1,b}\over \Lambda_{\rm QCD}^6}\right]^{1\over 8}$
& $1.56,~1.49\times \left[{\hat g_{3\times1,b}\over \Lambda_{\rm QCD}^6}\right]^{1\over 8}$
\\
$C_{Q^6L^2}^{S,(S)}$
& $2.80,~2.69,~2.81\times \left[{\hat g_{3\times1,a}\over \Lambda_{\rm QCD}^6}\right]^{1\over 8}$
& $2.25,~2.23,~1.48\times \left[{\hat g_{3\times1,a}\over \Lambda_{\rm QCD}^6}\right]^{1\over 8}$
& $1.70,~1.63\times \left[{\hat g_{3\times1,a}\over \Lambda_{\rm QCD}^6}\right]^{1\over 8}$
\\
$C_{u^4d^2e^21}^{S,(S)}$
& $2.21,~2.12,~2.21\times \left[{\hat g_{1\times3,a}\over \Lambda_{\rm QCD}^6}\right]^{1\over 8}$
&  \multicolumn{1}{|c |}{-} &  \multicolumn{1}{|c |}{-}
\\
$C_{u^4d^2e^22}^{S,(S)}$
& $2.35,~2.26,~2.36\times \left[{\hat g_{1\times3,a}\over \Lambda_{\rm QCD}^6}\right]^{1\over 8}$
&  \multicolumn{1}{|c |}{-} &  \multicolumn{1}{|c |}{-}
\\
$C_{u^3dQ^2e^2}^{S,(S)}$
& $2.57,~2.46,~2.57\times \left[{\hat g_{1\times3,b}\over \Lambda_{\rm QCD}^6}\right]^{1\over 8}$
&  \multicolumn{1}{|c |}{-} &  \multicolumn{1}{|c |}{-}
\\
$C_{u^2Q^4e^2}^{S,(S)}$ 
& $2.80,~2.69,~2.81\times \left[{\hat g_{1\times3,c}\over \Lambda_{\rm QCD}^6}\right]^{1\over 8}$
&  \multicolumn{1}{|c |}{-} &  \multicolumn{1}{|c |}{-}
\\\hline
$C_{u^3d^2QeL1,3}^{V}$
& $0.808,~1.58,~1.58\times \left[{\hat g_{2\times2,d}\over \Lambda_{\rm QCD}^6}\right]^{1\over 8}$
& $1.81,~1.79,~1.22\times \left[{g_{2\times2,d}\over \Lambda_{\rm QCD}^6}\right]^{1\over 8}$
&  \multicolumn{1}{|c |}{-}
\\
$C_{u^3d^2QeL2}^{V}$
& $0.704,~1.37,~1.37\times \left[{\hat g_{2\times2,d}\over \Lambda_{\rm QCD}^6}\right]^{1\over 8}$
& $1.58,~1.56,~1.06\times \left[{g_{2\times2,d}\over \Lambda_{\rm QCD}^6}\right]^{1\over 8}$
&  \multicolumn{1}{|c |}{-}
\\
$C_{u^2dQ^3eL1}^{V}$
& $0.837,~1.63,~1.63\times \left[{\hat g_{2\times2,c}\over \Lambda_{\rm QCD}^6}\right]^{1\over 8}$
& $1.88,~1.86,~1.27\times \left[{g_{2\times2,c}\over \Lambda_{\rm QCD}^6}\right]^{1\over 8}$
&  \multicolumn{1}{|c |}{-}
\\
$C_{u^2dQ^3eL2}^{V}$
& $0.881,~1.72,~1.72\times \left[{\hat g_{2\times2,b}\over \Lambda_{\rm QCD}^6}\right]^{1\over 8}$
& $1.98,~1.96,~1.33\times \left[{ g_{2\times2,b}\over \Lambda_{\rm QCD}^6}\right]^{1\over 8}$
&  \multicolumn{1}{|c |}{-}
\\
$C_{u^2dQ^3eL3}^{V}$
& $0.881,~1.72,~1.72\times \left[{\hat g_{2\times2,c}\over \Lambda_{\rm QCD}^6}\right]^{1\over 8}$
& $1.98,~1.96,~1.33\times \left[{ g_{2\times2,c}\over \Lambda_{\rm QCD}^6}\right]^{1\over 8}$
& \multicolumn{1}{|c |}{-}
\\
$C_{uQ^5eL}^{V}$
& $0.961,~1.88,~1.88\times \left[{\hat g_{2\times2,a}\over \Lambda_{\rm QCD}^6}\right]^{1\over 8}$
& $2.16,~2.13,~1.45\times \left[{g_{2\times2,a}\over \Lambda_{\rm QCD}^6}\right]^{1\over 8}$
&  \multicolumn{1}{|c |}{-}
\\\hline
$C_{u^2d^2Q^2L^21,3}^{T,(S)}$
&  \multicolumn{1}{|c |}{-} 
& $2.25,~2.23,~1.35\times \left[{g_{1\times1,c}^r\over \Lambda_{\rm QCD}^6}\right]^{1\over 8}$
& \multicolumn{1}{|c |}{-}
\\
$C_{u^2d^2Q^2L^22}^{T,(A)}$
& \multicolumn{1}{|c |}{-}
& $1.89,~1.87,~1.33\times \left[{g_{3\times3,b}^r\over \Lambda_{\rm QCD}^6}\right]^{1\over 8}$
& \multicolumn{1}{|c |}{-}
\\
$C_{udQ^4L^21}^{T,(A)}$
&  \multicolumn{1}{|c |}{-}
& $2.07,~2.04,~1.45\times \left[{g_{3\times3,a}^r\over \Lambda_{\rm QCD}^6}\right]^{1\over 8}$
& \multicolumn{1}{|c |}{-}
\\
$C_{udQ^4L^22}^{T,(S)}$
&  \multicolumn{1}{|c |}{-}
& $2.46,~2.43,~1.72\times \left[{g_{1\times1,b}^r\over \Lambda_{\rm QCD}^6}\right]^{1\over 8}$
&  \multicolumn{1}{|c |}{-}
\\
$C_{Q^6L^2}^{T,(S)}$
&  \multicolumn{1}{|c |}{-}
& $2.68,~2.65,~1.88\times \left[{g_{1\times1,a}^r\over \Lambda_{\rm QCD}^6}\right]^{1\over 8}$
&  \multicolumn{1}{|c |}{-}
\\\hline
\end{tabular}
}
\caption{The constraint on the effective NP scale from the current experimental data in Tab.~\ref{tab:exp}. The flavor index is suppressed and can be easily recognized in terms of the transition mode. One should keep in mind that the SMEFT Wilson coefficients with a superscript `(A)' vanish for identical lepton flavors.}
\label{tab:SMEFTWC}
\end{table}

Next, we consider the above limits on the implications of SMEFT Wilson coefficients and the relevant NP scales. Based on the matching results in Eqs.~(\ref{h11}-\ref{h20}) in Appendix~\ref{app:commatching}, and assuming one term active in the matching result at a time, then the limits in Tab.~\ref{tab:constraint} translate into limits on the SMEFT Wilson coefficients as shown in Tab.~\ref{tab:SMEFTWC}.  In obtaining the results, we have taken our previous estimation of the LECs $g_i\sim \Lambda_{\rm QCD}^6\sim 6.4\times10^{-5}~\rm GeV^6$ as our benchmark value, therefore, the factor $(g_i/\Lambda_{\rm QCD}^6)^{1/8}$ is ${\cal O}(1)$. Up to the ${\cal O}(1)$ hadronic LECs ratio, the associated NP scale is found to be around $0.8-2.7~\rm TeV$ for all relevant operators. Here we see that, even the effective interactions are at dim 12, the matter instability puts a stringent limit on the NP scale. Similarly, we can set constraints on the LEFT operators. However, taking the assumption of the NP scale much higher than $\Lambda_{\rm EW}$, the above constraints on the SMEFT interactions are more illuminating in connection with NP scenarios, and thus we do not show the constraints on the LEFT interactions here for brevity.

Furthermore, we consider the contribution to the transitions from the dim-12 operators ${\cal O}_{Q^6L^2}^{S,(S)}$, ${\cal O}_{u^2d^2Q^2L^21,3}^{S,(S)}$ and ${\cal O}_{udQ^4L^22}^{S,(S)}$ containing purely left-handed lepton fields. By weak isospin symmetry, they can contribute to both three transition modes, and in particular, they are the only possible operators contributing to the $nn\to\bar\nu\bar\nu^\prime$ modes at leading order. From the previous discussion, the transition rates become
\begin{align}\nonumber
\tilde \Gamma_{pp\to \ell_\alpha^+\ell_\beta^+}&={S\over\pi}\left(1-\delta_\alpha-\delta_\beta\right)\sqrt{1-2(\delta_\alpha+\delta_\beta)+(\delta_\alpha-\delta_\beta)^2}m_N^2\rho_NC_{\alpha\beta}^2\;,
\\\nonumber
\tilde \Gamma_{pn\to \ell_\alpha^+  \bar \nu_\beta}&={1\over \pi}(1-\delta_\alpha)^2m_N^2\rho_NC_{\alpha\beta}^2\;,
\\
\tilde \Gamma_{nn\to  \bar\nu_\alpha\bar \nu_\beta}&={S\over \pi}m_N^2\rho_NC_{\alpha\beta}^2\;,
\label{decayrate}
\end{align}
where we use a `tilde' to represent such special contributions, and 
\begin{align*}
C_{\alpha\beta}=4\hat g_{3\times1,a}C_{Q^6L^2}^{S,(S),\alpha\beta}+2\hat g_{3\times1,b}C_{udQ^4L^22}^{S,(S),\alpha\beta}+\hat g_{3\times1,c}\big(C_{u^2d^2Q^2L^21}^{S,(S),\alpha\beta}+C_{u^2d^2Q^2L^23}^{S,(S),\alpha\beta}\big)
+{\hat g_{3\times5}\over 6}C_{u^2d^2Q^2L^21}^{S,(S),\alpha\beta}\;.
\end{align*}
Here we have added the lepton flavor indices for a more careful treatment\footnote{For $pn\to\ell^+_\alpha\bar\nu_\beta$ mode, the last term should take a minus sign. We neglect this sign difference here.}. By the weak isospin symmetry, we see the three transitions are related to each other.  From Eq.~\eqref{decayrate}, we obtain
\begin{align}\nonumber
\tilde \Gamma_{pn\to \ell_\alpha^+  \bar \nu_\beta}^{-1}&={S\left(1-\delta_\alpha-\delta_\beta\right)\sqrt{1-2(\delta_\alpha+\delta_\beta)+(\delta_\alpha-\delta_\beta)^2}\over (1-\delta_\alpha)^2 }\tilde \Gamma_{pp\to \ell_\alpha^+\ell_\beta^+}^{-1}\;,
\\\nonumber
\tilde \Gamma_{nn\to  \bar\nu_\alpha\bar \nu_\beta}^{-1}&=\left(1-\delta_\alpha-\delta_\beta\right)\sqrt{1-2(\delta_\alpha+\delta_\beta)+(\delta_\alpha-\delta_\beta)^2}\tilde\Gamma_{pp\to \ell_\alpha^+\ell_\beta^+}^{-1}\;,
\\
\tilde\Gamma_{nn\to  \bar\nu_\alpha\bar \nu_\beta}^{-1}&=S^{-1}(1-\delta_\alpha)^2\tilde\Gamma_{pn\to \ell_\alpha^+  \bar \nu_\beta}^{-1}\;.
\label{raterelation}
\end{align}  
Due to the stronger experimental limits on $pp\to \ell^+_\alpha\ell^+_\beta$ and 
$pn\to \ell^+_\alpha\bar\nu_\beta$, through the above relations, we can set new stronger limits on the neutral modes $nn\to \bar\nu_\alpha\bar\nu_\beta$. Taking the experimental limits for the charged modes into consideration and requiring $\tilde \Gamma_{pp\to \ell_\alpha^+\ell_\beta^+}^{-1}\gtrsim\tau_{pp\to \ell_\alpha^+\ell_\beta^+}^{\rm exp}$  and  $\tilde \Gamma_{pn\to \ell^+_\alpha\bar\nu_\beta}^{-1}\gtrsim\tau_{pn\to \ell^+_\alpha\bar\nu_\beta}^{\rm exp}$,  we obtain
\begin{align}
&\{\tilde\Gamma^{-1}_{nn\to \bar\nu_e\bar\nu_e},\tilde\Gamma^{-1}_{nn\to \bar\nu_e\bar\nu_\mu},\tilde\Gamma^{-1}_{nn\to \bar\nu_\mu\bar\nu_\mu}\} \gtrsim 4\times 10^{33}{~\rm yr}\;,
&&\{\tilde\Gamma^{-1}_{nn\to \bar\nu_e\bar\nu_\tau},\tilde\Gamma^{-1}_{nn\to \bar\nu_\mu\bar\nu_\tau}\} \gtrsim 2\times 10^{32}{~\rm yr}\;.
\end{align}
One can see the limits on $nn\to \bar\nu_\alpha\bar\nu_\beta$ are improved by $2-3$ orders of magnitude than the direct experimental search in Tab.~\ref{tab:exp}. On the other hand, if we assume the charged modes also exclusively mediated by the same operators, then the experimental bounds on $pp\to \ell^+_\alpha\ell^+_\beta$ imply the following new bounds on $pn\to \ell^+_\alpha\bar\nu_\beta$ for $\alpha,\beta=e,\mu$ flavors,
\begin{align}
&\{ \tilde\Gamma^{-1}_{pn\to e^+\bar\nu_e}, \tilde\Gamma^{-1}_{pn\to \mu^+\bar\nu_\mu}\} \gtrsim 8\times 10^{33}{~\rm yr}\;,
&&\{\tilde\Gamma^{-1}_{pn\to e^+\bar\nu_\mu}, \tilde\Gamma^{-1}_{pn\to \mu^+\bar\nu_e}\} \gtrsim 4\times 10^{33}{~\rm yr}\;,
\end{align}
which are also stronger than the current experimental bounds by at least an order of magnitude. Conversely, the experimental limit on $pn\to \ell^+_e\bar\nu_\tau$ can further translate into a bound on the transition $pp\to e^+\tau^+$, which is also kinetically allowed but has not yet been searched for experimentally. Based on Eq.~\eqref{raterelation} and the limit on $pn\to \ell^+_e\bar\nu_\tau$ in Tab.~\ref{tab:exp}, we obtain
\begin{align}
\tilde \Gamma^{-1}_{pp\to e^+\tau^+}\gtrsim 2\times 10^{34}~\rm yr\;,
\end{align} 
we see this bound is even more stronger than any other ones due to the small phase space. 

\begin{figure}
\centering
\includegraphics[width=10cm]{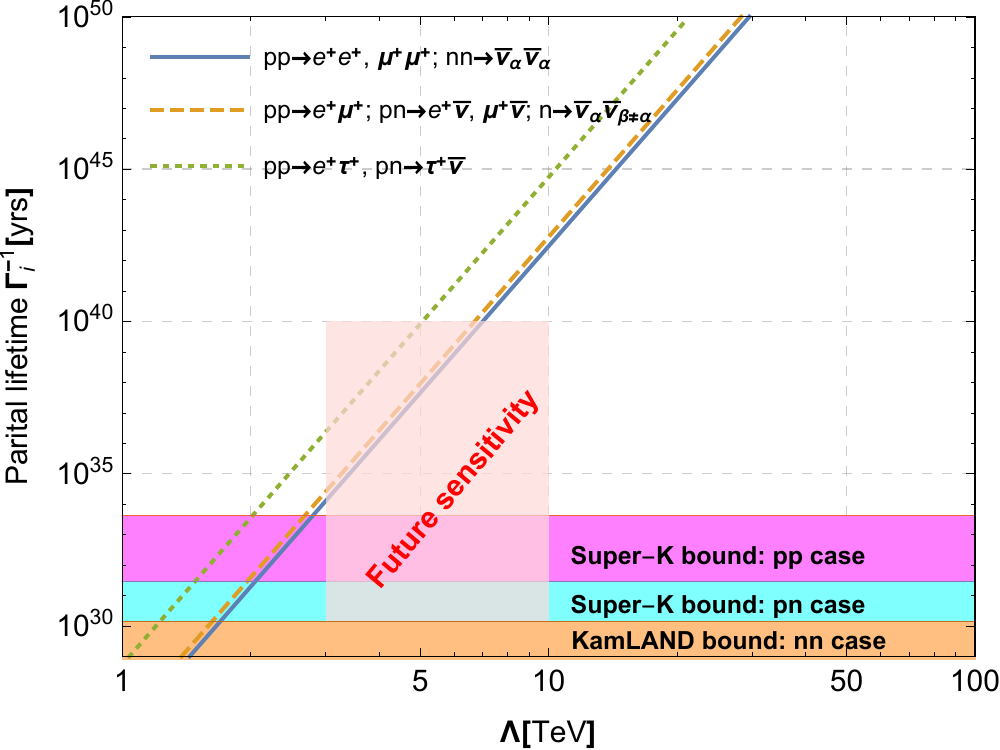}
\caption{The partial lifetime of the dinucleon to dilepton transitions as a function of the NP scale in the SMEFT. Where we assume the contribution from operator ${\cal O}_{Q^6L^2}^{S,(S)}$ and set $\Lambda\equiv \left[C_{Q^6L^2}^{S,(S)}\right]^{-1/8}$.  }
\label{Fig2}
\end{figure}

In Fig.~\ref{Fig2} we show the dependence of the partial lifetime on the NP scale. For simplicity, we only consider the contribution from operator ${\cal O}_{Q^6L^2}^{S,(S)}$ and take $\Lambda\equiv \left[C_{Q^6L^2}^{S,(S)}\right]^{-1/8}$. The relevant hadronic LEC $\hat g_{1\times 3,a}$ is set equal to $6.4\times 10^{-5}~\rm GeV$. From the figure we see the partial lifetime is very sensitive to the NP scale,  because of the large power dependence($\propto \Lambda^{16}$). For a future experimental sensitivity about $10^{40}$ yr the NP scale is pushed towards $5~{\rm TeV}$ or so. 

Finally, we make a brief comments on the concrete NP models and the collider signals. For a given NP model, one can integrate out the heavy new physics states and match onto the dim-12 SMEFT operators. In literature, there exist models contributing to dinucleon to dilepton transitions with $\Delta B=\Delta L=2$ but not giving rise to the $\Delta B=1$ nucleon decays or $\Delta B=2$ neutron-antineutron oscillation. In Refs.~\cite{Arnold:2012sd,Gardner:2018azu}, the authors considered a class of such models which involve new scalar-fermion and scalar-quartic interactions, meanwhile, the left-right symmetric models with extra-dimensions considered in Refs.~\cite{Girmohanta:2019fsx,Girmohanta:2020eav} can also do the job. On the other hand, from the above limit of NP scale we have set, one may expect the search of the $\Delta B=\Delta L=2$ NP signals at the current/future high energy colliders to be interesting. Ref.~\cite{Bramante:2014uda} has made such a try by studying the process $pp\to e^+e^{+}+4~{\rm jets}$ based on a dim-12 SMEFT operator (similar to the operator ${\cal O}_{Q^6L^2}^{S,(S)}$ in our basis). However, the eight fermion operator considered in~\cite{Bramante:2014uda} cannot yet be processed by the FeynRules to MG5AMC framework, and the authors take a “stand-in” operator for estimation. Such a procedure could yield large uncertainty, and we would like to come back to the collider signals in the future for a more precise analysis.   

\section{Conclusion and outlook}
\label{sec:con}

In this work we have made a thorough investigation on the baryon and lepton number violating dinucleon to dilepton decays ($pp\to \ell^+\ell^{\prime+}, pn\to \ell^+\bar\nu^\prime, nn\to \bar\nu\bar\nu^\prime$) with $\Delta B=\Delta L=-2$ in the framework of effective field theory. We first construct a basis of dim-12 operators mediating such processes in the low energy effective field theory (LEFT) below the electroweak scale. Such a basis not only contribute to the dinucleon decays studied in this work, it also serves as a starting point for model independent study of the hydrogen-antihydrogen oscillation and the low energy baryon number violating conversions $e^-p\to e^+\bar p, \bar\nu \bar n, e^- n\to \bar\nu\bar p$ in electron-deuteron scattering. Then we consider their leading-order standard model effective field theory (SMEFT) completions upwards and obtain the SMEFT basis mediating such processes at dim 12. We find the SMEFT gauge symmetry has a strong constraint on the structure of the interactions. The dim-12 SMEFT operators are suitable for the high energy signature on colliders like LHC to search the excess of events with four jets plus two same-sign charged leptons.    

Next, we analyze the chiral structure of the LEFT operators and make a non-perturbative matching through the baryon chiral perturbation theory (B$\chi$PT). In doing so, we construct a chiral basis in which each operator belongs to an irreducible representation of the two-flavor chiral group $SU(2)_L^f\times SU(2)_R^f$, and then we construct the corresponding hadronic operators through the spurion techniques.
Last, we express the dinucleon to dilepton decay rates in terms of the Wilson coefficients associated with the dim-12 operators in the LEFT/SMEFT and the low energy constants pertinent to B$\chi$PT. Our result is general in that it does not depend on dynamical details of physics at a high scale that induce the effective interactions in the SMEFT and in that it does not appeal to any hadronic models. We find the current experimental limits push the associated new physics scale larger than a few TeV, a scale appealing to the future experimental searches. Due to the weak isospin symmetry, based on the experimental limits on $pp\to \ell^+_\alpha\ell^+_\beta,pn\to \ell^+_\alpha\bar\nu_\beta$, we improve the lower limits on the partial lifetimes of the neutral transition modes $nn\to\bar\nu_\alpha\bar\nu_\beta$ (except the $(\alpha,\beta)=(\tau,\tau)$ case) by $2-3$ orders of magnitude than their current experimental sensitivity.  Furthermore, assuming these transitions dominantly generated by the similar dim-12 SMEFT operators, we find the limits on the partial lifetime of $pp\to e^+e^+,e^+\mu^+,\mu^+\mu^+$ are also transformed into stronger limits on $pn\to \ell^+_\alpha\bar\nu_\beta$ ($\alpha,\beta=e,\mu$) than their existing lower bounds.

Our operator basis obtained in this work is a starting point for further investigation on the related processes with $\Delta B=\Delta L=2$ signals, in which the hydrogen-antihydrogen oscillation and the collider signals $pp\to \ell^+\ell^{\prime+}+4~\rm jets$ and $e^-p\to \ell^++5~\rm jet$ are the most interesting ones. Both of these processes can be systematically studied in the current LEFT/SMEFT framework, and we will come back to these processes in the future publications.

\section*{Acknowledgement}
The authors acknowledge G. Valencia for his valuable comments on the manuscript and useful discussion concerning the collider aspect. The authors also thank J. Bramante for providing us with their FeynRules code to clarify what they had done in~\cite{Bramante:2014uda}. XDM would like to thank F-S Yu for his invitation as a visitor at Lanzhou Uni. where part of this work has been done.  This work was supported in part by the MOST (Grants No. 109-2112-M-002-017-MY3 and 109-2811-M-002-535), and in part by NSFC (Grants 11735010, 11975149, 12090064), by Key Laboratory for Particle Physics, Astrophysics and Cosmology, Ministry of Education, and Shanghai Key Lab- oratory for Particle Physics and Cosmology (Grant No. 15DZ2272100). 

\begin{appendices}
\numberwithin{equation}{section}
\setcounter{equation}{0}

\section{Independent color tensors}
\label{app:colortensor}

In this appendix, we give details the independent color tensors to contract with six quarks in a color $SU(3)_{\rm C}$ invariant way. Denoting a general six-quark field configuration as 
\begin{align}
{\cal O}^{ijklmn}=q^i_Iq^j_Jq^k_Kq^l_Lq^m_Mq^n_N\;,
\end{align}
where the superscripts $\{i,j,k,l,m,n\}$ are the color indices in fundamental representation of $SU(3)_{\rm C}$ while the subscripts $\{I,J,K,L,M,N\}$ encode the flavor and chiral information for each quark field. To form a color invariant operator, the color indices must be contracted by a color tensor $T_{ijklmn}$ such that ${\cal O}^{ijklmn}T_{ijklmn}$ is invariant under $SU(3)_{\rm C}$ . Since all quark fields belong to the fundamental representation of the color group, the color tensor $T_{ijklmn}$ must be a linear combination of two rank-3 totally anti-symmetric tensor $\epsilon_{xyz}$s. By the Schouten identities (SI)~\cite{Liao:2020jmn}
\begin{align}\nonumber
\epsilon_{imn}\epsilon_{jkl}&=-\epsilon_{ijm}\epsilon_{kln}+\epsilon_{ikm}\epsilon_{jln}-\epsilon_{ilm}\epsilon_{jkn}\;,
\\
\epsilon_{ijm}\epsilon_{kln}&=\epsilon_{ikm}\epsilon_{jln}+\epsilon_{ikn}\epsilon_{jlm}-\epsilon_{ilm}\epsilon_{jkn}-\epsilon_{iln}\epsilon_{jkm}-\epsilon_{ijn}\epsilon_{klm}\;, 
\label{SI:r3}
\end{align}
we can split the $m$-index and $n$-index into two epsilon tensors via the first SI, in turn, there are six independent combinations remained and the second SI further reduces them into five independent ones. By symmetrizing or anti-symmetrizing pairs of indices $(ij), (kl), (mn)$, we can choose the following five independent color tensors:
\begin{align}\nonumber
T^{SSS}_{\{ij\}\{kl\}\{mn\}}&=\epsilon_{ikm}\epsilon_{jln}+\epsilon_{ikn}\epsilon_{jlm}+\epsilon_{ilm}\epsilon_{jkn}+\epsilon_{iln}\epsilon_{jkm}\;,
\\\nonumber
T^{SAA}_{\{ij\}[kl][mn]}&=\epsilon_{imn}\epsilon_{jkl}+\epsilon_{ikl}\epsilon_{jmn}=\epsilon_{ikm}\epsilon_{jln}-\epsilon_{ikn}\epsilon_{jlm}-\epsilon_{ilm}\epsilon_{jkn}+\epsilon_{iln}\epsilon_{jkm}\;,
\\\nonumber
T^{SAA}_{\{kl\}[mn][ij]}&=\epsilon_{ijk}\epsilon_{mnl}+\epsilon_{ijl}\epsilon_{mnk}=\epsilon_{ikm}\epsilon_{jln}-\epsilon_{ikn}\epsilon_{jlm}+\epsilon_{ilm}\epsilon_{jkn}-\epsilon_{iln}\epsilon_{jkm}\;,
\\\nonumber
T^{SAA}_{\{mn\}[ij][kl]}&=\epsilon_{ijm}\epsilon_{kln}+\epsilon_{ijn}\epsilon_{klm}
=\epsilon_{ikm}\epsilon_{jln}+\epsilon_{ikn}\epsilon_{jlm}-\epsilon_{ilm}\epsilon_{jkn}-\epsilon_{iln}\epsilon_{jkm}\;,
\\\nonumber
T^{AAA}_{[ij][kl][mn]}&={1\over 3}\left(\epsilon_{ijm}\epsilon_{kln}-\epsilon_{ijn}\epsilon_{klm}-\epsilon_{ijk}\epsilon_{mnl}+\epsilon_{ijl}\epsilon_{mnk}+\epsilon_{ikl}\epsilon_{jmn}-\epsilon_{imn}\epsilon_{jkl}\right)
\\
&=\epsilon_{ijm}\epsilon_{kln}-\epsilon_{ijn}\epsilon_{klm}\;.
\end{align}
where the subscripts in curly bracket $\{ij\}$ and squared bracket $[kl]$ indicate separately the symmetrization and anti-symmetrization under the exchange of two color indices $i\leftrightarrow j$ within. In the above, the $T^{SSS}_{\{ij\}\{kl\}\{mn\}}$ and $T^{AAA}_{[ij][kl][mn]}$ are separately totally symmetric and anti-symmetric under the exchange of any pairs of the arguments, and $T^{SAA}_{\{ij\}[mn][kl]}$ is symmetric for the latter two pairs of indices. In addition, we have the constraints for exchanging two indices among two different pairs of indices
\begin{align}\nonumber
2T^{SSS}_{\{ik\}\{jl\}\{mn\}}&=-T^{SSS}_{\{ij\}\{kl\}\{mn\}}+3T^{SAA}_{\{mn\}[ij][kl]}\;,
\\\nonumber
2T^{SAA}_{\{ij\}[km][ln]}&=T^{SSS}_{\{ij\}\{kl\}\{mn\}}+T^{SAA}_{\{ij\}[kl][mn]}\;,
\\\nonumber
2T^{SAA}_{\{ik\}[jl][mn]}&=-T^{SAA}_{\{ij\}[kl][mn]}+T^{SAA}_{\{kl\}[mn][ij]}+2T^{AAA}_{[ij][kl][mn]}\;,
\\\nonumber
2T^{SAA}_{\{in\}[kl][mj]}&=-T^{SAA}_{\{ij\}[kl][mn]}-T^{SAA}_{\{mn\}[ij][kl]}-2T^{AAA}_{[ij][kl][mn]}\;,
\\
2T^{AAA}_{[ik][jl][mn]}&=T^{SAA}_{\{ij\}[kl][mn]}+T^{SAA}_{\{kl\}[mn][ij]}\;.
\label{relation:TSA1}
\end{align}
These relations are useful to reduce redundant operators, and will be used repeatedly in latter sections to reach the minimal basis for the  dim-12 $\Delta B=\Delta L=-2$ operators both in the LEFT and in the SMEFT . 

\section{LEFT operators}
\label{app:LEFTope}

The full list for dim-12 operators inducing dinucleon to dilepton transitions.

\noindent
{\small$\scriptscriptstyle\blacksquare$ \small\bf\boldmath Dim-12 operators contributing to $pp\to\ell^+\ell^{\prime+}$}\\ 
For the operators with a scalar current $j_{S,\pm}^{\ell\ell^\prime}$, we find there are 28 independent operators which can be parametrized as follows
\begin{align}\nonumber
{\cal Q}_{1LLL,a}^{(pp)S,\pm}&=(u_L^{i\T} Cu_L^j)(u_L^{k\T} Cd_L^l)(u_L^{m\T} Cd_L^n)j_{S,\pm}^{\ell\ell^\prime}T^{SSS}_{\{ij\}\{kl\}\{mn\}}\;,
\\\nonumber
{\cal Q}_{1LLL,b}^{(pp)S,\pm}&=(u_L^{i\T} Cu_L^j)(u_L^{k\T} Cd_L^l)(u_L^{m\T} Cd_L^n)j_{S,\pm}^{\ell\ell^\prime}T^{SAA}_{\{ij\}[kl][mn]}\;,
\\\nonumber
{\cal Q}_{2LLR,a}^{(pp)S,\pm}&=(u_L^{i\T} Cu_L^j)(u_L^{k\T} Cd_L^l)(u_R^{m\T} Cd_R^n)j_{S,\pm}^{\ell\ell^\prime}T^{SSS}_{\{ij\}\{kl\}\{mn\}}\;,
\\\nonumber
{\cal Q}_{2LLR,b}^{(pp)S,\pm}&=(u_L^{i\T} Cu_L^j)(u_L^{k\T} Cd_L^l)(u_R^{m\T} Cd_R^n)j_{S,\pm}^{\ell\ell^\prime}T^{SAA}_{\{ij\}[kl][mn]}\;,
\\\nonumber
{\cal Q}_{3LLR,a}^{(pp)S,\pm}&=(u_L^{i\T} Cd_L^j)(u_L^{k\T} Cd_L^l)(u_R^{m\T} Cu_R^n)j_{S,\pm}^{\ell\ell^\prime}T^{SSS}_{\{ij\}\{kl\}\{mn\}}\;,
\\\nonumber
{\cal Q}_{3LLR,b}^{(pp)S,\pm}&=(u_L^{i\T} Cd_L^j)(u_L^{k\T} Cd_L^l)(u_R^{m\T} Cu_R^n)j_{S,\pm}^{\ell\ell^\prime}T^{SAA}_{\{mn\}[ij][kl]}\;,
\\
{\cal Q}_{4LLR}^{(pp)S,\pm}&=(u_L^{i\T} Cu_L^j)(u_L^{k\T} Cu_L^l)(d_R^{m\T} Cd_R^n)j_{S,\pm}^{\ell\ell^\prime}T^{SSS}_{\{ij\}\{kl\}\{mn\}}\;,
\label{ope:scalarf}
\end{align}
together with their parity partners with $L\leftrightarrow R$. 

For the operators with a vector lepton current $j_{V}^{\ell\ell^\prime,\mu}$, there are 19 independent operators which are chosen to take
\begin{align}\nonumber
{\cal Q}^{(pp)V}_{1LL,a}&=(u_L^{i\T} Cd_L^j)(u_L^{k\T} Cd_L^l)(u_L^{m\T} C\gamma_\mu u_R^n)j_{V}^{\ell\ell^\prime,\mu}T^{SSS}_{\{ij\}\{kl\}\{mn\}}\;,
\\\nonumber
{\cal Q}^{(pp)V}_{1LL,b}&=(u_L^{i\T} Cd_L^j)(u_L^{k\T} Cd_L^l)(u_L^{m\T} C\gamma_\mu u_R^n)j_{V}^{\ell\ell^\prime,\mu}T^{SAA}_{\{ij\}[kl][mn]}\;,
\\\nonumber
{\cal Q}^{(pp)V}_{1LL,c}&=(u_L^{i\T} Cd_L^j)(u_L^{k\T} Cd_L^l)(u_L^{m\T} C\gamma_\mu u_R^n)j_{V}^{\ell\ell^\prime,\mu}T^{SAA}_{\{mn\}[kl][ij]}\;,
\\\nonumber
{\cal Q}^{(pp)V}_{2LL,a}&=(u_L^{i\T} Cu_L^j)(u_L^{k\T} Cd_L^l)(u_L^{m\T} C\gamma_\mu d_R^n)j_{V}^{\ell\ell^\prime,\mu}T^{SSS}_{\{ij\}\{kl\}\{mn\}}\;,
\\\nonumber
{\cal Q}^{(pp)V}_{2LL,b}&=(u_L^{i\T} Cu_L^j)(u_L^{k\T} Cd_L^l)(u_L^{m\T} C\gamma_\mu d_R^n)j_{V}^{\ell\ell^\prime,\mu} T^{SAA}_{\{ij\}[kl][mn]}\;,
\\\nonumber
{\cal Q}^{(pp)V}_{3LR,a}&=(u_L^{i\T} Cu_L^j)(u_R^{k\T} Cd_R^l)(u_L^{m\T} C\gamma_\mu d_R^n)j_{V}^{\ell\ell^\prime,\mu} T^{SSS}_{\{ij\}\{kl\}\{mn\}}\;,
\\\nonumber
{\cal Q}^{(pp)V}_{3LR,b}&=(u_L^{i\T} Cu_L^j)(u_R^{k\T} Cd_R^l)(u_L^{m\T} C\gamma_\mu d_R^n)j_{V}^{\ell\ell^\prime,\mu}T^{SAA}_{\{ij\}[kl][mn]}\;,
\\\nonumber
{\cal Q}^{(pp)V}_{4LR,a}&=(u_L^{i\T} Cd_L^j)(u_R^{k\T} Cd_R^l)(u_L^{m\T} C\gamma_\mu u_R^n)j_{V}^{\ell\ell^\prime,\mu} T^{SSS}_{\{ij\}\{kl\}\{mn\}}\;,
\\\nonumber
{\cal Q}^{(pp)V}_{4LR,b}&=(u_L^{i\T} Cd_L^j)(u_R^{k\T} Cd_R^l)(u_L^{m\T} C\gamma_\mu u_R^n)j_{V}^{\ell\ell^\prime,\mu}T^{SAA}_{\{ij\}[kl][mn]}\;,
\\\nonumber
{\cal Q}^{(pp)V}_{4LR,c}&=(u_L^{i\T} Cd_L^j)(u_R^{k\T} Cd_R^l)(u_L^{m\T} C\gamma_\mu u_R^n)j_{V}^{\ell\ell^\prime,\mu}T^{SAA}_{\{kl\}[mn][ij]}\;,
\\\nonumber
{\cal Q}^{(pp)V}_{4LR,d}&=(u_L^{i\T} Cd_L^j)(u_R^{k\T} Cd_R^l)(u_L^{m\T} C\gamma_\mu u_R^n)j_{V}^{\ell\ell^\prime,\mu} T^{SAA}_{\{mn\}[ij][kl]}\;,
\\
{\cal Q}^{(pp)V}_{4LR,e}&=(u_L^{i\T} Cd_L^j)(u_R^{k\T} Cd_R^l)(u_L^{m\T} C\gamma_\mu u_R^n)j_{V}^{\ell\ell^\prime,\mu} T^{AAA}_{[ij][kl][mn]}\;,
\label{ope:vectorf}
\end{align}
together with the parity partners for operators ${\cal Q}^{V}_{1-3}$ with $L\leftrightarrow R$. 

For the operators with a tensor lepton current $j_{T}^{\ell\ell^\prime,\mu\nu}$, we find there are 16 independent operators which are chosen to take
\begin{align}\nonumber
{\cal Q}_{1LLL}^{(pp)T,-}&=(u_L^{i\T} C\sigma_{\mu\nu}u_L^j)(u_L^{k\T} Cd_L^l)(u_L^{m\T} Cd_L^n)j_{T,-}^{\ell\ell^\prime,\mu\nu}T^{SAA}_{\{mn\}[ij][kl]}\;,
\\\nonumber
{\cal Q}_{2LLR,a}^{(pp)T,-}&=(u_L^{i\T} C\sigma_{\mu\nu}u_L^j)(u_L^{k\T} Cd_L^l)(u_R^{m\T} Cd_R^n)j_{T,-}^{\ell\ell^\prime,\mu\nu}T^{SAA}_{\{kl\}[mn][ij]}\;,
\\\nonumber
{\cal Q}_{2LLR,b}^{(pp)T,-}&=(u_L^{i\T} C\sigma_{\mu\nu}u_L^j)(u_L^{k\T} Cd_L^l)(u_R^{m\T} Cd_R^n)j_{T,-}^{\ell\ell^\prime,\mu\nu}T^{SAA}_{\{mn\}[ij][kl]}\;,
\\\nonumber
{\cal Q}_{2LLR,c}^{(pp)T,-}&=(u_L^{i\T} C\sigma_{\mu\nu}u_L^j)(u_L^{k\T} Cd_L^l)(u_R^{m\T} Cd_R^n)j_{T,-}^{\ell\ell^\prime,\mu\nu}T^{AAA}_{[ij][kl][mn]}\;,
\\\nonumber
{\cal Q}_{2LLR,a}^{(pp)T,+}&=(u_L^{i\T} Cu_L^j)(u_L^{k\T} Cd_L^l)(u_R^{m\T} C\sigma_{\mu\nu}d_R^n)j_{T,+}^{\ell\ell^\prime,\mu\nu}T^{SSS}_{\{ij\}\{kl\}\{mn\}}\;,
\\\nonumber
{\cal Q}_{2LLR,b}^{(pp)T,+}&=(u_L^{i\T} Cu_L^j)(u_L^{k\T} Cd_L^l)(u_R^{m\T} C\sigma_{\mu\nu}d_R^n)j_{T,+}^{\ell\ell^\prime,\mu\nu}T^{SAA}_{\{ij\}[kl][mn]}\;,
\\\nonumber
{\cal Q}_{3LLR}^{(pp)T,-}&=(u_L^{i\T} Cd_L^j)(u_L^{k\T} C\sigma_{\mu\nu}d_L^l)(u_R^{m\T} Cu_R^n)j_{T,-}^{\ell\ell^\prime,\mu\nu}T^{SAA}_{\{mn\}[ij][kl]}\;,
\\
{\cal Q}_{3LLR}^{(pp)T,+}&=(u_L^{i\T} Cd_L^j)(u_L^{k\T} Cd_L^l)(u_R^{m\T} C\sigma_{\mu\nu}u_R^n)j_{T,+}^{\ell\ell^\prime,\mu\nu}T^{SAA}_{\{ij\}[kl][mn]}\;,
\label{pp:tensorf}
\end{align}
together with the parity partners for operators ${\cal Q}^{T}_{1-3}$ with $L\leftrightarrow R$ and $-\leftrightarrow+$. 

As a non-trivial example for the reduction of redundant operators, we consider the above tensor current operator ${\cal Q}_{3LRR}^{(pp)T,+}$ in Eq.~\eqref{pp:tensorf} with the replacement of the color tensor $T^{SAA}_{\{ij\}[kl][mn]}$ by $T^{SSS}_{\{ij\}\{kl\}\{mn\}}$, then the new operator is reduced as follows
\begin{align*}
&2(u_L^{i\T} Cu_L^j)(u_R^{k\T} Cd_R^l)(u_R^{m\T} C\sigma_{\mu\nu}d_R^n)j_{T,+}^{\ell\ell^\prime,\mu\nu}T^{SSS}_{\{ij\}\{kl\}\{mn\}}
\\
&\overset{\rm FI}{=}-2(u_L^{i\T} Cu_L^j)\left[ (u_R^{m\T} Cd_R^l)(u_R^{k\T} C\sigma_{\mu\nu}d_R^n)+(u_R^{k\T} Cu_R^m)(d_R^{l\T} C\sigma_{\mu\nu}d_R^n) \right]j_{T,+}^{\ell\ell^\prime,\mu\nu}T^{SSS}_{\{ij\}\{kl\}\{mn\}}
\\
&\overset{\rm SI}{=}(u_L^{i\T} Cu_L^j)(u_R^{k\T} Cd_R^l)(u_R^{m\T} C\sigma_{\mu\nu}d_R^n)j_{T,+}^{\ell\ell^\prime,\mu\nu}\left(T^{SSS}_{\{ij\}\{kl\}\{mn\}}+3T^{SAA}_{\{ij\}[kl][mn]}\right)
\\
&~~~~ +(u_L^{i\T} Cu_L^j)(u_R^{k\T} Cu_R^l)(d_R^{m\T} C\sigma_{\mu\nu}d_R^n)j_{T,+}^{\ell\ell^\prime,\mu\nu}\left(T^{SSS}_{\{ij\}\{kl\}\{mn\}}-3T^{SAA}_{\{ij\}[kl][mn]}\right)(=0)\;,
\end{align*}
where in the second step we have used the FIs in Appendix~\ref{app:LEFTreduction} and the third step the SI in Eq.~\eqref{relation:TSA1}, and the terms in last line vanish due to mismatched color symmetry. We see this new operator is equivalent to ${\cal Q}_{3LRR}^{(pp)T,+}$ and therefore redundant. All other operators with different color tensors or Lorentz structures beyond the above lists can be reduced in a similar manner.

\noindent
{\small$\scriptscriptstyle\blacksquare$ \small\bf\boldmath Dim-12 operators contributing to $pn\to\ell^+\nu^\prime$}\\
For the operators with a scalar current $j_{S}^{\ell\nu^\prime}$, there are 14 independent operators which can be parametrized as follows
\begin{align}\nonumber
{\cal Q}_{1LLL,a}^{(pn)S}&=(u_L^{i\T} Cd_L^j)(u_L^{k\T} Cd_L^l)(u_L^{m\T} Cd_L^n)j_{S}^{\ell\nu^\prime}T^{SSS}_{\{ij\}\{kl\}\{mn\}}\;,
\\\nonumber
{\cal Q}_{1LLL,b}^{(pn)S}&=(u_L^{i\T} Cd_L^j)(u_L^{k\T} Cd_L^l)(u_L^{m\T} Cd_L^n)j_{S}^{\ell\nu^\prime}T^{SAA}_{\{ij\}[kl][mn]}\;,
\\\nonumber
{\cal Q}_{2LLR}^{(pn)S}&=(u_L^{i\T} Cu_L^j)(u_L^{k\T} Cd_L^l)(d_R^{m\T} Cd_R^n)j_{S}^{\ell\nu^\prime}T^{SSS}_{\{ij\}\{kl\}\{mn\}}\;,
\\\nonumber
{\cal Q}_{3LLR,a}^{(pn)S}&=(u_L^{i\T} Cd_L^j)(u_L^{k\T} Cd_L^l)(u_R^{m\T} Cd_R^n)j_{S}^{\ell\nu^\prime}T^{SSS}_{\{ij\}\{kl\}\{mn\}}\;,
\\\nonumber
{\cal Q}_{3LLR,b}^{(pn)S}&=(u_L^{i\T} Cd_L^j)(u_L^{k\T} Cd_L^l)(u_R^{m\T} Cd_R^n)j_{S}^{\ell\nu^\prime}T^{SAA}_{\{ij\}[kl][mn]}\;,
\\\nonumber
{\cal Q}_{3LLR,c}^{(pn)S}&=(u_L^{i\T} Cd_L^j)(u_L^{k\T} Cd_L^l)(u_R^{m\T} Cd_R^n)j_{S}^{\ell\nu^\prime}T^{SAA}_{\{mn\}[ij][kl]}\;,
\\
{\cal Q}_{4LLR}^{(pn)S}&=(d_L^{i\T} Cd_L^j)(u_L^{k\T} Cd_L^l)(u_R^{m\T} Cu_R^n)j_{S}^{\ell\nu^\prime}T^{SSS}_{\{ij\}\{kl\}\{mn\}}\;,
\label{pn:scalarf}
\end{align}
together with their parity partners with $L\leftrightarrow R$. 

For the operators with a vector current $j_{V}^{\ell\nu^\prime,\mu}$, we find there are 24 independent operators which are parametrized as follows
\begin{align}\nonumber
{\cal Q}_{1LL,a}^{(pn)V}&=(u_L^{i\T} Cd_L^j)(u_L^{k\T} Cd_L^l)(u_L^{m\T} C\gamma_\mu d_R^n)j_{V}^{\ell\nu^\prime,\mu}T^{SSS}_{\{ij\}\{kl\}\{mn\}}\;,
\\\nonumber
{\cal Q}_{1LL,b}^{(pn)V}&=(u_L^{i\T} Cd_L^j)(u_L^{k\T} Cd_L^l)(u_L^{m\T} C\gamma_\mu d_R^n)j_{V}^{\ell\nu^\prime,\mu}T^{SAA}_{\{ij\}[kl][mn]}\;,
\\\nonumber
{\cal Q}_{1LL,c}^{(pn)V}&=(u_L^{i\T} Cd_L^j)(u_L^{k\T} Cd_L^l)(u_L^{m\T} C\gamma_\mu d_R^n)j_{V}^{\ell\nu^\prime,\mu}T^{SAA}_{\{mn\}[ij][kl]}\;,
\\\nonumber
{\cal Q}_{2LL,a}^{(pn)V}&=(u_L^{i\T} Cd_L^j)(u_L^{k\T} Cd_L^l)(d_L^{m\T} C\gamma_\mu u_R^n)j_{V}^{\ell\nu^\prime,\mu}T^{SSS}_{\{ij\}\{kl\}\{mn\}}\;,
\\\nonumber
{\cal Q}_{2LL,b}^{(pn)V}&=(u_L^{i\T} Cd_L^j)(u_L^{k\T} Cd_L^l)(d_L^{m\T} C\gamma_\mu u_R^n)j_{V}^{\ell\nu^\prime,\mu}T^{SAA}_{\{ij\}[kl][mn]}\;,
\\\nonumber
{\cal Q}_{2LL,c}^{(pn)V}&=(u_L^{i\T} Cd_L^j)(u_L^{k\T} Cd_L^l)(d_L^{m\T} C\gamma_\mu u_R^n)j_{V}^{\ell\nu^\prime,\mu}T^{SAA}_{\{mn\}[ij][kl]}\;,
\\\nonumber
{\cal Q}_{3LR}^{(pn)V}&=(u_L^{i\T} Cu_L^j)(d_R^{k\T} Cd_R^l)(u_L^{m\T} C\gamma_\mu d_R^n)j_{V}^{\ell\nu^\prime,\mu}T^{SSS}_{\{ij\}\{kl\}\{mn\}}\;,
\\\nonumber
{\cal Q}_{4LR,a}^{(pn)V}&=(u_L^{i\T} Cd_L^j)(u_R^{k\T} Cd_R^l)(u_L^{m\T} C\gamma_\mu d_R^n)j_{V}^{\ell\nu^\prime,\mu}T^{SSS}_{\{ij\}\{kl\}\{mn\}}\;,
\\\nonumber
{\cal Q}_{4LR,b}^{(pn)V}&=(u_L^{i\T} Cd_L^j)(u_R^{k\T} Cd_R^l)(u_L^{m\T} C\gamma_\mu d_R^n)j_{V}^{\ell\nu^\prime,\mu}T^{SAA}_{\{ij\}[kl][mn]}\;,
\\\nonumber
{\cal Q}_{4LR,c}^{(pn)V}&=(u_L^{i\T} Cd_L^j)(u_R^{k\T} Cd_R^l)(u_L^{m\T} C\gamma_\mu d_R^n)j_{V}^{\ell\nu^\prime,\mu}T^{SAA}_{\{kl\}[mn][ij]}\;,
\\\nonumber
{\cal Q}_{4LR,d}^{(pn)V}&=(u_L^{i\T} Cd_L^j)(u_R^{k\T} Cd_R^l)(u_L^{m\T} C\gamma_\mu d_R^n)j_{V}^{\ell\nu^\prime,\mu}T^{SAA}_{\{mn\}[ij][kl]}\;,
\\
{\cal Q}_{4LR,e}^{(pn)V}&=(u_L^{i\T} Cd_L^j)(u_R^{k\T} Cd_R^l)(u_L^{m\T} C\gamma_\mu d_R^n)j_{V}^{\ell\nu^\prime,\mu}T^{AAA}_{[ij][kl][mn]}\;,
\label{pn:vectorf}
\end{align}
together with their parity partners with $L\leftrightarrow R$. 

For the operators with a tensor current $j_{T}^{\ell\nu^\prime,\mu\nu}$, there are 13 independent operators which can be parametrized as follows
\begin{align}\nonumber
{\cal Q}_{LLL,a}^{(pn)T}&=(u_L^{i\T} Cd_L^j)(u_L^{k\T} Cd_L^l)(u_L^{m\T}C\sigma_{\mu\nu}d_L^n)j_{T}^{\ell\nu^\prime,\mu\nu}T^{SAA}_{\{ij\}[kl][mn]}\;,
\\\nonumber
{\cal Q}_{LLL,b}^{(pn)T}&=(u_L^{i\T} Cd_L^j)(u_L^{k\T} Cd_L^l)(u_L^{m\T}C\sigma_{\mu\nu}d_L^n)j_{T}^{\ell\nu^\prime,\mu\nu}T^{SAA}_{\{mn\}[ij][kl]}\;,
\\\nonumber
{\cal Q}_{2LLR}^{(pn)T}&=(u_L^{i\T} C\sigma_{\mu\nu}u_L^j)(u_L^{k\T} Cd_L^l)(d_R^{m\T} Cd_R^n)j_{T}^{\ell\nu^\prime,\mu\nu}T^{SAA}_{\{mn\}[ij][kl]}\;,
\\\nonumber
{\cal Q}_{3LLR,a}^{(pn)T}&=(u_L^{i\T} Cd_L^j)(u_L^{k\T} C\sigma_{\mu\nu}d_L^l)(u_R^{m\T} Cd_R^n)j_{T}^{\ell\nu^\prime,\mu\nu}T^{SAA}_{\{ij\}[kl][mn]}\;,
\\\nonumber
{\cal Q}_{3LLR,b}^{(pn)T}&=(u_L^{i\T} Cd_L^j)(u_L^{k\T} C\sigma_{\mu\nu}d_L^l)(u_R^{m\T} Cd_R^n)j_{T}^{\ell\nu^\prime,\mu\nu}T^{SAA}_{\{kl\}[mn][ij]}\;,
\\\nonumber
{\cal Q}_{3LLR,c}^{(pn)T}&=(u_L^{i\T} Cd_L^j)(u_L^{k\T} C\sigma_{\mu\nu}d_L^l)(u_R^{m\T} Cd_R^n)j_{T}^{\ell\nu^\prime,\mu\nu}T^{SAA}_{\{mn\}[ij][kl]}\;,
\\\nonumber
{\cal Q}_{3LLR,d}^{(pn)T}&=(u_L^{i\T} Cd_L^j)(u_L^{k\T} C\sigma_{\mu\nu}d_L^l)(u_R^{m\T} Cd_R^n)j_{T}^{\ell\nu^\prime,\mu\nu}T^{AAA}_{[ij][kl][mn]}\;,
\\\nonumber
{\cal Q}_{4LLR}^{(pn)T}&=(d_L^{i\T} C\sigma_{\mu\nu}d_L^j)(u_L^{k\T}Cd_L^l)(u_R^{m\T} Cu_R^n)j_{T}^{\ell\nu^\prime,\mu\nu}T^{SAA}_{\{mn\}[ij][kl]}\;,
\\\nonumber
{\cal Q}_{2RRL}^{(pn)T}&=(u_R^{i\T} Cu_R^j)(u_R^{k\T} Cd_R^l)(d_L^{m\T} C\sigma_{\mu\nu} d_L^n)j_{T}^{\ell\nu^\prime,\mu\nu}T^{SAA}_{\{ij\}[kl][mn]}\;,
\\\nonumber
{\cal Q}_{3RRL,a}^{(pn)T}&=(u_R^{i\T} Cd_R^j)(u_R^{k\T} Cd_R^l)(u_L^{m\T} C\sigma_{\mu\nu}d_L^n)j_{T}^{\ell\nu^\prime,\mu\nu}T^{SSS}_{\{ij\}\{kl\}\{mn\}}\;,
\\\nonumber
{\cal Q}_{3RRL,b}^{(pn)T}&=(u_R^{i\T} Cd_R^j)(u_R^{k\T} Cd_R^l)(u_L^{m\T} C\sigma_{\mu\nu}d_L^n)j_{T}^{\ell\nu^\prime,\mu\nu}T^{SAA}_{\{ij\}[kl][mn]}\;,
\\\nonumber
{\cal Q}_{3RRL,c}^{(pn)T}&=(u_R^{i\T} Cd_R^j)(u_R^{k\T} Cd_R^l)(u_L^{m\T} C\sigma_{\mu\nu}d_L^n)j_{T}^{\ell\nu^\prime,\mu\nu}T^{SAA}_{\{mn\}[ij][kl]}\;,
\\
{\cal Q}_{4RRL}^{(pn)T}&=(d_R^{i\T} Cd_R^j)(u_R^{k\T} Cd_R^l)(u_L^{m\T} C\sigma_{\mu\nu}u_L^n)j_{T}^{\ell\nu^\prime,\mu\nu}T^{SAA}_{\{ij\}[kl][mn]}\;.
\label{pn:tensorf}
\end{align}

\noindent
{\small$\scriptscriptstyle\blacksquare$ \small\bf\boldmath Dim-12 operators contributing to $nn\to\bar\nu\bar\nu^\prime$}\\
For the operators with a scalar current $j_{S}^{\nu\nu^\prime}$, there are 14 independent operators which are parametrized as follows
\begin{align}\nonumber
{\cal Q}_{1LLL,a}^{(nn)S}&=(d_L^{i\T} Cd_L^j)(d_L^{k\T} Cu_L^l)(d_L^{m\T} Cu_L^n)j_{S}^{\nu\nu^\prime}T^{SSS}_{\{ij\}\{kl\}\{mn\}}\;,
\\\nonumber
{\cal Q}_{1LLL,b}^{(nn)S}&=(d_L^{i\T} Cd_L^j)(d_L^{k\T} Cu_L^l)(d_L^{m\T} Cu_L^n)j_{S}^{\nu\nu^\prime}T^{SAA}_{\{ij\}[kl][mn]}\;,
\\\nonumber
{\cal Q}_{2LLR,a}^{(nn)S}&=(d_L^{i\T} Cd_L^j)(d_L^{k\T} Cu_L^l)(d_R^{m\T} Cu_R^n)j_{S}^{\nu\nu^\prime}T^{SSS}_{\{ij\}\{kl\}\{mn\}}\;,
\\\nonumber
{\cal Q}_{2LLR,b}^{(nn)S}&=(d_L^{i\T} Cd_L^j)(d_L^{k\T} Cu_L^l)(d_R^{m\T} Cu_R^n)j_{S}^{\nu\nu^\prime}T^{SAA}_{\{ij\}[kl][mn]}\;,
\\\nonumber
{\cal Q}_{3LLR,a}^{(nn)S}&=(d_L^{i\T} Cu_L^j)(d_L^{k\T} Cu_L^l)(d_R^{m\T} Cd_R^n)j_{S}^{\nu\nu^\prime}T^{SSS}_{\{ij\}\{kl\}\{mn\}}\;,
\\\nonumber
{\cal Q}_{3LLR,b}^{(nn)S}&=(d_L^{i\T} Cu_L^j)(d_L^{k\T} Cu_L^l)(d_R^{m\T} Cd_R^n)j_{S}^{\nu\nu^\prime}T^{SAA}_{\{mn\}[ij][kl]}\;,
\\
{\cal Q}_{4LLR}^{(nn)S}&=(d_L^{i\T} Cd_L^j)(d_L^{k\T} Cd_L^l)(u_R^{m\T} Cu_R^n)j_{S}^{\nu\nu^\prime}T^{SSS}_{\{ij\}\{kl\}\{mn\}}\;,
\label{nn:scalarf}
\end{align}
together with their parity partners with $L\leftrightarrow R$.  

And for the operators with a tensor neutrino current $j_{T}^{\nu\nu^\prime,\mu\nu}$, there are only 8 independent operators which are parametrized as follows 
\begin{align}\nonumber
{\cal Q}_{1LLL}^{(nn)T}&=(d_L^{i\T} C\sigma_{\mu\nu}d_L^j)(d_L^{k\T} Cu_L^l)(d_L^{m\T} Cu_L^n)j_{T}^{\nu\nu^\prime,\mu\nu}T^{SAA}_{\{mn\}[kl][ij]}\;,
\\\nonumber
{\cal Q}_{2LLR,a}^{(nn)T}&=(d_L^{i\T} C\sigma_{\mu\nu}d_L^j)(d_L^{k\T} Cu_L^l)(d_R^{m\T} Cu_R^n)j_{T}^{\nu\nu^\prime,\mu\nu}T^{SAA}_{\{kl\}[mn][ij]}\;,
\\\nonumber
{\cal Q}_{2LLR,b}^{(nn)T}&=(d_L^{i\T} C\sigma_{\mu\nu}d_L^j)(d_L^{k\T} Cu_L^l)(d_R^{m\T} Cu_R^n)j_{T}^{\nu\nu^\prime,\mu\nu}T^{SAA}_{\{mn\}[ij][kl]}\;,
\\\nonumber
{\cal Q}_{2LLR,c}^{(nn)T}&=(d_L^{i\T} C\sigma_{\mu\nu}d_L^j)(d_L^{k\T} Cu_L^l)(d_R^{m\T} Cu_R^n)j_{T}^{\nu\nu^\prime,\mu\nu}T^{AAA}_{[ij][kl][mn]}\;,
\\\nonumber
{\cal Q}_{3LLR}^{(nn)T}&=(d_L^{i\T} Cu_L^j)(d_L^{k\T} C\sigma_{\mu\nu}u_L^l)(d_R^{m\T} Cd_R^n)j_{T}^{\nu\nu^\prime,\mu\nu}T^{SAA}_{\{mn\}[ij][kl]}\;,
\\\nonumber
{\cal Q}_{2RRL,a}^{(nn)T}&=(d_R^{i\T} Cd_R^j)(d_R^{k\T} Cu_R^l)(d_L^{m\T} C\sigma_{\mu\nu}u_L^n)j_{T}^{\nu\nu^\prime,\mu\nu}T^{SSS}_{\{ij\}\{kl\}\{mn\}}\;,
\\\nonumber
{\cal Q}_{2RRL,b}^{(nn)T}&=(d_R^{i\T} Cd_R^j)(d_R^{k\T} Cu_R^l)(d_L^{m\T} C\sigma_{\mu\nu}u_L^n)j_{T}^{\nu\nu^\prime,\mu\nu}T^{SAA}_{\{ij\}[kl][mn]}\;,
\\
{\cal Q}_{3RRL}^{(nn)T}&=(d_R^{i\T} Cu_R^j)(d_R^{k\T} Cu_R^l)(d_L^{m\T} C\sigma_{\mu\nu}d_L^n)j_{T}^{\nu\nu^\prime,\mu\nu}T^{SAA}_{\{ij\}[kl][mn]}\;.
\label{nn:tensorf}
\end{align}

\section{Reduction of the redundant operators in the LEFT}
\label{app:LEFTreduction}

The dim-12 operators contributing to ${\rm H-}\bar{\rm H}$ oscillation and $pp\to e^+e^+$ transitions in the LEFT were given first in Ref.~\cite{Caswell:1982qs}. For the operators with the scalar lepton current, their results are consistent with ours, and the 28 operators in their paper can be easily identified with the results shown in Eq.~\eqref{ope:scalarf}. For the operators with a vector current, they count 32 operators and 13 of them are redundant and will be reduced in the following. In doing so, we first notice that the color tensors in Ref.~\cite{Caswell:1982qs} have the following one-to-one correspondence with our notation
\begin{align}
&(T_S)_{ijklmn}=T^{SSS}_{\{ij\}\{kl\}\{mn\}}\;, 
&&(T_A)_{ijklmn}=T^{SAA}_{\{mn\}[kl][ij]}\;,
&&(\tilde T_A)_{ijklmn}=T^{AAA}_{[ij][kl][mn]}\;.
\end{align}
The relations in Eq.~\eqref{relation:TSA1} imply the following corresponding relations
\begin{align}\nonumber
2(T_S)_{ikjlmn}&=3(T_A)_{ijklmn}-(T_S)_{ijklmn}\;,
\\
2(T_A)_{mjklin}&=-2(\tilde T_A)_{ijklmn}-(T_A)_{ijklmn}-(T_A)_{mnklij}\;.
\label{relation:TSA2}
\end{align}
Since the two lepton fields in the vector current have different chirality, Ref.~\cite{Caswell:1982qs} parametrized all such operators using four scalar fermion bilinears in which the two lepton fields are combined separately with two quark fields to make scalar currents. By the Fierz identity
\begin{align}
(q_L^{m\T}C\ell_L)(q_R^{n\T}C\ell_R)=-{1\over 2} (q_L^{m\T}C\gamma_\mu q_R^{n})(\ell_R^\T C\gamma^\mu \ell_L)\;, 
\end{align}
we can rewrite the operators with a pair of $(\ell_L, \ell_R)$ in Ref.~\cite{Caswell:1982qs} to have a factorized vector lepton current as follows
\begin{align}\nonumber
(O_{H1}^{H\bar H})_{\chi_1\chi_2 LR}&=-{1\over2}(u^{i\T}_{\chi_1}Cu^j_{\chi_1})(u^{k\T}_{\chi_2}Cu^l_{\chi_2}) (d_L^{m\T}C\gamma_\mu d_R^n)(\ell_R^\T C\gamma^\mu \ell_L)(T_S)_{ijklmn}\;,
\\\nonumber
(O_{H2}^{H\bar H})_{\chi_1\chi_2 LR}&=-{1\over2}(u^{i\T}_{\chi_1}Cu^j_{\chi_1})(d^{k\T}_{\chi_2}Cd^l_{\chi_2}) (u_L^{m\T}C\gamma_\mu u_R^n)(\ell_R^\T C\gamma^\mu \ell_L)(T_S)_{ijklmn}\;,
\\\nonumber
(O_{H3}^{H\bar H})_{\chi_1\chi_2 LR}&=-{1\over2}(u^{i\T}_{\chi_1}Cu^j_{\chi_1})(u^{k\T}_{\chi_2}Cd^l_{\chi_2}) (u_L^{m\T}C\gamma_\mu d_R^n)(\ell_R^\T C\gamma^\mu \ell_L)(T_{S})_{mnklij}\;,
\\\nonumber
(O_{H4}^{H\bar H})_{\chi_1\chi_2 LR}&=-{1\over2}(u^{i\T}_{\chi_1}Cu^j_{\chi_1})(u^{k\T}_{\chi_2}Cd^l_{\chi_2}) (u_L^{m\T}C\gamma_\mu d_R^n)(\ell_R^\T C\gamma^\mu \ell_L)(T_{A})_{mnklij}\;,
\\\nonumber
(O_{H3}^{H\bar H})_{\chi_1\chi_2 RL}&=-{1\over2}(u^{i\T}_{\chi_1}Cu^j_{\chi_1})(u^{k\T}_{\chi_2}Cd^l_{\chi_2}) (d_L^{m\T}C\gamma_\mu u_R^n)(\ell_R^\T C\gamma^\mu \ell_L)(T_{S})_{mnklij}\;,
\\\nonumber
(O_{H4}^{H\bar H})_{\chi_1\chi_2 RL}&=+{1\over2}(u^{i\T}_{\chi_1}Cu^j_{\chi_1})(u^{k\T}_{\chi_2}Cd^l_{\chi_2}) (d_L^{m\T}C\gamma_\mu u_R^n)(\ell_R^\T C\gamma^\mu \ell_L)(T_{A})_{mnklij}\;,
\\\nonumber
(O_{H5}^{H\bar H})_{\chi_1\chi_2 LR}&=-{1\over2}(u^{i\T}_{\chi_1}Cd^j_{\chi_1})(u^{k\T}_{\chi_2}Cd^l_{\chi_2}) (u_L^{m\T}C\gamma_\mu u_R^n)(\ell_R^\T C\gamma^\mu \ell_L)(T_{S})_{ijklmn}\;,
\\\nonumber
(O_{H6}^{H\bar H})_{\chi_1\chi_2 LR}&=-{1\over2}(u^{i\T}_{\chi_1}Cd^j_{\chi_1})(u^{k\T}_{\chi_2}Cd^l_{\chi_2}) (u_L^{m\T}C\gamma_\mu u_R^n)(\ell_R^\T C\gamma^\mu \ell_L)(T_{A})_{ijklmn}\;,
\\\nonumber
(O_{H7}^{H\bar H})_{\chi_1\chi_2 LR}&=-{1\over2}(u^{i\T}_{\chi_1}Cd^j_{\chi_1})(u^{k\T}_{\chi_2}Cd^l_{\chi_2}) (u_L^{m\T}C\gamma_\mu u_R^n)(\ell_R^\T C\gamma^\mu \ell_L)(\tilde T_{A})_{ijklmn}\;,
\\
(O_{H8}^{H\bar H})_{\chi_1\chi_2 LR}&=-{1\over2}(u^{i\T}_{\chi_1}Cd^j_{\chi_1})(u^{k\T}_{\chi_2}Cd^l_{\chi_2}) (u_L^{m\T}C\gamma_\mu u_R^n)(\ell_R^\T C\gamma^\mu \ell_L)(T_{A})_{mnklij}\;,
\label{OH8}
\end{align}
where the convention for the operators is taken from Ref.~\cite{Caswell:1982qs} with a minor change $e\leftrightarrow \ell$. Under the exchange of the chirality $L\leftrightarrow R$, we see easily that $(O_{H i}^{H\bar H})_{\chi_1\chi_2 RL}=(O_{H i}^{H\bar H})_{\chi_1\chi_2 LR}$ for $i=1,2,5,6$ and $(O_{H j}^{H\bar H})_{\chi_1\chi_2 RL}=-(O_{H j}^{H\bar H})_{\chi_1\chi_2 LR}$ for $j=7,8$. Considering $(\chi_1\chi_2)\in \{LL,LR,RL,RR\}$, there are totally 40 operators in the above. The 32 operators counted in~\cite{Caswell:1982qs} can be obtained after taking into account the following eight obvious relations
\begin{align}\nonumber
(O_{H i}^{H\bar H})_{LR LR}&=(O_{H i}^{H\bar H})_{RL LR}\;, i=1,5,6\;,~ 
\\\nonumber
(O_{H 7}^{H\bar H})_{\chi_1\chi_2LR}&=-(O_{H 7}^{H\bar H})_{\chi_2\chi_1LR}\;, \chi_{1,2}=L, R\;,
\\
(O_{H 2}^{H\bar H})_{\chi\chi LR}&=(O_{H 5}^{H\bar H})_{\chi\chi LR}-3(O_{H 6}^{H\bar H})_{\chi\chi LR}\;, \chi=L, R\;,
\label{relation:ope1}
\end{align}
where the last relation is obtained by exploiting the first relation in Eq.~\eqref{relation:TSA2} and the Fierz identity
\begin{align*}
(\psi^\T_{1\chi}C\psi_{2\chi})(\psi_{3\chi}^\T C\psi_{4\chi})&=-(\psi^\T_{1\chi}C\psi_{3\chi})(\psi_{2\chi}^\T C\psi_{4\chi})-(\psi^\T_{1\chi}C\psi_{4\chi})(\psi_{3\chi}^\T C\psi_{2\chi})\;,\chi=L,R\;.
\end{align*}

Now we show that the remaining 32 operators in Eq.~\eqref{OH8} after modulo the relations in Eq.~\eqref{relation:ope1} can be further reduced into the 19 operators shown in Eqs.~\eqref{ope:vectorf}. Using the following Fierz identities 
\begin{align}
(\psi^\T_{1L}C\psi_{2L})(\psi_{3L}^\T C\gamma^\mu \psi_{4R})&=-(\psi^\T_{1L}C\psi_{3L})(\psi_{2L}^\T C\gamma^\mu \psi_{4R})-(\psi^\T_{2L}C\psi_{3L})(\psi_{1L}^\T C\gamma^\mu \psi_{4R})\;,
\label{Fierz1}
\\\nonumber
(\psi^\T_{1R}C\psi_{2R})(\psi_{3L}^\T C\gamma^\mu \psi_{4R})&=-(\psi^\T_{1R}C\psi_{4R})(\psi_{3L}^\T C\gamma^\mu \psi_{2R})-(\psi^\T_{2R}C\psi_{4R})(\psi_{3L}^\T C\gamma^\mu \psi_{1R})\;,
\\
&
=+(\psi^\T_{1R}C\psi_{4R})(\psi_{2R}^\T C\gamma^\mu \psi_{3L})+(\psi^\T_{2R}C\psi_{4R})(\psi_{1R}^\T C\gamma^\mu \psi_{3L})\;,
\label{Fierz2}
\end{align}
where the third equality is obtained due to $\psi_{aL}^{\T}C\gamma^{\mu}\psi_{bR}=-\psi_{bR}^{\T}C\gamma^{\mu}\psi_{aL}$~\cite{Liao:2016hru}, and the relations in Eq.~\eqref{relation:TSA2}, we finally obtain the following relations among the remaining 32 operators
\begin{align}\nonumber
(O_{H 1}^{H\bar H})_{\chi L LR}&=(O_{H 3}^{H\bar H})_{\chi L LR}-3(O_{H 4}^{H\bar H})_{\chi L LR}\;,\chi=L,R\;,
\\\nonumber
(O_{H 1}^{H\bar H})_{\chi R LR}&=(O_{H 3}^{H\bar H})_{\chi R RL}-3(O_{H 4}^{H\bar H})_{\chi R RL}\;,\chi=L,R\;,
\\\nonumber
(O_{H 2}^{H\bar H})_{\chi L LR}&=(O_{H 3}^{H\bar H})_{\chi L RL}-3(O_{H 4}^{H\bar H})_{\chi L RL}\;,\chi=L,R\;,
\\\nonumber
(O_{H 2}^{H\bar H})_{\chi R LR}&=(O_{H 3}^{H\bar H})_{\chi R LR}-3(O_{H 4}^{H\bar H})_{\chi R LR}\;,\chi=L,R\;,
\\\nonumber
(O_{H 3}^{H\bar H})_{R\chi LR}&=(O_{H 5}^{H\bar H})_{R\chi LR}-3(O_{H 8}^{H\bar H})_{\chi R LR}\;,\chi=L,R\;,
\\\nonumber
(O_{H 3}^{H\bar H})_{L\chi RL}&=(O_{H 5}^{H\bar H})_{L\chi LR}+3(O_{H 8}^{H\bar H})_{\chi L  LR}\;,\chi=L,R\;,
\\\nonumber
(O_{H 4}^{H\bar H})_{R\chi LR}&=+(O_{H 6}^{H\bar H})_{R\chi LR}+2(O_{H 7}^{H\bar H})_{R\chi LR}+(O_{H 8}^{H\bar H})_{R\chi LR}\;,\chi=L,R\;,
\\
(O_{H 4}^{H\bar H})_{L\chi RL}&=-(O_{H 6}^{H\bar H})_{L\chi LR}+2(O_{H 7}^{H\bar H})_{L\chi LR}+(O_{H 8}^{H\bar H})_{L\chi LR}\;,\chi=L,R\;,
\label{relation:ope2}
\end{align}
On top of the relations in Eq.~\eqref{relation:ope1}, the above relations give 13 new constraints which therefore reduce the 32 operators into 19 independent operators as we claimed.  Choosing the following 19  independent operators
\begin{align*}
&(O_{H 3,4}^{H\bar H})_{L\chi LR}\;, (O_{H 3,4}^{H\bar H})_{R\chi RL }\;, (O_{H 5,6,8}^{H\bar H})_{\chi\chi LR}\;, \chi=L,R\;;\;
(O_{H 5,6,7,8}^{H\bar H})_{LRLR}\;,(O_{H8}^{H\bar H})_{RLLR}\;.
\end{align*}
One can easily find they have a one-to-one correspondence with the ones given in Eq.~\eqref{ope:vectorf}:
\begin{align}\nonumber
&(O_{H 3,4}^{H\bar H})_{LLLR}+L \leftrightarrow R\Leftrightarrow {\cal Q}_{2LL,a,b}^{(pp)V}+L\leftrightarrow R\;,
\\\nonumber
&(O_{H 3,4}^{H\bar H})_{LRLR}+L \leftrightarrow R\Leftrightarrow {\cal Q}_{3LR,a,b}^{(pp)V}+L\leftrightarrow R\;,
\\\nonumber
&(O_{H 5,6,8}^{H\bar H})_{LLLR}+(O_{H 5,6,8}^{H\bar H})_{RRLR}\Leftrightarrow {\cal Q}_{1LL,a,b,c}^{(pp)V}+L \leftrightarrow R\;,
\\
&(O_{H 5,6,7,8}^{H\bar H})_{LRLR}+(O_{H8}^{H\bar H})_{RLLR}\Leftrightarrow {\cal Q}_{4LR,a,b,c,d,e}^{(pp)V}\;.
\end{align}

Last, in response to the statement at the beginning of section \ref{LEFT:ope}, here we take a constructive approach to reduce the lepton tensor current operators with a  quark scalar-vector-vector current structure into those with a scalar-scalar-tensor structure. In doing so, we only need to transform the vector-vector current into scalar-tensor ones. From the FIs in Eqs.~(\ref{Fierz1}, \ref{Fierz2}), we can replace $\psi_{2L}$ by $\gamma^{\nu}\psi_{2R}$ in Eq.~\eqref{Fierz1} and  $\psi_{1R}$ by $-\gamma^{\nu}\psi_{1L}$ in Eq.~\eqref{Fierz2} to obtain new FIs, such a manipulation is guaranteed by the fact that the FI actually is the algebraic identity of gamma matrix and independent of the specific representation of spinor fields. Then we combine the two new FIs from the above replacements, after noticing that $\psi_{1R}^{\T}C\to (-\gamma^{\nu}\psi_{1L})^{\T}C=-\psi_{1L}^{\T}(\gamma^{\nu})^{\T}C=+\psi_{1L}^{\T}C\gamma^{\nu}$, and obtain 
\begin{align*}
2(\psi^\T_{1L}C\gamma^{\nu}\psi_{2R})(\psi_{3L}^\T C\gamma^\mu \psi_{4R})
&=(\psi^\T_{1L}C\psi_{3L})(\psi_{2R}^\T C\gamma^{\nu}\gamma^\mu \psi_{4R})
+(\psi_{1L}^\T C\gamma^{\nu}\gamma^\mu \psi_{3L})(\psi^\T_{2R}C\psi_{4R})
\\
&+(\psi_{1L}^\T C\gamma^\mu \psi_{4R})(\psi_{2R}^\T C\gamma^{\nu}\psi_{3L})
+(\psi^\T_{1L}C\gamma^{\nu}\psi_{4R})(\psi_{2R}^\T C\gamma^\mu \psi_{3L})
\;,
\end{align*}
where the two terms in the second line are symmetric with each other under $\mu\leftrightarrow \nu$.  Anti-symmetrizing the two indices $\mu$ and $\nu$ from the contraction with the lepton tensor current, we reach
\begin{align*}
i(\psi^\T_{1L}C\gamma^{[\nu}\psi_{2R})(\psi_{3L}^\T C\gamma^{\mu]} \psi_{4R})
=(\psi^\T_{1L}C\psi_{3L})(\psi_{2R}^\T C\sigma^{\nu\mu} \psi_{4R})
+(\psi_{1L}^\T C\sigma^{\nu\mu} \psi_{3L})(\psi^\T_{2R}C\psi_{4R})\;,
\end{align*}
where $[\nu,\;\cdots,\mu]$ means anti-symmetrization of the two indices. The other chiral structures can be shown in a similar manner.  Thus, the above finishes the proof of the equivalence.

\section{SMEFT operators}
\label{app:SMEFTope}

For operators with a scalar lepton current,  there are 12 independent operators chosen as follows 
\begin{align}\nonumber
{\cal O}_{u^3d^3L^21}^{S,(A)}&=(u_R^{i\T} Cd_R^j)(u_R^{k\T} Cd_R^l)(u_R^{m\T} Cd_R^n)(L^\T_a CL_b^\prime)\epsilon_{ab}T^{SSS}_{\{ij\}\{kl\}\{mn\}}\;,
\\\nonumber
{\cal O}_{u^3d^3L^22}^{S,(A)}&=(u_R^{i\T} Cd_R^j)(u_R^{k\T} Cd_R^l)(u_R^{m\T} Cd_R^n)(L^\T_a CL_b^\prime)\epsilon_{ab}T^{SAA}_{\{ij\}[kl][mn]}\;, 
\\\nonumber
{\cal O}_{u^2d^2Q^2L^21}^{S,(S)}&=(u_R^{i\T} Cd_R^j)(u_R^{k\T} Cd_R^l)(Q^{m\T}_a CQ^n_b)(L^\T_c CL_d^\prime)\epsilon_{ac}\epsilon_{bd}T^{SSS}_{\{ij\}\{kl\}\{mn\}}\;,
\\\nonumber
{\cal O}_{u^2d^2Q^2L^22}^{S,(A)}&=(u_R^{i\T} Cd_R^j)(u_R^{k\T} Cd_R^l)(Q^{m\T}_a CQ^n_b)(L^\T_c CL_d^\prime)\epsilon_{ab}\epsilon_{cd}T^{SAA}_{\{ij\}[kl][mn]}\;, 
\\\nonumber
{\cal O}_{u^2d^2Q^2L^23}^{S,(S)}&=(u_R^{i\T} Cd_R^j)(u_R^{k\T} Cd_R^l)(Q^{m\T}_a CQ^n_b)(L^\T_c CL_d^\prime)\epsilon_{ac}\epsilon_{bd}T^{SAA}_{\{mn\}[kl][ij]}\;,
\\\nonumber
{\cal O}_{udQ^4L^21}^{S,(A)}&=(u_R^{i\T} Cd_R^j)(Q^{k\T}_a CQ^l_b)(Q^{m\T}_c CQ^n_d)(L^\T_e CL_f^\prime)\epsilon_{ab}\epsilon_{cd}\epsilon_{ef}T^{SAA}_{\{ij\}[kl][mn]}\;,
\\\nonumber
{\cal O}_{udQ^4L^22}^{S,(S)}&=(u_R^{i\T} Cd_R^j)(Q^{k\T}_a CQ^l_b)(Q^{m\T}_c CQ^n_d)(L^\T_e CL_f^\prime)\epsilon_{ab}\epsilon_{ce}\epsilon_{df}T^{SAA}_{\{mn\}[kl][ij]}\;,
\\\nonumber
{\cal O}_{Q^6L^2}^{S,(S)}&=(Q^{i\T}_a CQ^j_b)(Q^{k\T}_c CQ^l_d)(Q^{m\T}_e CQ^n_f)(L^\T_g CL_h^\prime)\epsilon_{ab}\epsilon_{cd}\epsilon_{eg}\epsilon_{fh}T^{SAA}_{\{mn\}[kl][ij]}\;,
\\\nonumber
{\cal O}_{u^4d^2e^21}^{S,(S)}&=(u_R^{i\T} Cu_R^j)(u_R^{k\T} Cd_R^l)(u_R^{m\T} Cd_R^n)(e^\T_R Ce_R^\prime)T^{SSS}_{\{ij\}\{kl\}\{mn\}}\;,
\\\nonumber
{\cal O}_{u^4d^2e^22}^{S,(S)}&=(u_R^{i\T} Cu_R^j)(u_R^{k\T} Cd_R^l)(u_R^{m\T} Cd_R^n)(e^\T_R Ce_R^\prime)T^{SAA}_{\{ij\}[kl][mn]}\;,
\\\nonumber
{\cal O}_{u^3dQ^2e^2}^{S,(S)}&=(u_R^{i\T} Cu_R^j)(u_R^{k\T} Cd_R^l)(Q^{m\T}_a C Q^{n}_b)(e^\T_R Ce_R^\prime)\epsilon_{ab}T^{SAA}_{\{ij\}[kl][mn]}\;, 
\\
{\cal O}_{u^2Q^4e^2}^{S,(S)}&=(u_R^{i\T} Cu_R^j)(Q^{k\T}_a CQ^l_b)(Q^{m\T}_c CQ^n_d)(e^\T_R Ce_R^\prime)\epsilon_{ab}\epsilon_{cd}T^{SAA}_{\{ij\}[kl][mn]}\;,
\label{NN:scalarf}
\end{align}
where the first superscript `S' ( and the following `V' and `T' ) is used to represent the relevant operator with a scalar (vector and tensor for the following ones) lepton current, and the bracket superscripts `(S)/(A)' indicate the flavor symmetric/anti-symmetric property of the lepton current under the exchange of the two lepton fields. 

For operators with a vector lepton current, we find there are 7 independent operators chosen as follows
\begin{align}\nonumber
{\cal O}_{u^3d^2QeL1}^V&=(u_R^{i\T} Cd_R^j)(u_R^{k\T} Cd_R^l)(Q^{m\T}_a C\gamma_\mu u^n_R)(e^\T_R C\gamma^\mu L_b^\prime)\epsilon_{ab}T^{SSS}_{\{ij\}\{kl\}\{mn\}}\;, 
\\\nonumber
{\cal O}_{u^3d^2QeL2}^V&=(u_R^{i\T} Cd_R^j)(u_R^{k\T} Cd_R^l)(Q^{m\T}_a C\gamma_\mu u^n_R)(e^\T_R C\gamma^\mu L_b^\prime)\epsilon_{ab}T^{SAA}_{\{ij\}[kl][mn]}\;,
\\\nonumber
{\cal O}_{u^3d^2QeL3}^V&=(u_R^{i\T} Cd_R^j)(u_R^{k\T} Cd_R^l)(Q^{m\T}_a C\gamma_\mu u^n_R)(e^\T_R C\gamma^\mu L_b^\prime)\epsilon_{ab}T^{SAA}_{\{mn\}[kl][ij]}\;,
\\\nonumber
{\cal O}_{u^2dQ^3eL1}^V&=(u_R^{i\T} Cd_R^j)(Q^{k\T}_a CQ^l_b)(Q^{m\T}_c C\gamma_\mu u^n_R)(e^\T_R C\gamma^\mu L_d^\prime)\epsilon_{ab}\epsilon_{cd}T^{SAA}_{\{ij\}[kl][mn]}\;,
\\\nonumber
{\cal O}_{u^2dQ^3eL2}^V&=(u_R^{i\T} Cd_R^j)(Q^{k\T}_a CQ^l_b)(Q^{m\T}_c C\gamma_\mu u^n_R)(e^\T_R C\gamma^\mu L_d^\prime)\epsilon_{ab}\epsilon_{cd}T^{SAA}_{\{mn\}[kl][ij]}\;,
\\\nonumber
{\cal O}_{u^2dQ^3eL3}^V&=(u_R^{i\T} Cd_R^j)(Q^{k\T}_a CQ^l_b)(Q^{m\T}_c C\gamma_\mu u^n_R)(e^\T_R C\gamma^\mu L_d^\prime)\epsilon_{ab}\epsilon_{cd}T^{AAA}_{[ij][kl][mn]}\;,
\\
{\cal O}_{uQ^5eL}^V&=(Q^{i\T}_a CQ^j_b)(Q^{k\T}_c CQ^l_d)(Q^{m\T}_e C\gamma_\mu u^n_R)(e^\T_R C\gamma^\mu L_f^\prime)\epsilon_{ab}\epsilon_{cd}\epsilon_{ef}T^{SAA}_{\{mn\}[kl][ij]}\;,
\label{NN:vectorf}
\end{align}
where the lepton current mixes the lepton doublet field $L$ and singlet field $e_R$ and therefore no any flavor symmetry property. 

For operators with a tensor lepton current, there are 10  independent operators which are chosen as follows
\begin{align}\nonumber
{\cal O}_{u^2d^2Q^2L^21}^{T,(S)}&=(u_R^{i\T} Cd_R^j)(u_R^{k\T} Cd_R^l)(Q^{m\T}_a C\sigma_{\mu\nu}Q^n_b)(L^\T_c C\sigma^{\mu\nu}L_d^\prime)\epsilon_{ab}\epsilon_{cd}T^{SSS}_{\{ij\}\{kl\}\{mn\}}\;,
\\\nonumber
{\cal O}_{u^2d^2Q^2L^22}^{T,(A)}&=(u_R^{i\T} Cd_R^j)(u_R^{k\T} Cd_R^l)(Q^{m\T}_a C\sigma_{\mu\nu}Q^n_b)(L^\T_c C\sigma^{\mu\nu}L_d^\prime)\epsilon_{ac}\epsilon_{bd}T^{SAA}_{\{ij\}[kl][mn]}\;, 
\\\nonumber
{\cal O}_{u^2d^2Q^2L^23}^{T,(S)}&=(u_R^{i\T} Cd_R^j)(u_R^{k\T} Cd_R^l)(Q^{m\T}_a C\sigma_{\mu\nu}Q^n_b)(L^\T_c C\sigma^{\mu\nu}L_d^\prime)\epsilon_{ab}\epsilon_{cd}T^{SAA}_{\{mn\}[kl][ij]}\;,
\\\nonumber
{\cal O}_{udQ^4L^21}^{T,(A)}&=(u_R^{i\T} Cd_R^j)(Q^{k\T}_a CQ^l_b)(Q^{m\T}_c C\sigma_{\mu\nu}Q^n_d)(L^\T_e C\sigma^{\mu\nu}L_f^\prime)\epsilon_{ab}\epsilon_{ce}\epsilon_{df}T^{SAA}_{\{ij\}[kl][mn]}\;,
\\\nonumber
{\cal O}_{udQ^4L^22}^{T,(S)}&=(u_R^{i\T} Cd_R^j)(Q^{k\T}_a CQ^l_b)(Q^{m\T}_c C\sigma_{\mu\nu}Q^n_d)(L^\T_e C\sigma^{\mu\nu}L_f^\prime)\epsilon_{ab}\epsilon_{cd}\epsilon_{ef}T^{SAA}_{\{mn\}[kl][ij]}\;,
\\\nonumber
{\cal O}_{udQ^4L^23}^{T,(A)}&=(u_R^{i\T} Cd_R^j)(Q^{k\T}_a CQ^l_b)(Q^{m\T}_c C\sigma_{\mu\nu}Q^n_d)(L^\T_e C\sigma^{\mu\nu}L_f^\prime)\epsilon_{ab}\epsilon_{ce}\epsilon_{df}T^{AAA}_{[ij][kl][mn]}\;,
\\\nonumber
{\cal O}_{Q^6L^2}^{T,(S)}&=(Q^{i\T}_a CQ^j_b)(Q^{k\T}_cCQ^l_d)(Q^{m\T}_eC\sigma_{\mu\nu}Q^n_f)(L^\T_g C\sigma^{\mu\nu}L_h^\prime)\epsilon_{ab}\epsilon_{cd}\epsilon_{ef}\epsilon_{gh}T^{SAA}_{\{mn\}[kl][ij]}\;,
\\\nonumber
{\cal O}_{u^4d^2e^2}^{T,(A)}&=(u_R^{i\T} C\sigma_{\mu\nu}u_R^j)(u_R^{k\T} Cd_R^l)(u_R^{m\T} Cd_R^n)(e^\T_R C\sigma^{\mu\nu}e_R^\prime)T^{SAA}_{\{mn\}[ij][kl]}\;,
\\\nonumber
{\cal O}_{u^3dQ^2e^21}^{T,(A)}&=(u_R^{i\T} C\sigma_{\mu\nu} u_R^j)(u_R^{k\T} Cd_R^l)(Q^{m\T}_a C Q^{n}_b)(e^\T_R C\sigma^{\mu\nu}e_R^\prime)\epsilon_{ab}T^{SAA}_{\{kl\}[mn][ij]}\;,
\\
{\cal O}_{u^3dQ^2e^22}^{T,(A)}&=(u_R^{i\T} C\sigma_{\mu\nu}u_R^j)(u_R^{k\T} Cd_R^l)(Q^{m\T}_a C Q^{n}_b)(e^\T_R C\sigma^{\mu\nu}e_R^\prime)\epsilon_{ab}T^{AAA}_{[ij][kl][mn]}\;.
\label{NN:tensorf}
\end{align}

\section{Reduction of the redundant operators in the SMEFT }
\label{app:SMEFTreduction}
Ref.~\cite{Girmohanta:2019fsx} provides a bunch of the dim-12 operators contributing to dinucleon to dilepton transitions in the SMEFT, but the operators are neither complete nor independent as a basis. By using the color tensor relations in~Eq.~\eqref{relation:TSA1} and the SI for the $SU(2)_{\rm L}$ group and FIs in Appendix~\ref{app:LEFTreduction}, in the following, we translate their operators as linear combinations of operators given in Appendix~\ref{app:SMEFTope} so that one can easily recognize the redundancy and incompleteness in~\cite{Girmohanta:2019fsx}:
\begin{align}\nonumber
{\cal O}_{1}^{(pp)}&=+{\cal O}_{u^4d^2e^21}^{S,(S)}-3{\cal O}_{u^4d^2e^22}^{S,(S)}\;,\;\;
\\\nonumber
{\cal O}_{2}^{(pp)}&=+{\cal O}_{u^4d^2e^21}^{S,(S)}\;,\;\;
\\
{\cal O}_{3}^{(pp)}&=+{\cal O}_{u^4d^2e^22}^{S,(S)}\;;
\\
{\cal O}_{4}^{(pp)}&=+{\cal O}_{u^3dQ^2e^2}^{S,(S)}\;;\;\;
\\\nonumber
{\cal O}_{5}^{(pp)}&=+{\cal O}_{u^2Q^4e^2}^{S,(S)}\;,\;\; 
\\
{\cal O}_{6}^{(pp)}&=-3{\cal O}_{u^2Q^4e^2}^{S,(S)}\;;
\\\nonumber
{\cal O}_{7}^{(pp,np)}&=-{1\over 8}\left(C_{u^3d^2QeL1}^V-3C_{u^3d^2QeL3}^V\right)\;,
\\\nonumber
{\cal O}_{8}^{(pp,np)}&=-{1\over 8}\left(C_{u^3d^2QeL1}^V-6C_{u^3d^2QeL2}^V+3C_{u^3d^2QeL3}^V\right)\;,
\\\nonumber
{\cal O}_{9}^{(pp,np)}&=-{1\over 8}\left(C_{u^3d^2QeL1}^V-3C_{u^3d^2QeL3}^V\right)\;,
\\\nonumber
{\cal O}_{10}^{(pp,np)}&=-{1\over 8}\left(C_{u^3d^2QeL1}^V+4C_{u^3d^2QeL2}^V+C_{u^3d^2QeL3}^V\right)\;,
\\
{\cal O}_{11}^{(pp,np)}&=-{1\over 8}\left(C_{u^3d^2QeL1}^V-2C_{u^3d^2QeL2}^V-C_{u^3d^2QeL3}^V\right)\;;
\\\nonumber
{\cal O}_{12}^{(pp,np)}&=-{1\over 2}\left(C_{u^2dQ^3eL1}^V+C_{u^2dQ^3eL2}^V+C_{u^2dQ^3eL3}^V \right)\;,
\\\nonumber
{\cal O}_{13}^{(pp,np)}&=-{1\over 2}C_{u^2dQ^3eL1}^V\;,
\\\nonumber
{\cal O}_{14}^{(pp,np)}&=-{1\over 2}C_{u^2dQ^3eL3}^V\;,
\\\nonumber
{\cal O}_{16}^{(pp,np)}&=+{3\over 2}\left(C_{u^2dQ^3eL1}^V+C_{u^2dQ^3eL2}^V+2C_{u^2dQ^3eL3}^V \right)\;,
\\
{\cal O}_{17}^{(pp,np)}&=+{3\over 2}C_{u^2dQ^3eL1}^V\;;
\\\nonumber
{\cal O}_{15}^{(pp,np)}&=-{1\over 2}{\cal O}_{uQ^5eL}^S\;,
\\
{\cal O}_{24}^{(pp,np)}&=+{3\over 2}{\cal O}_{uQ^5eL}^S\;;
\\\nonumber
{\cal O}_{18}^{(pp,np,nn)}&=-{1\over 16}\left(8{\cal O}_{u^2d^2Q^2L^21}^{S,(S)}-24{\cal O}_{u^2d^2Q^2L^23}^{S,(S)}-{\cal O}_{u^2d^2Q^2L^21}^{T,(S)}+3{\cal O}_{u^2d^2Q^2L^23}^{T,(S)}\right)\;,
\\\nonumber
{\cal O}_{19}^{(pp,np,nn)}&=-{1\over 16}\left(8{\cal O}_{u^2d^2Q^2L^21}^{S,(S)}-{\cal O}_{u^2d^2Q^2L^21}^{T,(S)}\right)\;,
\\\nonumber
{\cal O}_{20}^{(pp,np,nn)}&=-{1\over 16}\left(8{\cal O}_{u^2d^2Q^2L^23}^{S,(S)}-{\cal O}_{u^2d^2Q^2L^23}^{T,(S)}\right)\;,
\\\nonumber
{\cal O}_{22}^{(pp,np,nn)}&=+{1\over 16}\left(8{\cal O}_{u^2d^2Q^2L^21}^{S,(S)}-24{\cal O}_{u^2d^2Q^2L^23}^{S,(S)}+3{\cal O}_{u^2d^2Q^2L^21}^{T,(S)}-9{\cal O}_{u^2d^2Q^2L^23}^{T,(S)}\right)\;,
\\
{\cal O}_{23}^{(pp,np,nn)}&=+{1\over 16}\left(8{\cal O}_{u^2d^2Q^2L^21}^{S,(S)}+3{\cal O}_{u^2d^2Q^2L^21}^{T,(S)}\right)\;;
\\\nonumber
{\cal O}_{21}^{(pp,np,nn)}&=-{1\over 16}\left(8{\cal O}_{udQ^4L^22}^{S,(S)}-{\cal O}_{udQ^4L^22}^{T,(S)}\right)\;,
\\\nonumber
{\cal O}_{25}^{(pp,np,nn)}&=+{3\over 8}\left(2{\cal O}_{udQ^4L^21}^{S,(A)}-{\cal O}_{udQ^4L^21}^{T,(A)}\right)\;,
\\\nonumber
{\cal O}_{26}^{(pp,np,nn)}&=+{1\over 16}\left(8{\cal O}_{udQ^4L^22}^{S,(S)}+3{\cal O}_{udQ^4L^22}^{T,(S)}\right)\;,
\\
{\cal O}_{27}^{(pp,np,nn)}&=+4{\cal O}_{udQ^4L^21}^{S,(A)}+{\cal O}_{udQ^4L^21}^{T,(A)}\;;
\\
{\cal O}_{28}^{(pp,np,nn)}&=-{1\over 16}\left( 8{\cal O}_{Q^6L^2}^{S,(S)}-{\cal O}_{Q^6L^2}^{T,(S)}\right)\;.
\end{align}

\section{Chiral basis construction}
\label{app:chiralope}

In Tabs.~\ref{tab:chiralirrep_pp}-\ref{tab:chiralirrep_nn}, we rewrite the operator basis in the LEFT into a chiral basis which has well-defined chiral transformation properties. This chiral basis is obtained from LEFT basis by  symmetrizing quark flavors with the same chirality modulo the anti-symmetric chiral singlet bilinear ${\cal D}_\chi^{ij}=\epsilon^{uv}(q_{\chi,u}^{i\T}\Gamma q_{\chi,v}^j)=(u_\chi^{i\T}\Gamma d_\chi^j)-i\leftrightarrow j$. In doing so, most of the operators with color tensors $T^{SAA}_{\{ij\}[kl][mn]}$ and $T^{AAA}_{[ij][kl][mn]}$ in the LEFT already belong to the chiral irreps under $SU(2)_{\rm L}^f\times SU(2)_R^f$ as shown in Tabs.~\ref{tab:chiralirrep_pp}-\ref{tab:chiralirrep_nn}. For the remaining operators, especially those with color tensor $T^{SSS}_{\{ij\}\{kl\}\{mn\}}$, the chiral irrep ones are defined in terms of the above flavor symmetrization procedures as follows:

\noindent
{\small$\scriptscriptstyle\blacksquare$ \small\bf\boldmath Dim-12 operators contributing to $pp\to\ell^+\ell^{\prime+}$}
{\footnotesize
\begin{align*}
P_{1,a}^{(pp)S,\pm}&={1\over 5}(u_L^{i\T} Cu_L^j)\left[4(u_L^{k\T} Cd_L^l)(u_L^{m\T} Cd_L^n)+(u_L^{k\T} Cu_L^l)(d_L^{m\T} Cd_L^n)\right]j_{S,\pm}^{\ell\ell^\prime}T^{SSS}_{\{ij\}\{kl\}\{mn\}}\;,
\\
P_{3,a}^{(pp)S,\pm}&={1\over 3}\left[2(u_L^{i\T} Cd_L^j)(u_L^{k\T} Cd_L^l)+(u_L^{i\T} Cu_L^j)(d_L^{k\T} Cd_L^l)\right](u_R^{m\T} Cu_R^n)j_{S,\pm}^{\ell\ell^\prime} T^{SSS}_{\{ij\}\{kl\}\{mn\}}\;,
\\
P_{1,a}^{(pp)V}&={1\over 5} \left[ 2(u_L^{i\T} Cd_L^j)(u_L^{k\T} Cd_L^l)(u_L^{m\T} C\gamma_\mu u_R^n)+(u_L^{i\T} Cu_L^j)(d_L^{k\T} Cd_L^l)(u_L^{m\T} C\gamma_\mu u_R^n)\right.
\\
&\left.+ 2(u_L^{i\T} Cu_L^j)(u_L^{k\T} Cd_L^l)(d_L^{m\T} C\gamma_\mu u_R^n) \right]j_{V}^{\ell\ell^\prime,\mu} T^{SSS}_{\{ij\}\{kl\}\{mn\}}\;,
\\
P_{1,b}^{(pp)V}&={1\over 3}(u_L^{k\T} Cd_L^l) \left[ 2(u_L^{i\T} Cd_L^j)(u_L^{m\T} C\gamma_\mu u_R^n)+(u_L^{i\T} Cu_L^j)(d_L^{m\T} C\gamma_\mu u_R^n)\right]j_{V}^{\ell\ell^\prime,\mu} T^{SAA}_{\{ij\}[kl][mn]}\;,
\\
P_{2,a}^{(pp)V}&={1\over 5}(u_L^{i\T} Cu_L^j) \left[ 4(u_L^{k\T} Cd_L^l)(u_L^{m\T} C\gamma_\mu d_R^n)+(u_L^{k\T} Cu_L^l)(d_L^{m\T} C\gamma_\mu d_R^n) \right]j_{V}^{\ell\ell^\prime,\mu} T^{SSS}_{\{ij\}\{kl\}\{mn\}}\;,
\\
P_{3,a}^{(pp)V}&={1\over 3}(u_L^{i\T} Cu_L^j)\left[2(u_R^{k\T} Cd_R^l)(u_L^{m\T} C\gamma_\mu d_R^n)+(d_R^{k\T} Cd_R^l)(u_L^{m\T} C\gamma_\mu u_R^n)\right]j_{V}^{\ell\ell^\prime,\mu} T^{SSS}_{\{ij\}\{kl\}\{mn\}}\;,
\\
P_{4,a}^{(pp)V}&={1\over 9}\left[4(u_L^{i\T} Cd_L^j)(u_R^{k\T} Cd_R^l)(u_L^{m\T} C\gamma_\mu u_R^n)+2(u_L^{i\T} Cu_L^j)(u_R^{k\T} Cd_R^l)(d_L^{m\T} C\gamma_\mu u_R^n)\right.
\\
&\left.+2(u_L^{i\T} Cd_L^j)(u_R^{k\T} Cu_R^l)(u_L^{m\T} C\gamma_\mu d_R^n)+(u_L^{i\T} Cu_L^j)(u_R^{k\T} Cu_R^l)(d_L^{m\T} C\gamma_\mu d_R^n) \right]j_{V}^{\ell\ell^\prime,\mu} T^{SSS}_{\{ij\}\{kl\}\{mn\}}\;,
\\
P_{4,b}^{(pp)V}&={1\over 3}(u_R^{k\T} Cd_R^l)\left[ 2(u_L^{i\T} Cd_L^j)(u_L^{m\T} C\gamma_\mu u_R^n)+(u_L^{i\T} Cu_L^j)(d_L^{m\T} C\gamma_\mu u_R^n)\right]j_{V}^{\ell\ell^\prime,\mu}T^{SAA}_{\{ij\}[kl][mn]}\;,
\\
P_{4,c}^{(pp)V}&={1\over 3}(u_L^{i\T} Cd_L^j)\left[ 2(u_R^{k\T} Cd_R^l)(u_L^{m\T} C\gamma_\mu u_R^n)+(u_R^{k\T} Cu_R^l)(u_L^{m\T} C\gamma_\mu d_R^n)\right]j_{V}^{\ell\ell^\prime,\mu} T^{SAA}_{\{kl\}[mn][ij]}\;,
\\
P_{2,a}^{(pp)T,-}&={1\over 2}\left[ (u_L^{i\T} C\sigma_{\mu\nu}u_L^j)(u_L^{k\T} Cd_L^l)+(u_L^{i\T} C\sigma_{\mu\nu}d_L^j)(u_L^{k\T} Cu_L^l) \right](u_R^{m\T} Cd_R^n)j_{T,-}^{\ell\ell^\prime,\mu\nu}T^{SAA}_{\{kl\}[mn][ij]}\;.
\end{align*}
}
\noindent
{\small$\scriptscriptstyle\blacksquare$ \small\bf\boldmath Dim-12 operators contributing to $pn\to\ell^+\bar\nu^{\prime}$}
{\footnotesize
\begin{align*}
P_{1,a}^{(pn)S}&={1\over 5}(u_L^{i\T} Cd_L^j)\left[2(u_L^{k\T} Cd_L^l)(u_L^{m\T} Cd_L^n)+3(u_L^{k\T} Cu_L^l)(d_L^{m\T} Cd_L^n)\right]j_{S}^{\ell\nu^\prime}T^{SSS}_{\{ij\}\{kl\}\{mn\}}\;,
\\
P_{3,a}^{(pn)S}&={1\over 3}\left[2(u_L^{i\T} Cd_L^j)(u_L^{k\T} Cd_L^l)+(u_L^{i\T} Cu_L^j)(d_L^{k\T} Cd_L^l)\right](u_R^{m\T} Cd_R^m)j_{S}^{\ell\nu^\prime}T^{SSS}_{\{ij\}\{kl\}\{mn\}}\;,
\\
P_{1,a}^{(pn)V}&={1\over 5} \left[ 2(u_L^{i\T} Cd_L^j)(u_L^{k\T} Cd_L^l)(u_L^{m\T} C\gamma_\mu d_R^n)+(u_L^{i\T} Cu_L^j)(d_L^{k\T} Cd_L^l)(u_L^{m\T} C\gamma_\mu d_R^n)\right.
\\
&\left.+ 2(u_L^{i\T} Cu_L^j)(u_L^{k\T} Cd_L^l)(d_L^{m\T} C\gamma_\mu d_R^n) \right]j_{V}^{\ell\nu^\prime,\mu} T^{SSS}_{\{ij\}\{kl\}\{mn\}}\;,
\\
P_{1,b}^{(pn)V}&={1\over 3}(u_L^{k\T} Cd_L^l) \left[ 2(u_L^{i\T} Cd_L^j)(u_L^{m\T} C\gamma_\mu d_R^n)+(u_L^{i\T} Cu_L^j)(d_L^{m\T} C\gamma_\mu d_R^n)\right]j_{V}^{\ell\nu^\prime,\mu}T^{SAA}_{\{ij\}[kl][mn]}\;,
\\
P_{2,a}^{(pn)V}&={1\over 5} \left[ 2(u_L^{i\T} Cd_L^j)(u_L^{k\T} Cd_L^l)(d_L^{m\T} C\gamma_\mu u_R^n)+(u_L^{i\T} Cu_L^j)(d_L^{k\T} Cd_L^l)(d_L^{m\T} C\gamma_\mu u_R^n)\right.
\\
&\left.+ 2(u_L^{i\T} Cd_L^j)(d_L^{k\T} Cd_L^l)(u_L^{m\T} C\gamma_\mu u_R^n) \right]j_{V}^{\ell\nu^\prime,\mu} T^{SSS}_{\{ij\}\{kl\}\{mn\}}\;,
\\
P_{2,b}^{(pn)V}&={1\over 3}(u_L^{k\T} Cd_L^l) \left[ 2(u_L^{i\T} Cd_L^j)(d_L^{m\T} C\gamma_\mu u_R^n)+(d_L^{i\T} Cd_L^j)(u_L^{m\T} C\gamma_\mu u_R^n)\right]j_{V}^{\ell\nu^\prime,\mu}T^{SAA}_{\{ij\}[kl][mn]}\;,
\\
P_{4,a}^{(pn)V}&={1\over 9}\left[4(u_L^{i\T} Cd_L^j)(u_R^{k\T} Cd_R^l)(u_L^{m\T} C\gamma_\mu d_R^n)+2(u_L^{i\T} Cu_L^j)(u_R^{k\T} Cd_R^l)(d_L^{m\T} C\gamma_\mu d_R^n)\right.
\\
&\left.+2(u_L^{i\T} Cd_L^j)(d_R^{k\T} Cd_R^l)(u_L^{m\T} C\gamma_\mu u_R^n)+(u_L^{i\T} Cu_L^j)(d_R^{k\T} Cd_R^l)(d_L^{m\T} C\gamma_\mu u_R^n) \right]j_{V}^{\ell\nu^\prime,\mu}T^{SSS}_{\{ij\}\{kl\}\{mn\}}\;,
\\
P_{4,b}^{(pn)V}&={1\over 3}(u_R^{k\T} Cd_R^l)\left[ 2(u_L^{i\T} Cd_L^j)(u_L^{m\T} C\gamma_\mu d_R^n)+(u_L^{i\T} Cu_L^j)(d_L^{m\T} C\gamma_\mu d_R^n)\right]j_{V}^{\ell\nu^\prime,\mu}T^{SAA}_{\{ij\}[kl][mn]}\;,
\\
P_{4,c}^{(pn)V}&={1\over 3}(u_L^{i\T} Cd_L^j)\left[ 2(u_R^{k\T} Cd_R^l)(u_L^{m\T} C\gamma_\mu d_R^n)+(d_R^{k\T} Cd_R^l)(u_L^{m\T} C\gamma_\mu u_R^n)\right]j_{V}^{\ell\nu^\prime,\mu}T^{SAA}_{\{kl\}[mn][ij]}\;,
\\
P_{1,a}^{(pn)T}&={1\over 6}(u_L^{k\T} Cd_L^l)\left[ 4(u_L^{i\T} Cd_L^j)(u_L^{m\T} C\sigma_{\mu\nu} d_L^n)+ (u_L^{i\T} Cu_L^j)(d_L^{m\T} C\sigma_{\mu\nu} d_L^n)\right.
\\
&\left.+(d_L^{i\T} Cd_L^j)(u_L^{m\T} C\sigma_{\mu\nu} u_L^n)\right]j_{T}^{\ell\nu^\prime,\mu\nu}T^{SAA}_{\{ij\}[kl][mn]}\;,
\\
P_{3,a}^{(pn)T}&={1\over 6}\left[4(u_L^{i\T} Cd_L^j)(u_L^{k\T} C\sigma_{\mu\nu} d_L^l) +(u_L^{i\T} Cu_L^j)(d_L^{k\T} C\sigma_{\mu\nu} d_L^l)+(d_L^{i\T} Cd_L^j)(u_L^{k\T} C\sigma_{\mu\nu} u_L^l)\right]
\\
&\times (u_R^{m\T}Cd_R^n) j_{T}^{\ell\nu^\prime,\mu\nu}T^{SAA}_{\{ij\}[kl][mn]}\;,
\\
\hat P_{3,a}^{(pn)T}&={1\over 3}\left[2(u_R^{i\T} Cd_R^j)(u_R^{k\T} Cd_R^l) +(u_R^{i\T} Cu_R^j)(d_R^{k\T} Cd_R^l)\right](u_L^{m\T}C\sigma_{\mu\nu}  d_L^n) j_{T}^{\ell\nu^\prime,\mu\nu}T^{SSS}_{\{ij\}\{kl\}\{mn\}}\;.
\end{align*}
}
\noindent
{\small$\scriptscriptstyle\blacksquare$ \small\bf\boldmath Dim-12 operators contributing to $nn\to\bar\nu\bar\nu^\prime$}
{\footnotesize
\begin{align*}
P_{1,a}^{(nn)S}&={1\over 5}(d_L^{i\T} Cd_L^j)\left[4(d_L^{k\T} Cu_L^l)(d_L^{m\T} Cu_L^n)+(d_L^{k\T} Cd_L^l)(u_L^{m\T} Cu_L^n)\right]j_{S}^{\nu\nu^\prime}T^{SSS}_{\{ij\}\{kl\}\{mn\}}\;,
\\
P_{3,a}^{(nn)S}&={1\over 3}\left[2(d_L^{i\T} Cu_L^j)(d_L^{k\T} Cu_L^l)+(d_L^{i\T} Cd_L^j)(u_L^{k\T} Cu_L^l)\right](d_R^{m\T} Cd_R^n)j_{S}^{\nu\nu^\prime} T^{SSS}_{\{ij\}\{kl\}\{mn\}}\;,
\\
P_{2,a}^{(nn)T}&=-{1\over 2}\left[ (d_L^{i\T} C\sigma_{\mu\nu}d_L^j)(d_L^{k\T} Cu_L^l)+(d_L^{i\T} C\sigma_{\mu\nu}u_L^j)(d_L^{k\T} Cd_L^l) \right](d_R^{m\T} Cu_R^n)j_{T}^{\nu\nu^\prime,\mu\nu}T^{SAA}_{\{kl\}[mn][ij]}\;.
\end{align*}
}
They are converted into the linear combinations of the LEFT operators shown in Tabs.~\ref{tab:chiralirrep_pp}-\ref{tab:chiralirrep_nn} via the Schouten identities in Eq.~\eqref{relation:TSA1} and the Fierz identities in Eqs.~(\ref{Fierz1}, \ref{Fierz2}) and their similar cousins.  


\begin{table}
\centering
\resizebox{\linewidth}{!}{
\begin{tabular}{|c | c | c | c|}
\hline
Chiral basis & LEFT basis & Chiral irrep. &  Chiral spurion  
\\\hline
$P_{1,a}^{(pp)S,\pm}$ & ${1\over 5}\left(5{\cal Q}_{1LLL,a}^{(pp)S,\pm}-3{\cal Q}_{1LLL,b}^{(pp)S,\pm}\right)$   &  $({\bf 7}_L,{\bf 1}_R)$ & $ \theta^{u_Lv_Lw_Lx_Ly_Lz_L}_{(111122)}$ 
\\
$P_{1,b}^{(pp)S,\pm}$ &  ${\cal Q}_{1LLL,b}^{(pp)S,\pm}$  & $({\bf 3}_L,{\bf 1}_R)|_a$ & $ \theta^{u_Lv_L}_{(11)} $ 
\\\hline
$P_{2,a}^{(pp)S,\pm}$ & ${\cal Q}_{2LLR,a}^{(pp)S,\pm}$ &  $({\bf 5}_L,{\bf 3}_R)$ & $\theta^{u_Lv_Lw_Lx_Ly_Rz_R}_{(1112)(12)}$
\\
$P_{2,b}^{(pp)S,\pm}$ & ${\cal Q}_{2LLR,b}^{(pp)S,\pm}$ & $({\bf 3}_L,{\bf 1}_R)|_b$  & $ \theta^{u_Lv_L}_{(11)} $ 
\\
$P_{3,a}^{(pp)S,\pm}$ & ${\cal Q}_{3LLR,a}^{(pp)S,\pm}-{\cal Q}_{3LLR,b}^{(pp)S,\pm}$ &  $({\bf 5}_L,{\bf 3}_R)$ &  $\theta^{u_Lv_Lw_Lx_Ly_Rz_R}_{(1122)(11)}$ 
\\
$P_{3,b}^{(pp)S,\pm}$ & ${\cal Q}_{3LLR,b}^{(pp)S,\pm}$ &  $({\bf 1}_L,{\bf 3}_R)|_c$  & $ \theta^{u_Rv_R}_{(11)} $ 
\\
$P_4^{(pp)S,\pm}$& ${\cal Q}_{4LLR}^{(pp)S,\pm}$ &  $({\bf 5}_L,{\bf 3}_R)$  & $\theta^{u_Lv_Lw_Lx_Ly_Rz_R}_{(1111)(22)}$
\\\hline\hline
$P_{1,a}^{(pp)V}$ & ${1\over 5}\left(5{\cal Q}_{1LL,a}^{(pp)V}-6{\cal Q}_{1LL,b}^{(pp)V}-3{\cal Q}_{1LL,c}^{(pp)V}\right) $ &  $({\bf 6}_L,{\bf 2}_R)$ & $ \theta^{u_Lv_Lw_Lx_Ly_Lz_R}_{(11122)1} $  
\\
$P_{1,b}^{(pp)V}$ &  ${1\over 3}\left(3{\cal Q}_{1LL,b}^{(pp)V}-{\cal Q}_{1LL,c}^{(pp)V}\right)$ &  $({\bf 4}_L,{\bf 2}_R)|_a$  & $ \theta^{u_Lv_Lw_Lx_R}_{(112)1}  $  
\\
$P_{1,c}^{(pp)V}$ & ${\cal Q}_{1LL,c}^{(pp)V}$   &  $({\bf 2}_L,{\bf 2}_R)|_a$  & $ \theta^{u_Lv_R}_{11} $
\\
$P_{2,a}^{(pp)V}$ & ${1\over 5}\left(5{\cal Q}_{2LL,a}^{(pp)V}-3{\cal Q}_{2LL,b}^{(pp)V}\right)$  &   $({\bf 6}_L,{\bf 2}_R)$  & $ \theta^{u_Lv_Lw_Lx_Ly_Lz_R}_{(11112)2} $ 
\\
$P_{2,b}^{(pp)V}$ & ${\cal Q}_{2LL,b}^{(pp)V}$ &   $({\bf 4}_L,{\bf 2}_R)|_a$ & $ \theta^{u_Lv_Lw_Lx_R}_{(111)2} $  
\\\hline
\cellcolor{gray!30}$P_{3,a}^{(pp)V}$ & ${\cal Q}_{3LR,a}^{(pp)V}-{\cal Q}_{3LR,b}^{(pp)V}$ &  $({\bf 4}_L,{\bf 4}_R)$  & $ \theta^{u_Lv_Lw_Lx_Ry_Rz_R}_{(111)(122)} $ 
\\
\cellcolor{gray!30}$\tilde P_{3,a}^{(pp)V}$ & $-\left({\cal Q}_{3RL,a}^{(pp)V}-{\cal Q}_{3RL,b}^{(pp)V}\right)$ &  $({\bf 4}_L,{\bf 4}_R)$  & $ \theta^{u_Lv_Lw_Lx_Ry_Rz_R}_{(122)(111)} $  
\\
\cellcolor{gray!30}$P_{3,b}^{(pp)V}$ & ${\cal Q}_{3LR,b}^{(pp)V}$ &   $({\bf 4}_L,{\bf 2}_R)|_b$  & $ \theta^{u_Lv_Lw_Lx_R}_{(111)2} $  
\\
\cellcolor{gray!30}$\tilde P_{3,b}^{(pp)V}$ & ${\cal Q}_{3RL,b}^{(pp)V}$ &   $({\bf 2}_L,{\bf 4}_R)|_b$  & $ \theta^{u_Lv_Rw_Rx_R}_{2(111)} $ 
\\
\cellcolor{gray!30}$P_{4,a}^{(pp)V}$ &~~~${1\over 3}\left( 3{\cal Q}_{4LR,a}^{(pp)V}+3{\cal Q}_{4LR,b}^{(pp)V}-3{\cal Q}_{4LR,c}^{(pp)V}-{\cal Q}_{4LR,d}^{(pp)V}+2{\cal Q}_{4LR,e}^{(pp)V}\right)$ ~~~&$({\bf 4}_L,{\bf 4}_R)$  &~~~$ \theta^{u_Lv_Lw_Lx_Ry_Rz_R}_{(112)(112)} $~~~  
\\
\cellcolor{gray!30}$P_{4,b}^{(pp)V}$ &${1\over3}\left(3{\cal Q}_{4LR,b}^{(pp)V}-{\cal Q}_{4LR,d}^{(pp)V}+2{\cal Q}_{4LR,e}^{(pp)V}\right)$  &$({\bf 4}_L,{\bf 2}_R)|_b$ & $ \theta^{u_Lv_Lw_Lx_R}_{(112)1} $ \\
\cellcolor{gray!30}$P_{4,c}^{(pp)V}$ & ${1\over3}\left(3{\cal Q}_{4LR,c}^{(pp)V}+{\cal Q}_{4LR,d}^{(pp)V}-2{\cal Q}_{4LR,e}^{(pp)V}\right)$
& $({\bf 2}_L,{\bf 4}_R)|_b$  & $ \theta^{u_Lv_Rw_Rx_R}_{1(112)} $  
\\
\cellcolor{gray!30}$P_{4,d}^{(pp)V}$  & ${\cal Q}_{4LR,d}^{(pp)V}$  &   $({\bf 2}_L,{\bf 2}_R)|_b$ & $ \theta^{u_Lv_R}_{11}$ 
\\
\cellcolor{gray!30}$P_{4,e}^{(pp)V}$    &${\cal Q}_{4LR,e}^{(pp)V}$  &$({\bf 2}_L,{\bf 2}_R)|_c$ & $ \theta^{u_Lv_R}_{11}$   
\\\hline\hline
$P_{1}^{(pp)T,-}$ & ${\cal Q}_{1LLL}^{(pp)T,-}$   &  $({\bf 5}_L,{\bf 1}_R)|_a$ & $ \theta^{u_Lv_Lw_Lx_L}_{(1112)} $ 
\\\hline
$P_{2,a}^{(pp)T,-}$ &${\cal Q}_{2LLR,a}^{(pp)T,-}+{\cal Q}_{2LLR,c}^{(pp)T,-}$  & $({\bf 5}_L,{\bf 1}_R)|_b$ &  $ \theta^{u_Lv_Lw_Lx_L}_{(1112)} $
\\
$P_{2,b}^{(pp)T,-}$ & ${\cal Q}_{2LLR,b}^{(pp)T,-}$  & $({\bf 3}_L,{\bf 3}_R)|_a$  &  $ \theta^{u_Lv_Lw_Rx_R}_{(11)(12)} $
\\
$P_{2,c}^{(pp)T,-}$ & ${\cal Q}_{2LLR,c}^{(pp)T,-}$ & $({\bf 3}_L,{\bf 1}_R)|_d$ & $ \theta^{u_Lv_L}_{(11)} $  \\
$P_{3}^{(pp)T,-}$ & ${\cal Q}_{3LLR}^{(pp)T,-}$  &$({\bf 3}_L,{\bf 3}_R)|_a$ & $ \theta^{u_Lv_Lw_Rx_R}_{(12)(11)} $  
\\\hline
$\hat P_{2,a}^{(pp)T,-}$ & ${\cal Q}_{2RRL,a}^{(pp)T,-}$  &$({\bf 1}_L,{\bf 5}_R)|_c$ & $\theta^{u_Rv_Rw_Rx_R}_{(1112)} $ 
\\
$\hat P_{2,b}^{(pp)T,-}$ &${\cal Q}_{2RRL,b}^{(pp)T,-}$  &$({\bf 3}_L,{\bf 3}_R)|_b$ & $ \theta^{u_Lv_Lw_Rx_R}_{(12)(11)} $  
\\
$\hat P_{3}^{(pp)T,-}$ & ${\cal Q}_{3RRL}^{(pp)T,-}$ & $({\bf 3}_L,{\bf 3}_R)|_b$ & $ \theta^{u_Lv_Lw_Rx_R}_{(11)(12)} $ 
\\\hline
\end{tabular}
}
\caption{The chiral basis and their chiral irreps under $G_\chi$ for the operators contributing to $pp\to\ell^+\ell^{\prime+}$.}
\label{tab:chiralirrep_pp}
\end{table}
\begin{table}
\centering
\resizebox{\linewidth}{!}{
\renewcommand{\arraystretch}{0.85}
\begin{tabular}{|c | c | c | c|}
\hline
~~Chiral basis~~ & LEFT basis & Chiral irrep. &  Chiral spurion  
\\\hline
$P_{1,a}^{(pn)S}$ & ${1\over 5}\left(5{\cal Q}_{1LLL,a}^{(pn)S}-9{\cal Q}_{1LLL,b}^{(pn)S}\right)$   &  $({\bf 7}_L,{\bf 1}_R)$ & $ \theta^{u_Lv_Lw_Lx_Ly_Lz_L}_{(111222)} $  
\\
$P_{1,b}^{(pn)S}$ &  ${\cal Q}_{1LLL,b}^{(pn)S}$  & $({\bf 3}_L,{\bf 1}_R)|_a$ & $ \theta^{u_Lv_L}_{(12)} $  
\\\hline
$P_{2}^{(pn)S}$ & ${\cal Q}_{2LLR}^{(pn)S}$ &  $({\bf 5}_L,{\bf 3}_R)$ & $\theta^{u_Lv_Lw_Lx_Ly_Rz_R}_{(1112)(22)}$  
\\
$P_{3,a}^{(pn)S}$ & ${\cal Q}_{3LLR,a}^{(pn)S}-{\cal Q}_{3LLR,c}^{(pn)S}$ &  $({\bf 5}_L,{\bf 3}_R)$ &  $\theta^{u_Lv_Lw_Lx_Ly_Rz_R}_{(1122)(12)}$   
\\
$P_{3,b}^{(pn)S}$ & ${\cal Q}_{3LLR,b}^{(pn)S}$ &  $({\bf 3}_L,{\bf 1}_R)|_b$  & $ \theta^{u_Lv_L}_{(12)} $ 
\\
$P_{3,c}^{(pn)S}$ & ${\cal Q}_{3LLR,c}^{(pn)S}$ &  $({\bf 1}_L,{\bf 3}_R)|_c$  & $ \theta^{u_Rv_R}_{(12)} $ 
\\
$P_4^{(pn)S}$& ${\cal Q}_{4LLR}^{(pn)S}$ &  $({\bf 5}_L,{\bf 3}_R)$  & $\theta^{u_Lv_Lw_Lx_Ly_Rz_R}_{(1222)(11)}$ 
\\\hline\hline
$P_{1,a}^{(pn)V}$ & ${1\over 5}\left(5{\cal Q}_{1LL,a}^{(pn)V}-6{\cal Q}_{1LL,b}^{(pn)V}-3{\cal Q}_{1LL,c}^{(pn)V}\right)$ &  $({\bf 6}_L,{\bf 2}_R)$ & $\theta^{u_Lv_Lw_Lx_Ly_Lz_R}_{(11122)2}$ 
\\
$P_{1,b}^{(pn)V}$ &  ${1\over 3}\left(3{\cal Q}_{1LL,b}^{(pn)V}-{\cal Q}_{1LL,c}^{(pn)V}\right)$ &  $({\bf 4}_L,{\bf 2}_R)|_a$  & $ \theta^{u_Lv_Lw_Lx_R}_{(112)2}$ 
\\
$P_{1,c}^{(pn)V}$ & ${\cal Q}_{1LL,c}^{(pn)V}$   &  $({\bf 2}_L,{\bf 2}_R)|_a$  & $ \theta^{u_Lv_R}_{12} $ 
\\
$P_{2,a}^{(pn)V}$ & ${1\over 5}\left(5{\cal Q}_{2LL,a}^{(pn)V}+6{\cal Q}_{2LL,b}^{(pn)V}-3{\cal Q}_{2LL,c}^{(pn)V}\right)$  &   $({\bf 6}_L,{\bf 2}_R)$  & $ \theta^{u_Lv_Lw_Lx_Ly_Lz_R}_{(11222)1}$ 
\\
$P_{2,b}^{(pn)V}$ & ${1\over 3}\left(3{\cal Q}_{2LL,b}^{(pn)V}+{\cal Q}_{2LL,c}^{(pn)V}\right)$ &   $({\bf 4}_L,{\bf 2}_R)|_a$ & $ \theta^{u_Lv_Lw_Lx_R}_{(122)1}$ 
\\
$P_{2,c}^{(pn)V}$ & ${\cal Q}_{2LL,c}^{(pn)V}$   &  $({\bf 2}_L,{\bf 2}_R)|_a$  & $ \theta^{u_Lv_R}_{21} $
\\\hline
\cellcolor{gray!30}$P_{3}^{(pn)V}$ & ${\cal Q}_{3LR}^{(pn)V}$ &   $({\bf 4}_L,{\bf 4}_R)$  & $ \theta^{u_Lv_Lw_Lx_Ry_Rz_R}_{(111)(222)}$ 
\\
\cellcolor{gray!30}$\tilde P_{3}^{(pn)V}$ & $-{\cal Q}_{3RL}^{(pn)V}$ &   $({\bf 4}_L,{\bf 4}_R)$  & $ \theta^{u_Lv_Lw_Lx_Ry_Rz_R}_{(222)(111)}$ 
\\
\cellcolor{gray!30}$P_{4,a}^{(pn)V}$ &~~~~~~
${1\over 3}\left( 3{\cal Q}_{4LR,a}^{(pn)V}-3{\cal Q}_{4LR,b}^{(pn)V}-3{\cal Q}_{4LR,c}^{(pn)V}+{\cal Q}_{4LR,d}^{(pn)V}-2{\cal Q}_{4LR,e}^{(pn)V}\right)$~~~~~~&$({\bf 4}_L,{\bf 4}_R)$&~~~~~~ 
$\theta^{u_Lv_Lw_Lx_Ry_Rz_R}_{(112)(122)}$ ~~~~~~
\\
\cellcolor{gray!30}$\tilde P_{4,a}^{(pn)V}$ &~~~
$-{1\over 3}\left( 3{\cal Q}_{4RL,a}^{(pn)V}-3{\cal Q}_{4RL,b}^{(pn)V}-3{\cal Q}_{4RL,c}^{(pn)V}+{\cal Q}_{4RL,d}^{(pn)V}-2{\cal Q}_{4RL,e}^{(pn)V}\right)$~~~&$({\bf 4}_L,{\bf 4}_R)$& ~~~
$\theta^{u_Lv_Lw_Lx_Ry_Rz_R}_{(122)(112)}$~~~ 
\\
\cellcolor{gray!30}$P_{4,b}^{(pn)V}$ & ${1\over3}\left(3 {\cal Q}_{4LR,b}^{(pn)V}-{\cal Q}_{4LR,d}^{(pn)V}+2{\cal Q}_{4LR,e}^{(pn)V} \right)$ &~~~~~$({\bf 4}_L,{\bf 2}_R)|_b$~~~~~& $ \theta^{u_Lv_Lw_Lx_R}_{(112)2}$ 
\\
\cellcolor{gray!30}$\tilde P_{4,b}^{(pn)V}$ & ${1\over3}\left(3 {\cal Q}_{4RL,b}^{(pn)V}-{\cal Q}_{4RL,d}^{(pn)V}+2{\cal Q}_{4RL,e}^{(pn)V} \right)$ &   $({\bf 2}_L,{\bf 4}_R)|_b$  & $ \theta^{u_Lv_Rw_Rx_R}_{2(112)}$ 
\\
\cellcolor{gray!30}$P_{4,c}^{(pn)V}$ & ${1\over3}\left(3 {\cal Q}_{4LR,c}^{(pn)V}-{\cal Q}_{4LR,d}^{(pn)V}+2{\cal Q}_{4LR,e}^{(pn)V} \right)$ &   $({\bf 2}_L,{\bf 4}_R)|_b$  & $ \theta^{u_Lv_Rw_Rx_R}_{1(122)}$
\\
\cellcolor{gray!30}$\tilde P_{4,c}^{(pn)V}$ & ${1\over3}\left(3 {\cal Q}_{4RL,c}^{(pn)V}-{\cal Q}_{4RL,d}^{(pn)V}+2{\cal Q}_{4RL,e}^{(pn)V} \right)$ &   $({\bf 4}_L,{\bf 2}_R)|_b$  & $ \theta^{u_Lv_Lw_Lx_R}_{(122)1}$ 
\\
\cellcolor{gray!30}$P_{4,d}^{(pn)V}$ & ${\cal Q}_{4LR,d}^{(pn)V}$ &   $({\bf 2}_L,{\bf 2}_R)|_b$  & $ \theta^{u_Lv_R}_{12}$ 
\\
\cellcolor{gray!30}$\tilde P_{4,d}^{(pn)V}$ & $-{\cal Q}_{4RL,d}^{(pn)V}$ &   $({\bf 2}_L,{\bf 2}_R)|_b$  & $ \theta^{u_Lv_R}_{21}$ 
\\
\cellcolor{gray!30}$P_{4,e}^{(pn)V}$ & ${\cal Q}_{4LR,e}^{(pn)V}$ &   $({\bf 2}_L,{\bf 2}_R)|_c$  & $ \theta^{u_Lv_R}_{12}$ 
\\
\cellcolor{gray!30}$\tilde P_{4,e}^{(pn)V}$ & $-{\cal Q}_{4RL,e}^{(pn)V}$ &   $({\bf 2}_L,{\bf 2}_R)|_c$  & $ \theta^{u_Lv_R}_{21}$
\\\hline\hline
\cellcolor{gray!30}$P_{1,a}^{(pn)T}$ & ${1\over 3}\left( 3{\cal Q}_{1LLL,a}^{(pn)T}-{\cal Q}_{1LLL,b}^{(pn)T}\right)$   &  $({\bf 5}_L,{\bf 1}_R)|_a$ & $ \theta^{u_Lv_Lw_Lx_L}_{(1122)} $
\\
\cellcolor{gray!30}$P_{1,b}^{(pn)T}$ & ${\cal Q}_{1LLL,b}^{(pn)T}$   &  $({\bf 1}_L,{\bf 1}_R)|_a$ & $ 1 $
\\\hline
\cellcolor{gray!30}$P_{2}^{(pn)T}$ &${\cal Q}_{2LLR}^{(pn)T}$  & $({\bf 3}_L,{\bf 3}_R)|_a$ & $\theta^{u_Lv_Lw_Rx_R}_{(11)(22)} $
\\
\cellcolor{gray!30} $P_{3,a}^{(pn)T}$ & ${1\over 3}\left( 3{\cal Q}_{3LLR,a}^{(pn)T}-{\cal Q}_{3LLR,b}^{(pn)T}\right)$ & $({\bf 5}_L,{\bf 1}_R)|_b$ &  $ \theta^{u_Lv_Lw_Lx_L}_{(1122)} $
\\
\cellcolor{gray!30} $P_{3,b}^{(pn)T}$ & ${\cal Q}_{3LLR,b}^{(pn)T}$  &$({\bf 1}_L,{\bf 1}_R)|_b$ & 1
\\
\cellcolor{gray!30} $P_{3,c}^{(pn)T}$ &${\cal Q}_{3LLR,c}^{(pn)T}$  &$({\bf 3}_L,{\bf 3}_R)|_a$ & 
$\theta^{u_Lv_Lw_Rx_R}_{(12)(12)} $
\\
\cellcolor{gray!30} $P_{3,d}^{(pn)T}$ & $-{\cal Q}_{3LLR,d}^{(pn)T}$  &$({\bf 3}_L,{\bf 1}_R)|_d$ & $\theta^{u_Lv_L}_{(12)}$
\\
\cellcolor{gray!30} $P_{4}^{(pn)T}$ & ${\cal Q}_{4LLR}^{(pn)T}$ & $({\bf 3}_L,{\bf 3}_R)|_a$ & $\theta^{u_Lv_Lw_Rx_R}_{(22)(11)} $
\\\hline
\cellcolor{gray!30} $\hat P_{2}^{(pn)T}$ & ${\cal Q}_{2RRL}^{(pn)T}$  & $({\bf 3}_L,{\bf 3}_R)|_b$  & $\theta^{u_Lv_Lw_Rx_R}_{(22)(11)} $
\\
\cellcolor{gray!30} $\hat P_{3,a}^{(pn)T}$ & ${\cal Q}_{3RRL,a}^{(pn)T}-{\cal Q}_{3RRL,c}^{(pn)T}$ & $({\bf 1}_L,{\bf 5}_R)|_c$ & $ \theta^{u_Rv_Rw_Rx_R}_{(1122)} $
\\
\cellcolor{gray!30} $\hat P_{3,b}^{(pn)T}$ & ${\cal Q}_{3RRL,b}^{(pn)T}$  &$({\bf 3}_L,{\bf 3}_R)|_b$ & $\theta^{u_Lv_Lw_Rx_R}_{(12)(12)} $
\\
\cellcolor{gray!30} $\hat P_{3,c}^{(pn)T}$ & ${\cal Q}_{3RRL,c}^{(pn)T}$  &$({\bf 1}_L,{\bf 1}_R)|_c$ & 1
\\
\cellcolor{gray!30} $\hat P_{4}^{(pn)T}$ & ${\cal Q}_{4RRL}^{(pn)T}$ & $({\bf 3}_L,{\bf 3}_R)|_b$ & $\theta^{u_Lv_Lw_Rx_R}_{(11)(22)} $ 
\\\hline
\end{tabular}
}
\caption{The chiral basis and their chiral irreps under $G_\chi$ for the operators contributing to $pn\to\ell^+\bar\nu^\prime$.}
\label{tab:chiralirrep_pn}
\end{table}
\begin{table}
\centering
\resizebox{\linewidth}{!}{
\begin{tabular}{|c | c | c | c| }
\hline
Chiral basis & LEFT basis & Chiral irrep. &  Chiral spurion 
\\\hline
$P_{1,a}^{(nn)S}$ & $~~~~~~~~~~~~{1\over 5}\left(5{\cal Q}_{1LLL,a}^{(nn)S}-3{\cal Q}_{1LLL,b}^{(nn)S}\right)~~~~~~~~~~~~$   &  $({\bf 7}_L,{\bf 1}_R)$ & $~~~~~\theta^{u_Lv_Lw_Lx_Ly_Lz_L}_{(112222)}~~~~~$ 
\\
$P_{1,b}^{(nn)S}$ &  ${\cal Q}_{1LLL,b}^{(nn)S}$  & $({\bf 3}_L,{\bf 1}_R)|_a$ & $ \theta^{u_Lv_L}_{(22)} $
\\\hline
$P_{2,a}^{(nn)S}$ & ${\cal Q}_{2LLR,a}^{(nn)S}$ &  $({\bf 5}_L,{\bf 3}_R)$ & $\theta^{u_Lv_Lw_Lx_Ly_Rz_R}_{(1222)(12)}$ 
\\
$P_{2,b}^{(nn)S}$ & ${\cal Q}_{2LLR,b}^{(nn)S}$ & $({\bf 3}_L,{\bf 1}_R)|_b$  & $ \theta^{u_Lv_L}_{(22)}  $  
\\
$P_{3,a}^{(nn)S}$ & ${\cal Q}_{3LLR,a}^{(nn)S}-{\cal Q}_{3LLR,b}^{(nn)S}$ &  $({\bf 5}_L,{\bf 3}_R)$ &  $ \theta^{u_Lv_Lw_Lx_Ly_Rz_R}_{(1122)(22)}$   
\\
$P_{3,b}^{(nn)S}$ & ${\cal Q}_{3LLR,b}^{(nn)S}$ &  $({\bf 1}_L,{\bf 3}_R)|_c$  & $ \theta^{u_Rv_R}_{(22)}  $
\\
$P_4^{(nn)S}$& ${\cal Q}_{4LLR}^{(nn)S}$ &  $({\bf 5}_L,{\bf 3}_R)$  & $\theta^{u_Lv_Lw_Lx_Ly_Rz_R}_{(2222)(11)}$ 
\\\hline\hline
\cellcolor{gray!30} $P_{1}^{(nn)T}$ & $-{\cal Q}_{1LLL}^{(nn)T}$   &  $({\bf 5}_L,{\bf 1}_R)|_a$ & $ \theta^{u_Lv_Lw_Lx_L}_{(1222)}  $ 
\\\hline
\cellcolor{gray!30} $P_{2,a}^{(nn)T}$ &  $-\left({\cal Q}_{2LLR,a}^{(nn)T}+{\cal Q}_{2LLR,c}^{(nn)T}\right)$  & $({\bf 5}_L,{\bf 1}_R)|_b$ &  $ \theta^{u_Lv_Lw_Lx_L}_{(1222)} $ 
\\
\cellcolor{gray!30} $P_{2,b}^{(nn)T}$ & $-{\cal Q}_{2LLR,b}^{(nn)T}$  & $({\bf 3}_L,{\bf 3}_R)|_a$  & $ \theta^{u_Lv_Lw_Rx_R}_{(22)(12)} $
\\
\cellcolor{gray!30} $P_{2,c}^{(nn)T}$ & ${\cal Q}_{2LLR,c}^{(nn)T}$ & $({\bf 3}_L,{\bf 1}_R)|_d$ & $ \theta^{u_Lv_L}_{(22)} $ 
\\
\cellcolor{gray!30} $P_{3}^{(nn)T}$ & $-{\cal Q}_{3LLR}^{(nn)T}$  &$({\bf 3}_L,{\bf 3}_R)|_a$ & $ \theta^{u_Lv_Lw_Rx_R}_{(12)(22)} $ 
\\\hline
\cellcolor{gray!30} $\hat P_{2,a}^{(nn)T}$ & $-{\cal Q}_{2RRL,a}^{(nn)T}$  & $({\bf 1}_L,{\bf 5}_R)|_c$ & $ \theta^{u_Rv_Rw_Rx_R}_{(1222)} $
\\
\cellcolor{gray!30} $\hat P_{2,b}^{(nn)T}$ & $-{\cal Q}_{2RRL,b}^{(nn)T}$  & $({\bf 3}_L,{\bf 3}_R)|_b$  & $ \theta^{u_Lv_Lw_Rx_R}_{(12)(22)} $ 
\\
\cellcolor{gray!30} $\hat P_{3}^{(nn)T}$ & $-{\cal Q}_{3RRL}^{(nn)T}$  &$({\bf 3}_L,{\bf 3}_R)|_b$ & $ \theta^{u_Lv_Lw_Rx_R}_{(22)(12)} $ 
\\\hline
\end{tabular}
}
\caption{The chiral basis and their chiral irreps under $G_\chi$ for the operators contributing to $nn\to\bar\nu\bar\nu^\prime$.}
\label{tab:chiralirrep_nn}
\end{table}
\clearpage
\newpage

\section{Chiral irreducible representations in terms of hadrons} 
\label{app:chiralmatching}

\begin{table}[!h]
\centering
\resizebox{\linewidth}{!}{
\renewcommand{\arraystretch}{1.35}
\begin{tabular}{|c | c | c| l  | }
\hline
Ope. type& Chi. irrep & Chi. order  & \multicolumn{1}{|c |}{Matching operator}   
\\\hline
\multirow{6}{*}{\rotatebox[origin=c]{90}{  Scalar  current: ${\cal O}^{S}_{\rm quark}\times j_S$}}
& $({\bf 3}_L,{\bf 1}_R)|_i$ & $p^0$
& $O^{S}_{3\times1,i}=\theta^{u_Lv_L}_{(\alpha\beta)}(u^\dagger)_{u_L a}(u^\dagger)_{v_L b}[\Psi^\T_{a}C(g_{3\times 1,i}+\hat g_{3\times 1,i}\gamma_5)\Psi_b]$
\\
& $({\bf 5}_L,{\bf 3}_R)$  & $p^0$
& $O^{S}_{5\times 3}=\theta^{u_Lv_Lw_Lx_Ly_Rz_R}_{(\alpha\beta\gamma\rho)(\sigma\tau)}(Ui\tau^2)_{y_Rw_L}(Ui\tau^2)_{z_Rx_L} (u^\dagger)_{u_L a}(u^\dagger)_{v_L b}[\Psi^\T_{a}C(g_{5\times 3}+\hat g_{5\times 3}\gamma_5)\Psi_b]$  
\\
& $({\bf 7}_L,{\bf 1}_R)$ & $p^2(\times)$
& $O^{S}_{7\times 1}=\theta^{u_Lv_Lw_Lx_Ly_Lz_L}_{(\alpha\beta\gamma\rho\sigma\tau)}(u^\dagger u_\mu u i\tau^2)_{w_Lx_L} (u^\dagger u^\mu u i\tau^2)_{y_Lz_L}(u^\dagger)_{u_L a}(u^\dagger)_{v_L b}[\Psi^\T_{a}C(g_{7\times 1}+\hat g_{7\times 1}\gamma_5)\Psi_b]$   
\\\cline{2-4}
& $({\bf 1}_L,{\bf 3}_R)|_i$ & $p^0$
& $\tilde O^{S}_{1\times 3,i}=\theta^{u_Rv_R}_{(\alpha\beta)}u_{u_R a}u_{v_R b}[\Psi^\T_{a}C(g_{1\times 3,i}+\hat g_{1\times 3,i}\gamma_5)\Psi_b]$ 
\\
& $({\bf 3}_L,{\bf 5}_R)$ & $p^0$ 
& $\tilde O^{S}_{3\times 5}=\theta^{u_Rv_Rw_Rx_Ry_Lz_L}_{(\alpha\beta\gamma\rho)(\sigma\tau)}(Ui\tau^2)_{w_Ry_L}(Ui\tau^2)_{x_Rz_L} u_{u_R a} u_{v_R b}[\Psi^\T_{a}C(g_{3\times 5}+\hat g_{3\times 5}\gamma_5)\Psi_b]$ 
\\
& $({\bf 1}_L,{\bf 7}_R)$ & $p^2(\times)$ 
& $\tilde O^{S}_{1\times 7}=\theta^{u_Rv_Rw_Rx_Ry_Rz_R}_{(\alpha\beta\gamma\rho\sigma\tau)}(u u_\mu u^\dagger i\tau^2)_{w_Rx_R} (u u^\mu u^\dagger i\tau^2)_{y_Rz_R}u_{u_R a}u_{v_R b}[\Psi^\T_{a}C(g_{1\times 7}+\hat g_{1\times 7}\gamma_5)\Psi_b]$    
\\\hline\hline
\multirow{6}{*}{\rotatebox[origin=c]{90}{ Vector current: ${\cal O}^{V,\mu}_{\rm quark}\times j_{V,\mu}$}}
& $({\bf 2}_L,{\bf 2}_R)|_i$ & $p^0$ 
& $O^{V,\mu}_{2\times2,i}=\theta^{u_Lv_R}_{\alpha\beta}(u^\dagger)_{u_L a}u_{v_R b}[\Psi^\T_{a}C\gamma^\mu(g_{2\times 2,i}+\hat g_{2\times 2,i}\gamma_5)\Psi_b]$   
\\
& $({\bf 4}_L,{\bf 2}_R)|_i$ & $p^0$ 
& $O^{V,\mu}_{4\times2,i}=g_{4\times 2,i}\theta^{u_Lv_Lw_Lx_R}_{(\alpha\beta\gamma)\rho}(Ui\tau^2)_{x_Rw_L} (u^\dagger)_{u_L a}(u^\dagger)_{v_L b}[\Psi^\T_{a}C\gamma^\mu\gamma_5 \Psi_b]$  
 \\
& $({\bf 4}_L,{\bf 4}_R)$ & $p^0$ 
& $O^{V,\mu}_{4\times4}=\theta^{u_Lv_Lw_Lx_Ry_Rz_R}_{(\alpha\beta\gamma)(\rho\sigma\tau)}(Ui\tau^2)_{y_Rv_L}(Ui\tau^2)_{z_Rw_L} (u^\dagger)_{u_L a}u_{x_R b}[\Psi^\T_{a}C\gamma^\mu(g_{4\times 4}+\hat g_{4\times 4}\gamma_5) \Psi_b]$   
\\
& $({\bf 6}_L,{\bf 2}_R)$ & $p^1(\times)$
& $O^{V,\mu}_{6\times2}=\theta^{u_Lv_Lw_Lx_Ly_Lz_R}_{(\alpha\beta\gamma\rho\sigma)\tau}(Ui\tau^2)_{z_Rw_L}(u^\dagger u^\mu u i\tau^2)_{x_Ly_L}  (u^\dagger)_{u_L a}(u^\dagger)_{v_L b}[\Psi^\T_{a}C(g_{6\times 2}+\hat g_{6\times 2}\gamma_5) \Psi_b]$   
\\\cline{2-4}
& $({\bf 2}_L,{\bf 4}_R)|_i$ & $p^0$ 
& $\tilde O^{V,\mu}_{2\times4,i}=-g_{2\times 4,i}\theta^{u_Rv_Rw_Rx_L}_{(\alpha\beta\gamma)\rho}(Ui\tau^2)_{w_Rx_L} u_{u_R a}u_{v_R b}(\Psi^\T_{a}C\gamma^\mu\gamma_5 \Psi_b)$    
\\
& $({\bf 2}_L,{\bf 6}_R)$ & $p^1(\times)$ 
& $\tilde O^{V,\mu}_{2\times6}=-\theta^{u_Rv_Rw_Rx_Ry_Rz_L}_{(\alpha\beta\gamma\rho\sigma)\tau}(Ui\tau^2)_{w_Rz_L}(u u^\mu u^\dagger i\tau^2)_{x_Ry_R}  u_{u_R a}u_{v_R b}[\Psi^\T_{a}C(g_{2\times 6}+\hat g_{2\times 6}\gamma_5) \Psi_b]$
\\\hline\hline
\multirow{6}{*}{\rotatebox[origin=c]{90}{ Tensor current: ${\cal O}^{T,\mu\nu}_{\rm quark}\times j_{T}^{\mu\nu}$}}
& $({\bf 1}_L,{\bf 1}_R)|_i$ & $p^0$ 
& $O^{T,\mu\nu}_{1\times1,i}={1\over2}\epsilon^{ab}[\Psi^\T_{a}C\sigma^{\mu\nu}(g_{1\times 1,i}+\hat g_{1\times 1,i} \gamma_5)\Psi_b]$   
\\
& $({\bf 3}_L,{\bf  1}_R)$ & $p^1(\times)$ 
& $O^{T,\mu\nu}_{3\times1}=\theta^{u_Lv_L}_{(\alpha\beta)}(u^\dagger u^\mu)_{u_La}(u^\dagger)_{v_L b}[\Psi^\T_{a}C\gamma^\nu(g_{3\times 1,T}+\hat g_{3\times 1,T}\gamma_5)\Psi_b]-\mu\leftrightarrow \nu$
 \\
& $({\bf 3}_L,{\bf 3}_R)|_i$ & $p^0$ 
& $O^{T,\mu\nu}_{3\times3,i}=\theta^{u_Lv_Lw_Rx_R}_{(\alpha\beta)(\gamma\rho)}(Ui\tau^2)_{x_Rv_L}(u^\dagger)_{u_L a}u_{w_R b}[\Psi^\T_{a}C\sigma^{\mu\nu}(g_{3\times 3,i}+\hat g_{3\times 3,i}\gamma_5)) \Psi_b]$
\\
& $({\bf 5}_L,{\bf 1}_R)|_i$ & $p^1(\times)$ 
& $O^{T,\mu\nu}_{5\times1,i}=g_{5\times 1,i}\theta^{u_Lv_Lw_Lx_L}_{(\alpha\beta\gamma\rho)}(u^\dagger u^\mu u i\tau^2)_{w_Lx_L}(u^\dagger)_{u_L a}(u^\dagger)_{v_L b}(\Psi^\T_{a}C\gamma^\nu\gamma_5\Psi_b)-\mu\leftrightarrow \nu$
\\\cline{2-4}
& $({\bf 1}_L,{\bf 3}_R)$ & $p^1(\times)$ 
& $\tilde O^{T,\mu\nu}_{1\times3}=\theta^{u_Rv_R}_{(\alpha\beta)}(u u^\mu)_{u_Ra}u_{v_R b}[\Psi^\T_{a}C\gamma^\nu(g_{1\times 3,T}+\hat g_{1\times 3,T}\gamma_5)\Psi_b]-\mu\leftrightarrow \nu$
\\
& $({\bf 1}_L,{\bf 5}_R)|_i$ & $p^1 (\times)$
& $\tilde O^{T,\mu\nu}_{1\times5,i}=g_{1\times 5,i}\theta^{u_Rv_Rw_Rx_R}_{(\alpha\beta\gamma\rho)}(u u^\mu u^\dagger i\tau^2)_{w_Rx_R}u_{u_R a}u_{v_R b}(\Psi^\T_{a}C\gamma^\nu\gamma_5\Psi_b)-\mu\leftrightarrow \nu$
\\\hline
\end{tabular}
}
\caption{The leading order chiral matching of the chiral irrep six-quark part in our assumed framework without including terms containing nucleon derivatives.  Where in the third column we show the leading chiral order of each matched operator, and the crosses $(\times)$ behind $p^{1,2}$ indicate the matched operators cannot contribute to the leading order dim-6 interactions in Eq.~\eqref{chpt:LO}.}
\label{tab:matchingchpt}
\end{table}

\newpage
\section{LEFT and SMEFT contributions to $C^{pp}$, $C^{pn}$ and $C^{nn}$}
\label{app:commatching}

Taking the specific expressions of the spurion fields in Tabs.~(\ref{tab:chiralirrep_pp}-\ref{tab:chiralirrep_nn}) into the leading order matching results in Eq.~\ref{chiralresult}, and combining with the relevant LEFT Wilson coefficients and lepton currents, we obtain the final matching results for the operators in Eqs.~(\ref{ppll}-\ref{nnnunu}) as the function of the LECs and the LEFT Wilson coefficients as follows
{\footnotesize
\begin{align}\nonumber
C_{R(L)}^{(pp)S}&=\left[g_{3\times1,a}\left({3\over5}C_{1LLL,a}^{(pp)S,\pm}+C_{1LLL,b}^{(pp)S,\pm}\right)+g_{3\times1,b}C_{2LLR,b}^{(pp)S,\pm}+g_{3\times1,c}\left(C_{3RRL,a}^{(pp)S,\pm}+C_{3RRL,b}^{(pp)S,\pm}\right)\right.
\\
&\left.+g_{5\times3}\left(-{ 1\over 4}C_{2LLR,a}^{(pp)S,\pm} +{1\over 6}C_{3LLR,a}^{(pp)S,\pm}+C_{4LLR}^{(pp)S,\pm}  \right)\right]+
\binom{L\leftrightarrow R}{g_{i\times j}\leftrightarrow g_{j\times i}}\;,
\\
C_{5R(L)}^{(pp)S}&=C_{R(L)}^{(pp)S}|_{g\to \hat g}\;,
\\\nonumber
C^{(pp)V}&=\left[\hat g_{2\times 2,a}\left(C_{1LL,a}^{(pp)V}+{1\over3}C_{1LL,b}^{(pp)V}+C_{1LL,c}^{(pp)V}\right)
+\binom{L\leftrightarrow R}{\hat g_{2\times 2,a}\leftrightarrow \hat g_{2\times 2,d}}
\right]
\\\nonumber
&+\hat g_{2\times 2,b}\left(C_{4LR,d}^{(pp)V}-{1\over3}\left(C_{4LR,a}^{(pp)V}-C_{4LR,b}^{(pp)V}+C_{4LR,c}^{(pp)V}\right)\right)
\\\nonumber
&+ \hat g_{2\times 2,c}\left(C_{4LR,e}^{(pp)V}+{2\over3}\left(C_{4LR,a}^{(pp)V}-C_{4LR,b}^{(pp)V}+C_{4LR,c}^{(pp)V}\right)\right)
\\\nonumber
&+\hat g_{4\times 4}\left({1\over 3}\left(C_{3LR,a}^{(pp)V}+C_{3RL,a}^{(pp)V}\right)-{2\over 9}C_{4LR,a}^{(pp)V}\right)
\\\nonumber
&+\left[g_{4\times 2,a}\left({1\over 3}\left({6\over5}C_{1LL,a}^{(pp)V}+C_{1LL,b}^{(pp)V}\right)-\left({3\over5}C_{2LL,a}^{(pp)V}+C_{2LL,b}^{(pp)V}\right)\right)
+\binom{L\leftrightarrow R}{g_{i\times j}\leftrightarrow g_{j\times i}}
\right]
\\\nonumber
&-g_{4\times 2,b}\left(\left(C_{3LR,a}^{(pp)V}+C_{3LR,b}^{(pp)V}\right)+ {1\over3}\left(C_{4LR,a}^{(pp)V}-C_{4LR,b}^{(pp)V}\right)\right)
\\
&-g_{2\times 4,b}\left(\left(C_{3RL,a}^{(pp)V}+C_{3RL,b}^{(pp)V}\right)- { 1\over 3}\left(C_{4LR,a}^{(pp)V}+C_{4LR,c}^{(pp)V}\right)\right)\;,
\\\nonumber
 C_L^{(pn)S}&=\left[ g_{3\times1,a}\left({9\over5}C_{1LLL,a}^{(pn)S}+C_{1LLL,b}^{(pn)S}\right)+g_{3\times1,b}C_{3LLR,b}^{(pn)S}+g_{3\times1,c}\left(C_{3RRL,a}^{(pn)S}+C_{3RRL,c}^{(pn)S}\right)\right.
\\
&\left.+g_{5\times3}\left({1\over 2}C_{2LLR}^{(pn)S} -{1\over 3}C_{3LLR,a}^{(pn)S}+{1\over 2}C_{4LLR}^{(pn)S}\right)\right]+
\binom{L\leftrightarrow R}{g_{i\times j}\leftrightarrow g_{j\times i}}\;,
\\
C_{5L}^{(pn)S}&=C_L^{(pn)S}|_{g\to \hat g}\;,
\\\nonumber
C_L^{(pn)V}&=
\left[g_{2\times 2,a}\left(C_{1LL,a}^{(pn)V}+{1\over3}C_{1LL,b}^{(pn)V}+C_{1LL,c}^{(pn)V}-C_{2LL,a}^{(pn)V}+{1\over3}C_{2LL,b}^{(pn)V}-C_{2LL,c}^{(pn)V}\right)
+\binom{L\leftrightarrow R}{ g_{2\times 2,a}\leftrightarrow g_{2\times 2,d}}
\right]
\\\nonumber
&+g_{2\times 2,b}\left[\left(C_{4LR,d}^{(pn)V}+{1\over3}\left(C_{4LR,a}^{(pn)V}+C_{4LR,b}^{(pn)V}+C_{4LR,c}^{(pn)V}\right)\right)+(L\leftrightarrow R)\right]
\\\nonumber
&+ g_{2\times 2,c}\left[\left(C_{4LR,e}^{(pn)V}-{2\over3}\left(C_{4LR,a}^{(pn)V}+C_{4LR,b}^{(pn)V}+C_{4LR,c}^{(pn)V}\right)\right)+(L\leftrightarrow R)\right]
\\
&+g_{4\times 4}\left(C_{3LR}^{(pn)V}+C_{3RL}^{(pn)V}-{1\over 3}\left(C_{4LR,a}^{(pn)V}+C_{4RL,a}^{(pn)V}\right)  \right)
\;,
\\\nonumber
C_{5L}^{(pn)V}&=
\left[\hat g_{2\times 2,a}\left(C_{1LL,a}^{(pn)V}+{1\over3}C_{1LL,b}^{(pn)V}+C_{1LL,c}^{(pn)V}+C_{2LL,a}^{(pn)V}-{1\over3}C_{2LL,b}^{(pn)V}+C_{2LL,c}^{(pn)V}\right)
+\binom{L\leftrightarrow R}{\hat g_{2\times 2,a}\leftrightarrow \hat g_{2\times 2,d}}
\right]
\\\nonumber
&+\hat g_{2\times 2,b}\left[\left(C_{4LR,d}^{(pn)V}+{1\over3}\left(C_{4LR,a}^{(pn)V}+C_{4LR,b}^{(pn)V}+C_{4LR,c}^{(pn)V}\right)\right)-(L\leftrightarrow R)\right]
\\\nonumber
&+\hat g_{2\times 2,c}\left[\left(C_{4LR,e}^{(pn)V}-{2\over3}\left(C_{4LR,a}^{(pn)V}+C_{4LR,b}^{(pn)V}+C_{4LR,c}^{(pn)V}\right)\right)-(L\leftrightarrow R)\right]
\\\nonumber
&+\hat  g_{4\times 4}\left(C_{3LR}^{(pn)V}-C_{3RL}^{(pn)V}-{1\over 9}\left(C_{4LR,a}^{(pn)V}-C_{4RL,a}^{(pn)V}\right)  \right)
\\\nonumber
&-{2\over 3}\left[g_{4\times 2,a}\left({6\over5}C_{1LL,a}^{(pn)V}+C_{1LL,b}^{(pn)V}+{6\over5}C_{2LL,a}^{(pn)V}-C_{2LL,b}^{(pn)V}\right)
+\binom{L\leftrightarrow R}{g_{i\times j}\leftrightarrow g_{j\times i}}
\right]
\\
&-{2\over 3}\left[g_{4\times 2,b}\left(C_{4LR,a}^{(pn)V}+C_{4LR,b}^{(pn)V}-C_{4RL,a}^{(pn)V}-C_{4RL,c}^{(pn)V}\right)
+\binom{L\leftrightarrow R}{g_{i\times j}\leftrightarrow g_{j\times i}}
\right]\;,
\\\nonumber
C^{(pn)T}&=g_{1\times 1,a}^r\left( {1\over3}C_{1LLL,a}^{(pn)T}+C_{1LLL,b}^{(pn)T}\right)
+g_{1\times 1,b}^r\left( {1\over3}C_{3LLR,a}^{(pn)T}+C_{3LLR,b}^{(pn)T}\right)
+g_{1\times 1,c}^r\left(C_{3RRL,a}^{(pn)T}+C_{3RRL,c}^{(pn)T}\right)
\\
&-g_{3\times 3,a}^r\left(C_{2LLR}^{(pn)T}-{1\over 2}C_{3LLR,c}^{(pn)T}+C_{4LLR}^{(pn)T} \right)
-g_{3\times 3,b}^r\left(C_{2RRL}^{(pn)T}-{1\over 2}C_{3RRL,b}^{(pn)T}+C_{4RRL}^{(pn)T} \right)\;,
\\\nonumber
C_L^{(nn)S}&=\left[g_{3\times1,a}\left({3\over5}C_{1LLL,a}^{(nn)S}+C_{1LLL,b}^{(nn)S}\right)+g_{3\times1,b}C_{2LLR,b}^{(nn)S}+g_{3\times1,c}\left(C_{3RRL,a}^{(nn)S}+C_{3RRL,b}^{(nn)S}\right)\right.
\\
&\left.+g_{5\times3}\left(-{1\over 4}C_{2LLR,a}^{(nn)S} +{1\over 6}C_{3LLR,a}^{(nn)S}+C_{4LLR}^{(nn)S}\right)\right]+
\binom{L\leftrightarrow R}{g_{i\times j}\leftrightarrow g_{j\times i}}\;,
\\
C_{5L}^{(nn)S}&=C_L^{(nn)S}|_{g\to \hat g}\;.
\end{align}}
~~After neglecting the QCD running effect and taking the matching result of the LEFT and SMEFT interactions in Tab.~\ref{tab:matchingLEFT} into consideration, we find the above results simplify considerably and take 
{\footnotesize
\begin{align}
C_L^{(pp)S}&=4g_{3\times1,a}C_{Q^6L^2}^{S,(S)}+2g_{3\times1,b}C_{udQ^4L^22}^{S,(S)}+g_{3\times1,c}\left(C_{u^2d^2Q^2L^21}^{S,(S)}+C_{u^2d^2Q^2L^23}^{S,(S)}\right)
+{g_{3\times5}\over 6}C_{u^2d^2Q^2L^21}^{S,(S)}\;,
\label{h11}
\\\nonumber
C_R^{(pp)S}
&=g_{1\times3,a}\left({3\over5}C_{u^4d^2e^21}^{S,(S)}+C_{u^4d^2e^22}^{S,(S)}\right)+2g_{1\times3,b}C_{u^3dQ^2e^2}^{S,(S)}+4g_{1\times3,c}C_{u^2Q^4e^2}^{S,(S)}\;,
\\
C_{5L(R)}^{(pp)S}&=C_{L(R)}^{(pp)S}|_{g\to \hat g}\;,
\\\nonumber
C^{(pp)V}
&=4\hat g_{2\times 2,a}C_{uQ^5eL}^{V}
-\hat g_{2\times 2,d}\left(C_{u^3d^2QeL1}^{V}-{1\over3}C_{u^3d^2QeL2}^{V}+C_{u^3d^2QeL3}^{V}\right)
\\\nonumber
&-2\hat g_{2\times 2,b}\left({1\over3}C_{u^2dQ^3eL1}^{V}-C_{u^2dQ^3eL2}^{V}\right)
+ 2\hat g_{2\times 2,c}\left({2\over3}C_{u^2dQ^3eL1}^{V}-C_{u^2dQ^3eL3}^{V}\right)
\\
&-{ g_{2\times 4,a}\over 3}\left({6\over5}C_{u^3d^2QeL1}^{V}-C_{u^3d^2QeL2}^{V}\right)
+ {2\over 3}g_{2\times 4,b}C_{u^2dQ^3eL1}^{V}\;,
\\\nonumber
 C_L^{(pn)S}&=-2\left(4g_{3\times1,a}C_{Q^6L^2}^{S,(S)}+2g_{3\times1,b}C_{udQ^4L^22}^{S,(S)}+g_{3\times1,c}\left(C_{u^2d^2Q^2L^21}^{S,(S)}+C_{u^2d^2Q^2L^23}^{S,(S)}\right)\right)
\\\nonumber
&
-2\left( g_{1\times3,a}\left({9\over5}C_{u^3d^3L^21}^{S,(A)}+C_{u^3d^3L^22}^{S,(A)}\right)
+2g_{1\times3,b}C_{u^2d^2Q^2L^22}^{S,(A)}+4g_{1\times3,c}C_{udQ^4L^21}^{S,(A)}\right)
\\
&
+{2\over 3}g_{3\times5}C_{u^2d^2Q^2L^21}^{S,(S)}\;,
\\
C_{5L}^{(pn)S}&=C_L^{(pn)S}|_{g\to \hat g}\;,
\\\nonumber
C_L^{(pn)V}&=4g_{2\times 2,a}C_{uQ^5eL}^{V}+g_{2\times 2,d}\left(C_{u^3d^2QeL1}^{V}-{1\over3}C_{u^3d^2QeL2}^{V}+C_{u^3d^2QeL3}^{V}\right)
\\
&
-2g_{2\times 2,b}\left({1\over3}C_{u^2dQ^3eL1}^{V}-C_{u^2dQ^3eL2}^{V}\right)
+2g_{2\times 2,c}\left({2\over3}C_{u^2dQ^3eL1}^{V}-C_{u^2dQ^3eL3}^{V}\right)\;,
\\\nonumber
C_{5L}^{(pn)V}&=
-4\hat g_{2\times 2,a}C_{uQ^5eL}^{V}
+\hat g_{2\times 2,d}\left(C_{u^3d^2QeL1}^{V}-{1\over3}C_{u^3d^2QeL2}^{V}+C_{u^3d^2QeL3}^{V}\right)
\\\nonumber 
&+2\hat g_{2\times 2,b}\left({1\over3}C_{u^2dQ^3eL1}^{V}-C_{u^2dQ^3eL2}^{V}\right)
-2\hat g_{2\times 2,c}\left({2\over3}C_{u^2dQ^3eL1}^{V}-C_{u^2dQ^3eL3}^{V}\right)\;,
\\
&-{2\over 3}g_{2\times 4,a}\left({6\over5}C_{u^3d^2QeL1}^{V}-C_{u^3d^2QeL2}^{V}\right)
+{4\over 3}g_{2\times 4,b}C_{u^2dQ^3eL1}^{V}\;,
\\\nonumber
C^{(pn)T}
&=-16g_{1\times 1,a}^rC_{Q^6L^2}^{T,(S)}
-8g_{1\times 1,b}^rC_{udQ^4L^22}^{T,(S)}
-4g_{1\times 1,c}^r\left(C_{u^2d^2Q^2L^21}^{T,(S)}+C_{u^2d^2Q^2L^23}^{T,(S)}\right)
\\
&-2g_{3\times 3,a}^rC_{udQ^4L^21}^{T,(A)}
-g_{3\times 3,b}^rC_{u^2d^2Q^2L^22}^{T,(A)}\;,
\\
C_L^{(nn)S}
&=4g_{3\times1,a}C_{Q^6L^2}^{S,(S)}+2g_{3\times1,b}C_{udQ^4L^22}^{S,(S)}+g_{3\times1,c}\left(C_{u^2d^2Q^2L^21}^{S,(S)}+C_{u^2d^2Q^2L^23}^{S,(S)}\right)
+{g_{3\times5}\over 6}C_{u^2d^2Q^2L^21}^{S,(S)}\;,
\\
C_{5L}^{(nn)S}&=C_L^{(nn)S}|_{g\to \hat g}\;.
\label{h20}
\end{align}}

\end{appendices}

\bibliography{refs}

\end{document}